\documentclass{article}
\usepackage{geometry}
\usepackage{graphicx}
\usepackage{hyperref}
\usepackage{subcaption}
\usepackage[numbers,comma,square,sort&compress]{natbib}
\graphicspath{{cll_transverse_images/}}
\begin{document}

\begin{center}
\LARGE Size and solvent effects on cellulose transverse\\
anisotropy and toughening design\\
\end{center}

\begin{center}
\renewcommand{\thefootnote}{\fnsymbol{footnote}}
Xu Dong$^{1,}$\footnote{Corresponding author, Email: donx@zuaa.zju.edu.cn}\\
$^{1}$\textit{Department of Engineering Mechanics, Zhejiang University, Hangzhou 310027, China}
\end{center}

\begin{center}
\end{center}

\section*{Abstract}
\addcontentsline{toc}{section}{\protect\numberline{}Abstract}\indent

Cellulose nanocrystals (CNCs) are a promising class of materials derived from the most abundant natural polymer resource on Earth.
Hydroxyl-induced polarity is a crucial advantage of CNCs, making them promising for advanced design and application.
Side chain hydroxyls, hydrogen bonds, and particular crystal structures of CNCs lead to unique anisotropy.
However, the nuanced anisotropy in the transverse section is not sufficiently stressed, and cannot be precisely described experimentally.
Although partially covered by previous studies, a quantitative explanation of size dependency and a systematic comparison of solvent influences are still lacking.
The manufacturing of cellulose materials requires a better understanding of anisotropy, size dependency, and solvent influences.
In this study, the anisotropic performance of CNCs in characteristic directions and a diverse array of solvent environments was carefully inspected and compared using molecular simulations.
Furthermore, a data-supported explanation for the size dependency, and transverse arrangement toughness-enhanced designs were both proposed.
These systematic comparisons and unique transverse arrangements could aid future applications of cellulose.

\section*{Introduction}
\addtocounter{section}{1}
\setcounter{subsection}{0}
\addcontentsline{toc}{section}{\protect\numberline{}Introduction}\indent

\subsection{Cellulose nanomaterials}\indent

Cellulose materials with large reserves\cite{wang2016recent}, low cost\cite{kim2015review}, easy accessibility\cite{moon2016overview}, environmental friendliness\cite{teo2020towards}, recyclability\cite{teo2020towards}, and biodegradability\cite{pandey2021pharmaceutical}, possess excellent biocompatibility\cite{pandey2021pharmaceutical} and mechanical properties\cite{moon2011cellulose}.
This widely available material has received extensive attention and has the potential to be applied in numerous fields\cite{wang2016recent,kim2015review,moon2011cellulose},
such as textiles\cite{felgueiras2021trends}, biopharmaceuticals\cite{du2019cellulose}, energy storage\cite{ruan2018carbonized,liu2021cellulose}, hydrogen storage\cite{dhar2018cellulose,conte2024tuning}, and smart materials\cite{zhu2020stimuli}.
For all applications, the mechanical properties play a crucial role.

Through more precise microscopic control, cellulose nanomaterials (including cellulose nanofibers\cite{liu2022cellulose} and nanocrystals\cite{kim2015review,rana2021cellulose}) possess excellent properties, such as high elastic modulus\cite{moon2011cellulose,nishiyama2009structure}, high transparency\cite{zhu2013transparent}, and low thermal expansion coefficient\cite{okahisa2009optically}, which traditional cellulose materials cannot achieve\cite{liu2022cellulose,huang2024flexible}.
Based on their superior performance, advanced materials with high strength, toughness\cite{chen2020exploring}, transparency\cite{cebrian2022development}, and sensing capabilities\cite{teodoro2021review} can be developed.
Between these two types, cellulose nanocrystals (CNCs), which exhibit a highly rigid three-dimensional structure and more potential for various application forms\cite{shojaeiarani2021cellulose}, are the focus of this study.

One significant advantage of cellulose nanomaterials is the ability of their abundant hydroxyl groups to form hydrogen bonds and interact with polar particles without chemical modification\cite{zhang2021hydrogen}.
This provides unique advantages for the development of composite materials, smart materials, and other advanced application fields\cite{kim2015review}.
The resulting materials possess special properties, such as antibacterial properties, conductivity, and sensing capabilities\cite{zhu2021highly,jiao2021highly}.
Sensors and biosensors based on cellulose nanomaterials (sensing pH\cite{wang2022active,saidi2017poly}, humidity\cite{wang2020flexible}, temperature\cite{yin2020flexible,kim2006discovery}, electromagnetic signals\cite{xie2020flexible,jiao2021highly}, etc.) are key frontiers of researcher focus\cite{hamidon2022cellulose}.
Cellulose nanomaterials fabricated using specific processes can be used for organic light-emitting diodes (OLEDs)\cite{okahisa2009optically}, battery separators, electrodes\cite{chen2018nanocellulose,wang2017cellulose}, and drug delivery\cite{zhang2011thermo}.
Owing to their excellent mechanical properties, cellulose nanomaterials are suitable reinforcing substrates for composite materials\cite{shojaeiarani2021cellulose,xie2022hydroxyl,zhang2020thermally}.
In addition, methods such as cross-linking\cite{tu2021superior} and alignment\cite{chen2020electric,li2021ionic} can be used to optimize the mechanical properties of cellulose nanomaterials\cite{zhang2019non,ray2021mechanics}.

\subsection{Cellulose nanocrystals: anisotropy and hydrogen bonds}\indent

CNCs typically exhibit anisotropic structures and mechanical properties\cite{kim2015review,moon2011cellulose}.
Previous studies have used X-ray diffraction\cite{nishiyama2002crystal} and nuclear magnetic resonance (NMR) spectroscopy\cite{atalla1984native} to measure the structure of CNCs and found that cellulose elementary fibrils have a segmented structure in the fiber direction, with crystalline and non-crystalline (amorphous) parts arranged sequentially\cite{nishiyama2009structure,mazeau2003molecular}, as shown in Figure \ref{fig:arrangement_patterns_and_characteristic_directions}.
Because of the flattened structure of cellulose monomers and hydrogen bonding between side chain hydroxyl groups, CNCs possess very high strength and elastic modulus\cite{moon2011cellulose}, whereas the amorphous region possesses important viscoelastic characteristics\cite{mazeau2003molecular}.
The major interaction type of cellulose along the chain direction is covalent bonds, and the dominant interaction for intra and inter layers in the laminar cross section are hydrogen bonds and van der Waals respectively\cite{wohlert2022cellulose}, leading to significant anisotropic mechanical properties.
Specifically, the bond energy of a covalent bond is approximately 300 $\rm{kJ/mol}$; and that of the hydrogen bond and van der Waals interaction are 10-40~$\rm{kJ/mol}$ and less than 10~$\rm{kJ/mol}$.
As a result, the axial modulus reaches 100-200~GPa and the transverse modulus can also reach 5-100~GPa\cite{moon2016overview}.
Owing to the higher strength of hydrogen bonding than van der Waals interactions, CNCs also possess significant anisotropy within the cross section.

There are various CNC configurations with different structures and properties, including I$\alpha$, I$\beta$, II, III, and IV configurations.
The most common type is Type I$\beta$, also known as natural cellulose\cite{kim2015review} and is widely available in plants.
When CNCs are mentioned in this study, I$\beta$-type CNCs are implied by default.
Researchers have measured the geometric data of different configurations of cellulose crystals\cite{nishiyama2002crystal,nishiyama2003crystal,langan1999revised} and noted that different types of cellulose crystals can be transformed into each other\cite{chundawat2011restructuring,el2011crystal,kobayashi2011crystal,miyamoto2015molecular,kugo2024elucidating}.
To accelerate dissociation and regulate product performance, researchers have tried using solutions\cite{chundawat2011restructuring,el2011crystal,kobayashi2011crystal,uto2019molecular} and loading\cite{chen2018ialpha} methods to transform Type I$\beta$ into other configurations to promote dissociation.

This study focused on the I$\beta$ and II type CNCs, which are the most abundant types in nature (Figure \ref{fig:arrangement_patterns_and_characteristic_directions}).
Because of insufficient experimental resolution, the molecular dynamics simulation approach is important for analyzing the microscopic mechanism of CNCs such as interfacial sliding\cite{wu2013atomistic}, loading direction\cite{zhang2021hydrogen,wu2014tensile}, hydrogen bonds\cite{zhang2021hydrogen}, and the impact of size-dependent edge effects\cite{sinko2014dimensions}.
Some molecular simulations have focused on CNCs and their interactions with other molecules, such as the wetting effects of water and organic molecules on the two surfaces of CNCs (with or without hydroxyl exposure)\cite{malaspina2019molecular},
the impact of the dehydration process on the superstructure and mechanics of water-swollen CNCs\cite{ogawa2020drying}, and the effects of solvents such as water\cite{miyamoto2015molecular,matthews2006computer,maurer2013molecular}.
Hydroxyl groups and hydrogen bonds play crucial roles in these phenomena.

Some previous studies are specifically related.
Diddens et al. used X-ray diffraction to measure the elastic modulus of I$\beta$ cellulose along the chain direction and perpendicular to the chain direction.
They found that the elastic modulus along the chain direction can reach 220~GPa, whereas the elastic modulus in the cross section is 15~GPa\cite{diddens2008anisotropic}.
Zhang et al. systematically simulated and detected the effects of the loading direction, interfacial humidity, and non-aligned structures on the mechanical behavior of the nanocrystal interface, and found a significant positive correlation between the hydrogen bond density and interaction energy\cite{zhang2021hydrogen}.
Based on molecular dynamics simulations, Sinko et al. concluded that width-dependent edge effects have a key influence on fracture energy.
The reduction in the fracture energy by edge effects, particularly for small crystal blocks, weakens as the size increases and can be ignored after the critical width.
However, the collective effect of cellulose stacking and stability caused by van der Waals interactions saturates at the critical crystal thickness.
Furthermore, they constructed an analytical relationship based on a physical model to predict fracture energy.
They found that CNCs reached the optimal fracture energy when the thickness was 4.8-5.6~nm (6-7 layers) and the width was 6.2-7.3~nm (6-7 cellulose chains), which are the dimensions of CNCs found in nature\cite{sinko2014dimensions}.
Wu et al. simulated the inter layer frictional sliding phenomenon between two layers of cellulose under different stretch speeds, normal stresses, and plane-to-plane angles, and found that the number of hydrogen bonds and inter layer distance are key factors affecting friction\cite{wu2013atomistic}.
Wu et al. also conducted research on the anisotropy of CNCs by simulating tension under different strain rates in three orthogonal directions, and found that mechanical properties such as the elastic modulus and Poisson's ratio are highly anisotropic and independent of the strain rate\cite{wu2014tensile}.

On the other hand, water and other solvents widely coexist with cellulose in nature\cite{kim2015review}, and play pivotal roles in the dissociation, extraction, regeneration, and performance of final products\cite{wang2020highly,he2021effects,swatloski2002dissolution,peng2021cellulose,zhang2019non,liu2009supramolecular}.
Although CNCs have been widely studied and applied, studies on the mechanical properties and behaviors of CNCs with different configurations and arrangements remain insufficient.
Systematic research on the mechanical properties of configurations, size-dependence, and a wide series of solvents for CNCs is still required for further structural design and utilization.

\section*{Methods}
\addtocounter{section}{1}
\setcounter{subsection}{0}
\addcontentsline{toc}{section}{\protect\numberline{}Methods}\indent

\subsection{Simulation setup}\indent

\begin{figure}[htbp]
    \centering
    \begin{subfigure}[b]{0.48\textwidth}
        \includegraphics[width=0.96\textwidth]{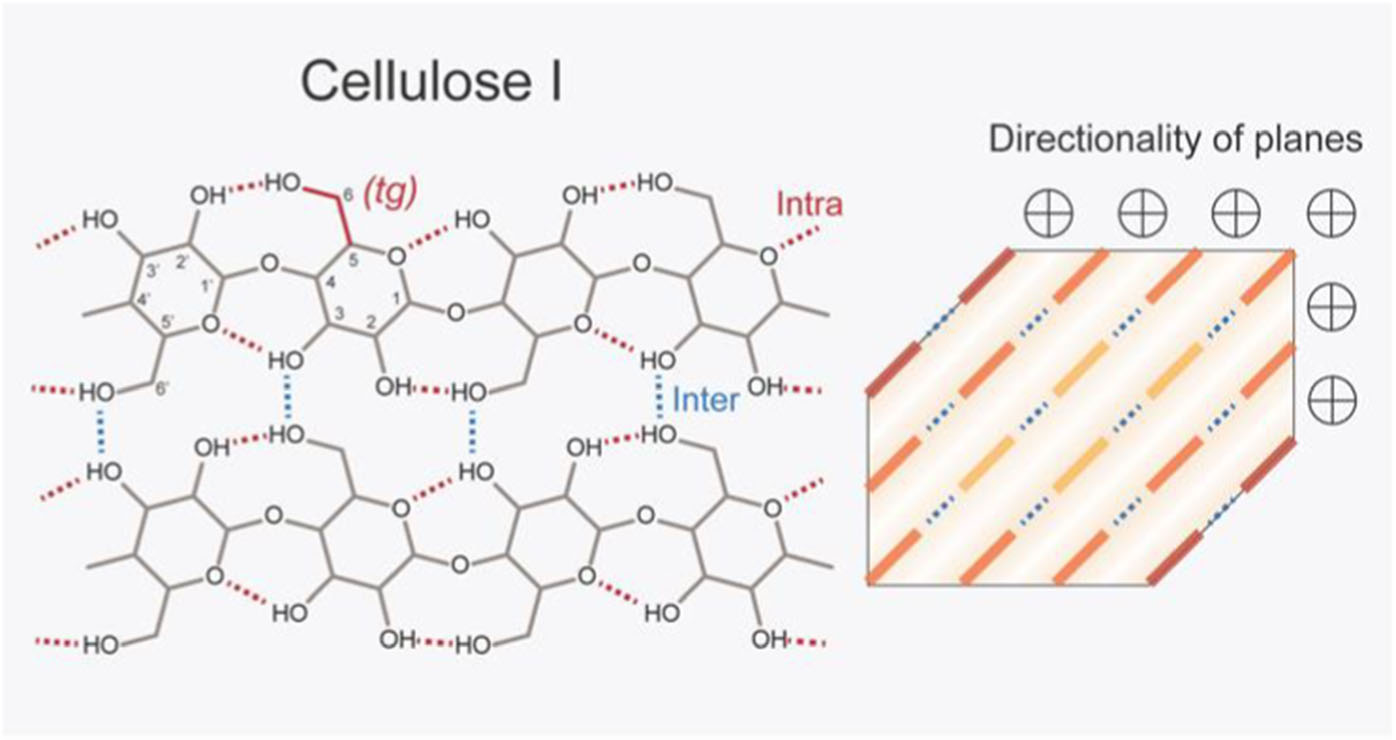}
        \subcaption{}
    \end{subfigure}
    \begin{subfigure}[b]{0.48\textwidth}
        \includegraphics[width=0.96\textwidth]{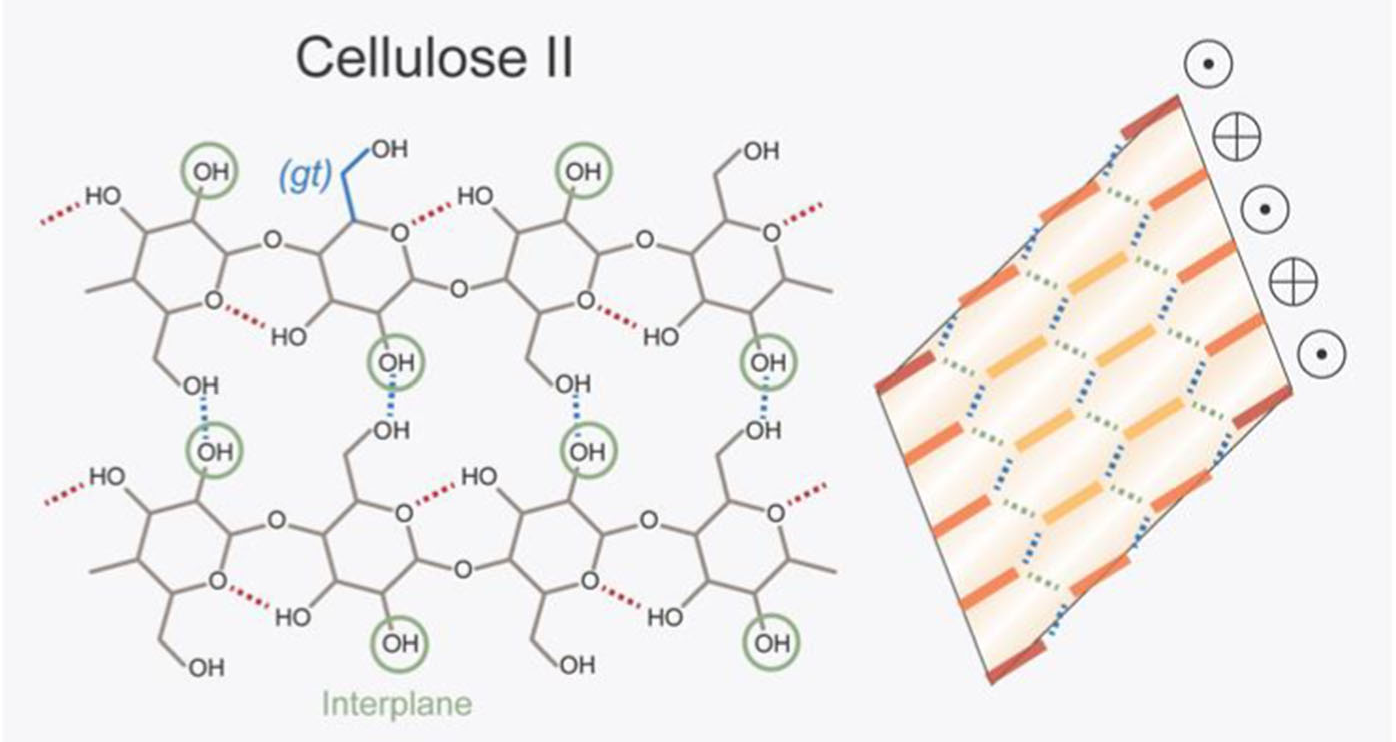}
        \subcaption{}
    \end{subfigure}
    \\
    \begin{subfigure}[b]{0.28\textwidth}
        \includegraphics[width=0.96\textwidth]{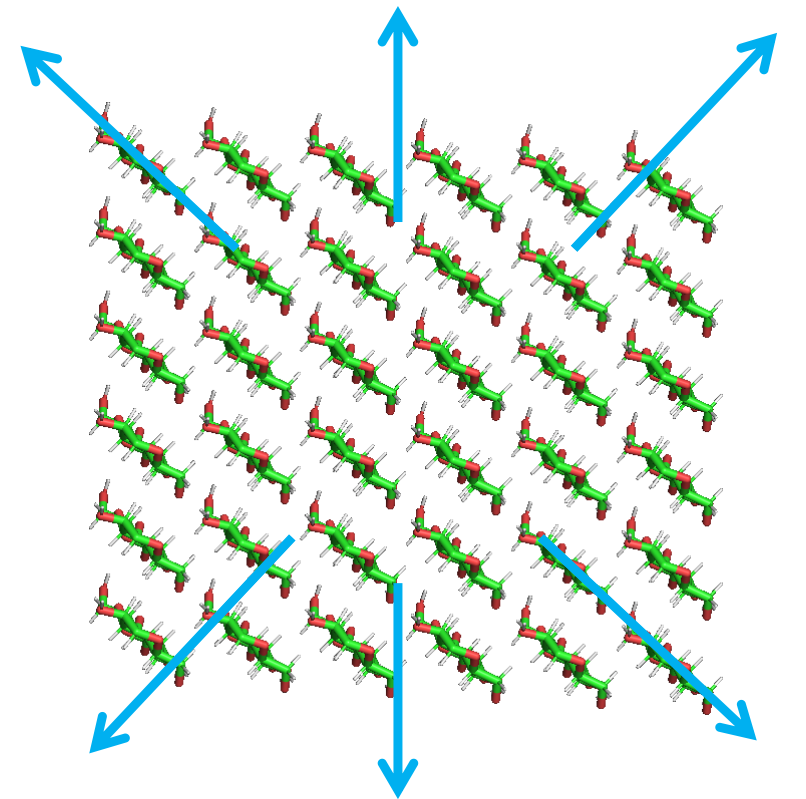}
        \subcaption{}
    \end{subfigure}
    \hspace{0.20\textwidth}
    \begin{subfigure}[b]{0.28\textwidth}
        \includegraphics[width=0.96\textwidth]{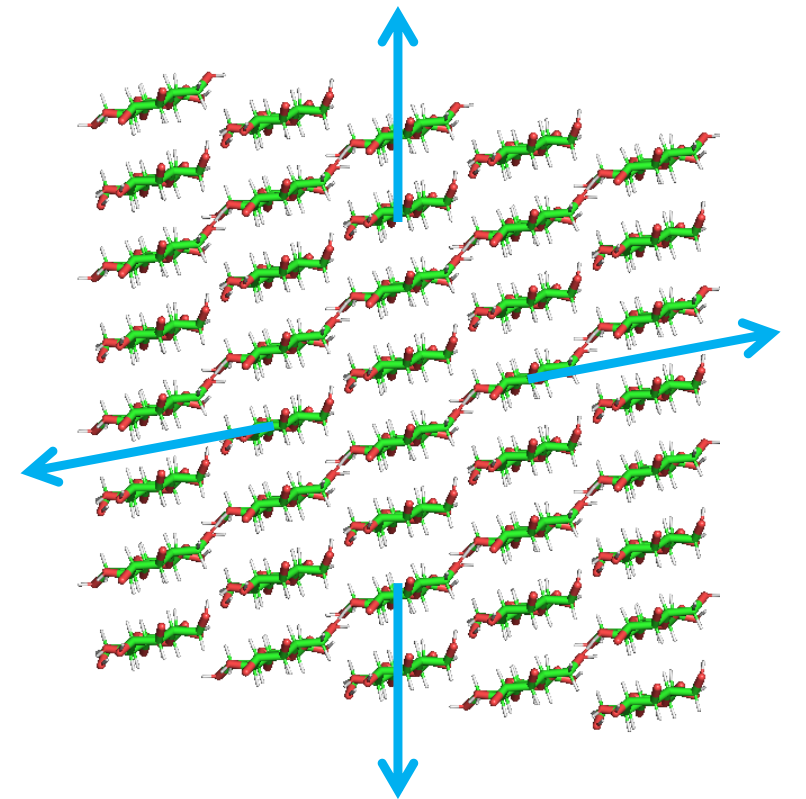}
        \subcaption{}
    \end{subfigure}
    \captionsetup{font=scriptsize}
    \caption{
    Arrangement patterns\cite{wohlert2022cellulose} and characteristic directions of CNCs.
    (a) Arrangement pattern of the Type I and (b) Type II CNCs\cite{wohlert2022cellulose}.
    This study focused on the two most common arrangement configurations, Type I and Type II, where the Type I nanocrystals are widely present in plants and are also known as natural cellulose crystals.
    These differences between the two arrangements lie in the direction and arrangement interval of the side chain hydroxyl groups, where the Type I is tg only, and Type II is an alternating arrangement of tg and gt.
    The gt type, whose side chain orientation is opposite to that of the tg type, stresses more on inter chain hydrogen bonds and provides fewer intra chain hydrogen bonds.
    Type I nanocrystals have more intra chain hydrogen bonds, whereas Type II possesses more inter chain hydrogen bonds.
    Unless explicitly specified, the CNCs mentioned in this study are Type I($\beta$) nanocrystals.
    (c) Characteristic directions of the Type I and (d) Type II considered in this study.
    Based on the hydrogen bonding patterns of Type I and Type II, three or two characteristic directions are considered in this study.
    The Type I models were named Vertical, Horizontal and Slant; and Type II models were named Vertical and Slant, which are also the directions relative to the hydrogen bonding layers.
    }
    \label{fig:arrangement_patterns_and_characteristic_directions}
\end{figure}

Considering the very high elastic modulus of covalent bonds along the chain and the technical difficulty in applying shear loads in molecular dynamics, models were built with different characteristic directions and stretched under the same vertical loads in the cross section.
Three or two characteristic directions of the Type I and Type II CNCs were considered in this study, as shown in Figure \ref{fig:arrangement_patterns_and_characteristic_directions}.
For all atom simulations at xyz periodic cells, models under the NPT ensemble at 0.1~MPa with a stretch load in the vertical direction were performed using GROMACS\cite{abraham2015gromacs,bjelkmar2010implementation} and CHARMM36\cite{huang2013charmm36,huang2017charmm36m} force fields (force field files were generated using CHARMM-GUI tools\cite{jo2008charmm,lee2016charmm}).
The full simulation process consisted of three procedures: energy minimization, relaxation, and vertical stretch loading.
During relaxation, positional restraints were applied to the carbon, oxygen, and nitrogen atoms, which were gradually released in four stages.
Vertical stretches were implemented using cell deformation loads at a constant speed 10.0~nm/ns.
The dependence on the stretch speed was validated via 100 replica all atom simulations with stretch speeds ranging from 0.1~nm/ns to 40.0~nm/ns, which is illustrated in the following sections.

\section*{Results}
\addtocounter{section}{1}
\setcounter{subsection}{0}
\addcontentsline{toc}{section}{\protect\numberline{}Results}\indent

\subsection{Anisotropic mechanical properties}\indent

\subsubsection{Cellulose nanocrystals Type I}\indent

According to the experimental data measured by Nishiyama et al.\cite{nishiyama2002crystal}, the three characteristic dimensions of the Type I lattice are a=0.778~nm, b=0.820~nm, c=1.038~nm, and the angle between the horizontal and vertical sides of the lattice is 96.55$^\circ$.
According to the X-ray data measured by Langan et al.\cite{langan2001x}, the three characteristic dimensions of the Type II lattice are a=0.810~nm, b=0.903~nm, c=1.031~nm, and the angle between the transverse and longitudinal sides of the lattice is 117.10$^{\circ}$.
Based on their data, Type I and Type II models were constructed and relaxed for further stretch simulations, as shown in Figure \ref{fig:cellulose_nanocrystal_structure_and_models} and Figure \ref{fig:cellulose_nanocrystal_ii_structure_and_models}.

As shown in Figure \ref{fig:cellulose_nanocrystal_relaxations} and Figure \ref{fig:cellulose_nanocrystal_ii_relaxations}, the structures of the different arrangement models did not undergo significant structural changes after the energy minimization and relaxation processes.
The presented models in these figures were processed for periodicity using the ``nojump'' method, meaning the coordinates of atoms do not change significantly relative to the initial structure by crossing periodic boundaries.
Therefore, the sliding of sheets in the vertical and horizontal models corresponds to actual displacement rather than display issues, which also demonstrates that the van der Waals interaction between sheets of CNCs is weaker than the hydrogen bonding within the sheets.
The root mean square deviation (RMSD) curves during relaxation quantitatively demonstrate the structural stability during the process.

\begin{figure}[htbp]
    \centering
    \scriptsize
    \begin{subfigure}[b]{0.96\textwidth}
        \centering
        \begin{minipage}[b]{0.20\textwidth}
            \centering
            \includegraphics[width=0.96\textwidth]{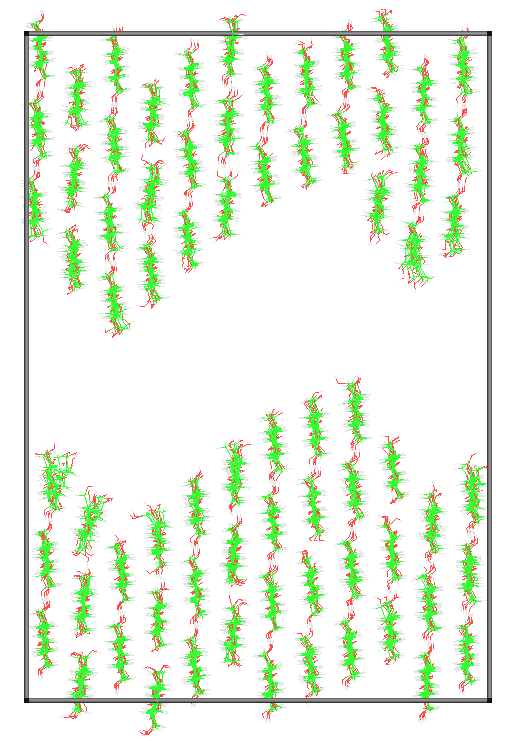}
            \\
            Vertical
        \end{minipage}
        \begin{minipage}[b]{0.25\textwidth}
            \centering
            \includegraphics[width=0.96\textwidth]{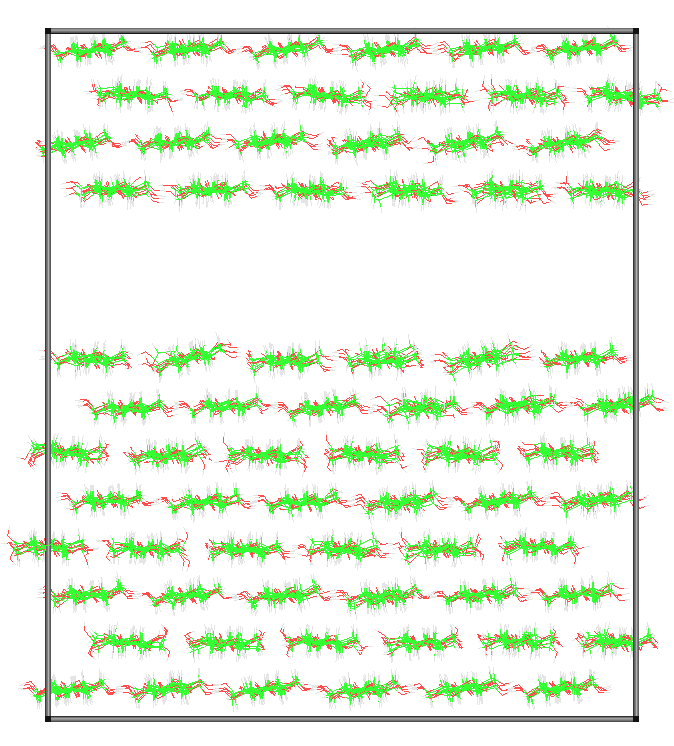}
            \\
            Horizontal
        \end{minipage}
        \begin{minipage}[b]{0.24\textwidth}
            \centering
            \includegraphics[width=0.96\textwidth]{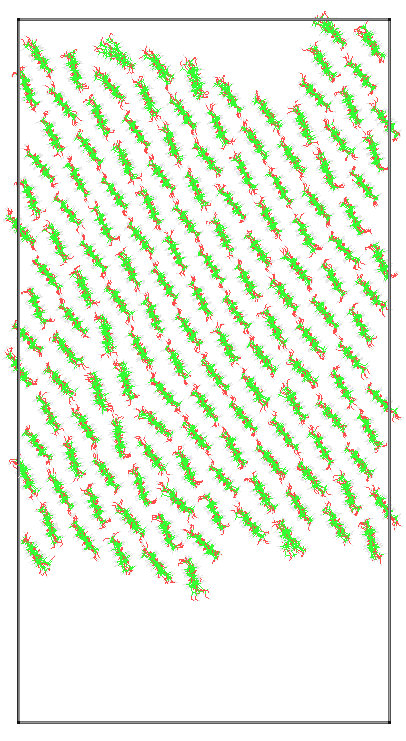}
            \\
            Slant
        \end{minipage}
        \subcaption{}
    \end{subfigure}
    \\
    \begin{subfigure}[b]{0.96\textwidth}
        \includegraphics[width=0.24\textwidth]{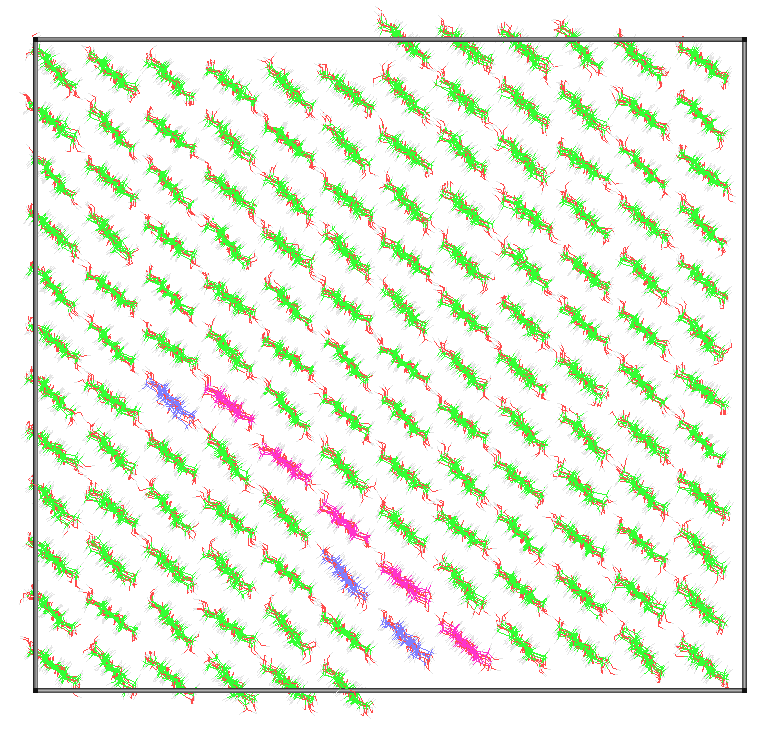}
        \includegraphics[width=0.24\textwidth]{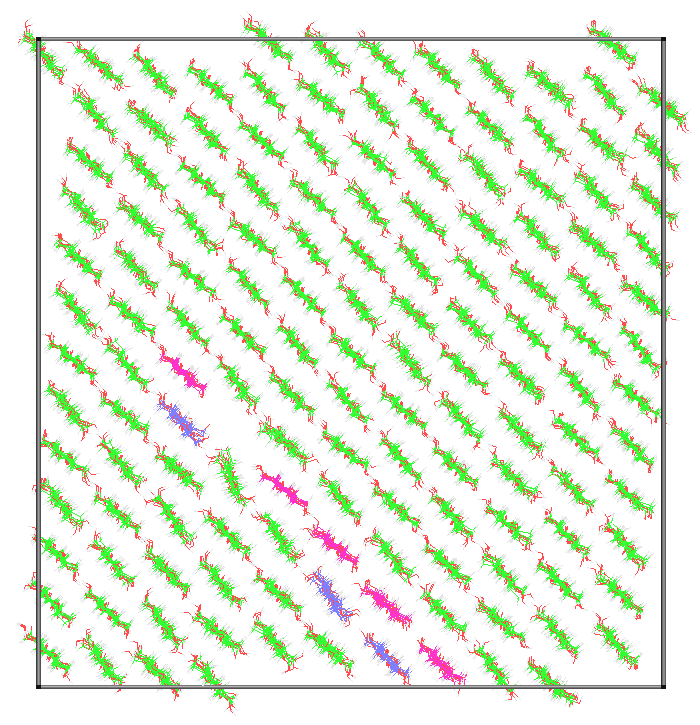}
        \includegraphics[width=0.24\textwidth]{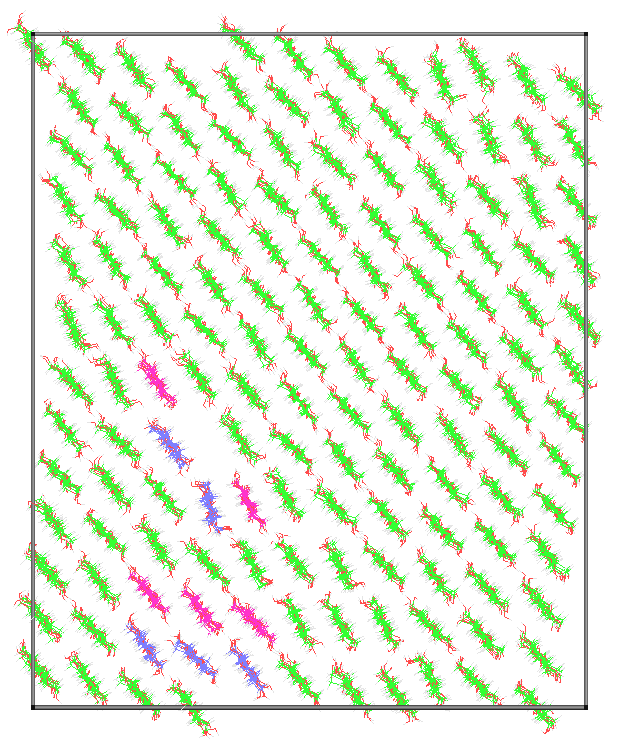}
        \includegraphics[width=0.24\textwidth]{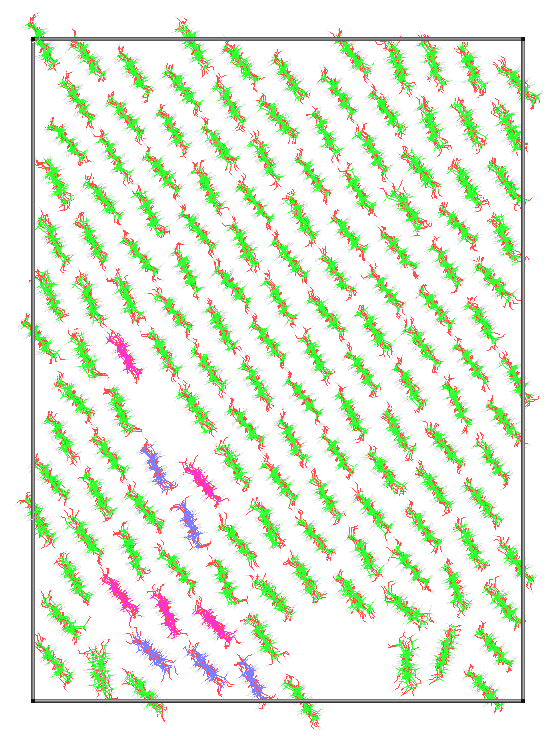}
        \subcaption{}
    \end{subfigure}
    \\
    \begin{subfigure}[b]{0.96\textwidth}
        \centering
        \includegraphics[width=0.32\textwidth]{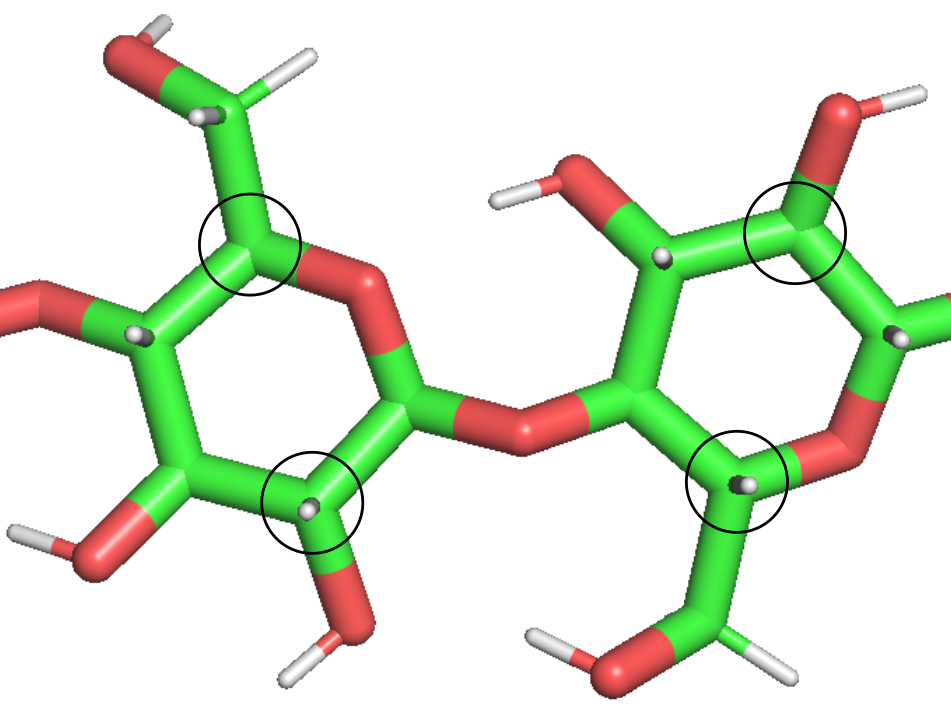}
        \includegraphics[width=0.24\textwidth]{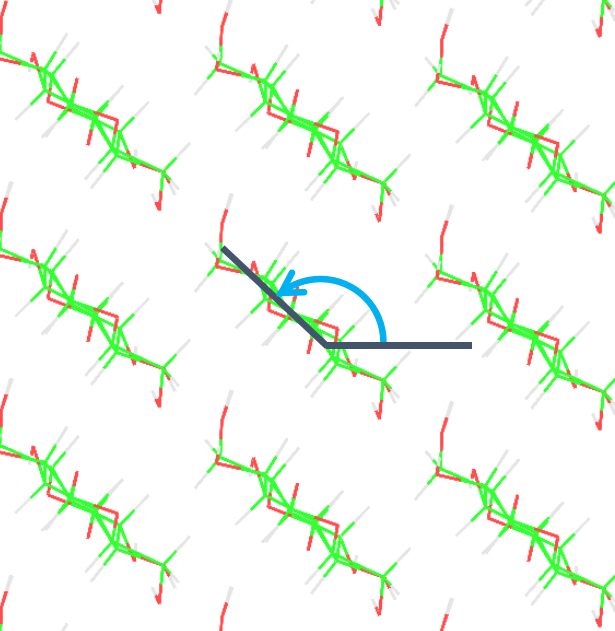}
        \includegraphics[width=0.40\textwidth]{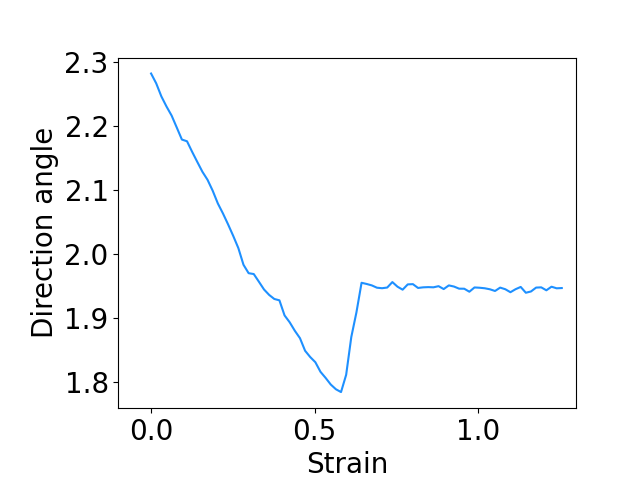}
        \subcaption{}
    \end{subfigure}
    \captionsetup{font=scriptsize}
    \caption{
    Fracture behaviors, structural transitions, and direction angle of the Type I CNCs.
    (a) Fracture behaviors.
    The vertical and horizontal models exhibited brittle fractures, whereas shear failure characterized by inter layer frictional sliding was presented by the slant model.
    The friction and overall rotation also illustrate the different strength of hydrogen bonding within a layer and the van der Waals interactions between layers.
    (b) Key frames of inter layer frictional sliding phenomenon in the slant model.
    The relative inter layer frictional sliding between adjacent layers is an important failure phenomenon in the slant model.
    (c) Definition of direction angle and the direction angle curve during the stretching of the slant model.
    To quantitatively characterize the inter layer frictional sliding and the resulting rotation in the slant model, the direction angle was introduced and defined as the projection angle of the carbon atoms on the ring.
    For the entire model, the direction angle was the average of all residues.
    The direction angle curve effectively illustrates the inter layer frictional sliding and rotation of the model, and the later recovery corresponds to the rebound after fracture.
    }
    \label{fig:cellulose_nanocrystal_fractures}
\end{figure}

As shown in Figure \ref{fig:cellulose_nanocrystal_fractures}, the vertical and horizontal models exhibited the expected brittle fracture behavior, whereas significant relative sliding between layers occurred in the slant model, leading to a rotation of the overall crystal sheets after sliding.
In the subsequent sections, this phenomenon is referred to as inter layer frictional sliding, which is further illustrated in Figure \ref{fig:cellulose_nanocrystal_fractures}.

\begin{figure}[htbp]
    \centering
    \scriptsize
    \begin{subfigure}[b]{0.96\textwidth}
        \centering
        \begin{minipage}[b]{0.32\textwidth}
            \centering
            \includegraphics[width=0.96\textwidth]{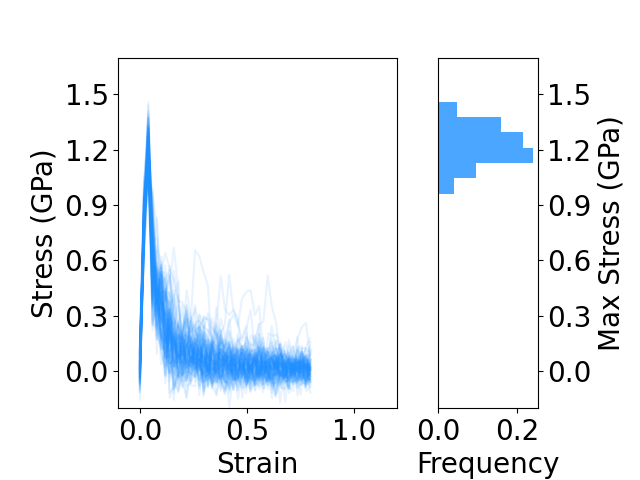}
        \end{minipage}
        \begin{minipage}[b]{0.32\textwidth}
            \centering
            \includegraphics[width=0.96\textwidth]{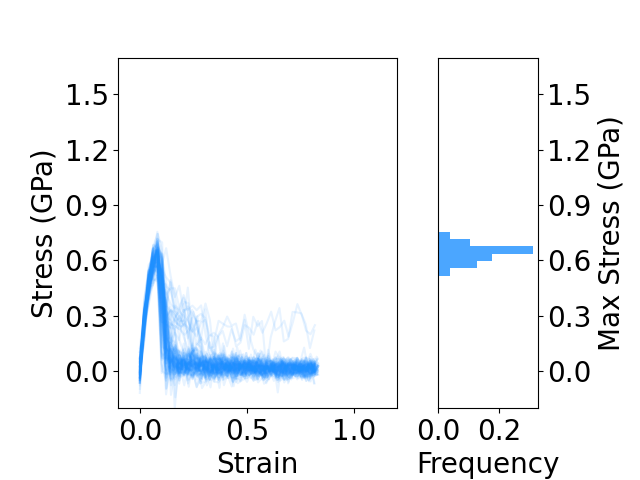}
        \end{minipage}
        \begin{minipage}[b]{0.32\textwidth}
            \centering
            \includegraphics[width=0.96\textwidth]{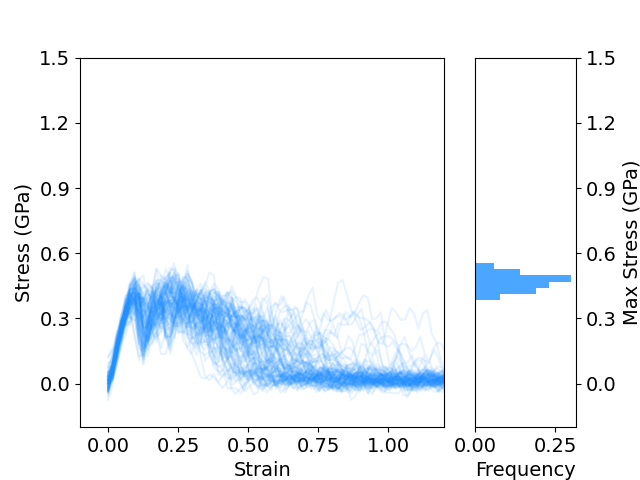}
        \end{minipage}
        \subcaption{}
    \end{subfigure}
    \\
    \begin{subfigure}[b]{0.96\textwidth}
        \centering
        \begin{minipage}[b]{0.32\textwidth}
            \centering
            \includegraphics[width=0.96\textwidth]{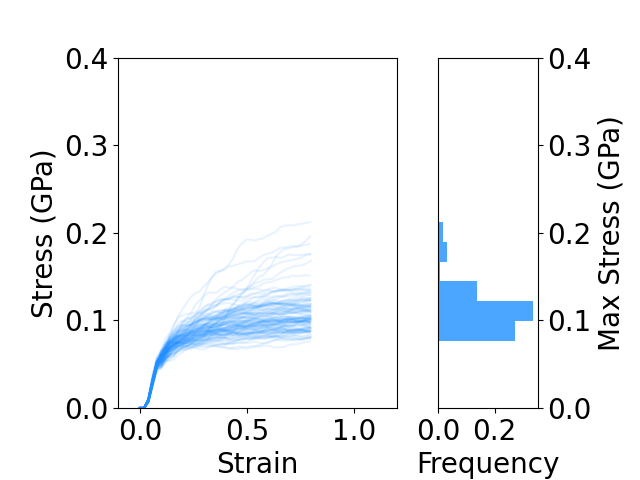}
            \\
            Vertical
        \end{minipage}
        \begin{minipage}[b]{0.32\textwidth}
            \centering
            \includegraphics[width=0.96\textwidth]{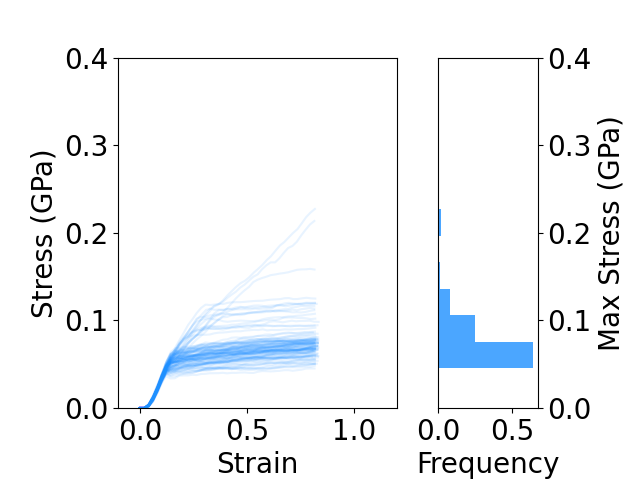}
            \\
            Horizontal
        \end{minipage}
        \begin{minipage}[b]{0.32\textwidth}
            \centering
            \includegraphics[width=0.96\textwidth]{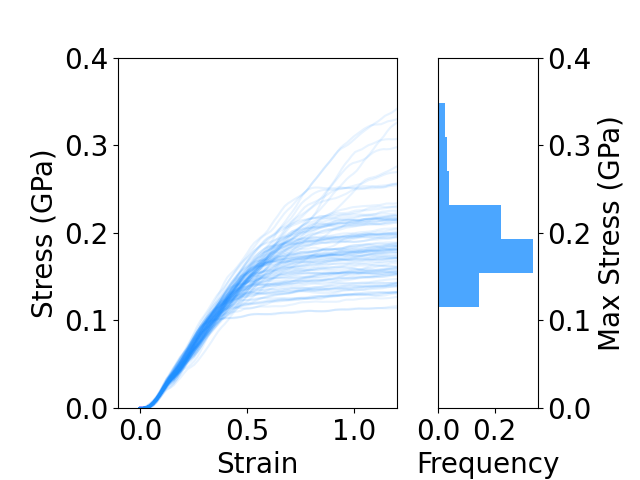}
            \\
            Slant
        \end{minipage}
        \subcaption{}
    \end{subfigure}
    \\
    \begin{subfigure}[b]{0.96\textwidth}
        \centering
        \includegraphics[width=0.48\textwidth]{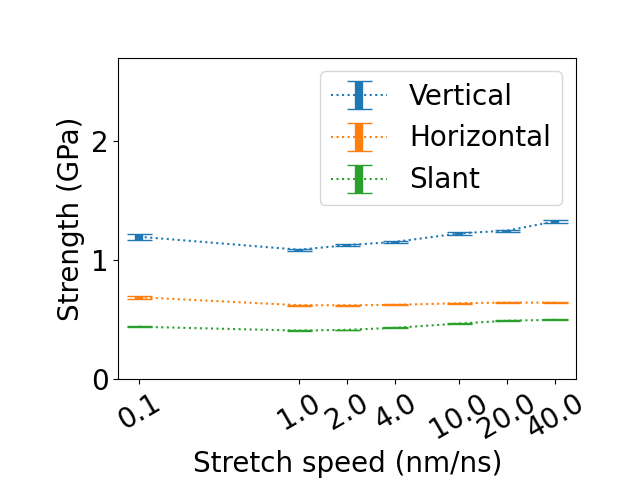}
        \includegraphics[width=0.48\textwidth]{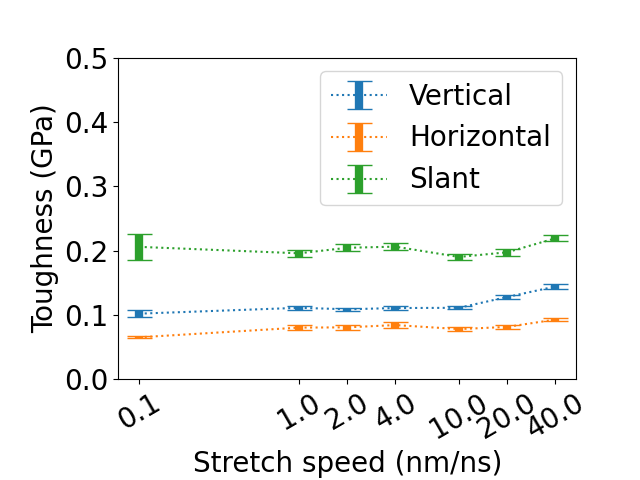}
        \subcaption{}
    \end{subfigure}
    \captionsetup{font=scriptsize}
    \caption{
    Stress curves and deviations of the Type I CNCs mechanical properties under varying stretch speeds.
    (a) Stress-strain curves and corresponding maximum stress distributions for the vertical, horizontal, and slant models under 100 replica simulations at 10.0~nm/ns.
    (b) Strain energy curves and corresponding maximum strain energy distributions for the vertical, horizontal, and slant models under 100 replica simulations at 10.0~nm/ns.
    (c) Mechanical properties of the vertical, horizontal, and slant models with respect to stretch speed.
    Each data point represents the mean value plus or minus one standard error.
    Except for the 0.1~nm/ns which was repeated 10 times, all other stretch speed data were the results of 100 replica simulations with distributions similar to those at 10.0~nm/ns.
    These data indicate that the mechanical properties of the Type I CNCs do not change significantly with stretch speed within the considered range.
    }
    \label{fig:cellulose_nanocrystal_stretch_collection_and_speed_dependency}
\end{figure}

The specific data for stress, strain energy, hydrogen bond number, and potential energy during the stretching of the vertical, horizontal, and slant models are shown in Figure \ref{fig:cellulose_nanocrystal_stretch_collection_and_speed_dependency} and Figure \ref{fig:cellulose_nanocrystal_stretch_hydrogen_bond_number_and_energy}.
From the perspectives of strength (maximum stress) and strain energy (maximum strain energy), the simulation data agree with the strength of hydrogen bonds and van der Waals interactions.
The strength of the vertical model, in which hydrogen bonds played a dominant role, also corresponded to the highest strength among the three arrangements.
Meanwhile, the slant arrangement model, which is affected by both hydrogen bonding and van der Waals interactions, exhibits a unique inter layer frictional sliding phenomenon, resulting in the highest toughness and ductility despite its lower strength.
During the stretching of the three models, the number of hydrogen bonds in the horizontal model did not change significantly (the number of hydrogen bonds shown is relative to each minimum for easier comparison),
whereas a significant decrease in hydrogen bonds was observed in the vertical and slant models involving hydrogen bonding layer fractures, reflecting the tight correlation between hydrogen bonds and mechanical behaviors.
However, the number of hydrogen bonds in the slant model gradually recovered after the decrease, which was related to the reconstruction of the hydrogen bond network after the inter layer frictional sliding of the slant model.
This also explains the higher toughness of the slant model from another perspective.

The dominant role of Coulomb interaction in the vertical models is further illustrated by the potential energy data (the potential energy is also relative to its respective minimum for easier comparison) in Figure \ref{fig:cellulose_nanocrystal_ii_stretch_hydrogen_bond_number_and_energy}.
Comparing the delta of the Coulomb potential energy during stretching, the vertical model was the only case in which the Coulomb potential energy increased.
This means that the Coulomb interaction plays an important role in resisting stretches only in the vertical model.
In addition, an increased van der Waals potential energy was observed in all three models, particularly for the slant model, which is related to its highest toughness.

The transition key frames for the unique inter layer frictional sliding phenomenon in CNCs are shown in Figure \ref{fig:cellulose_nanocrystal_fractures}.
Inter layer frictional sliding in the slant model began with the fracture of one layer of CNCs.
After that, significant sliding on the interface between both sides was observed, and the fractured edges may reconnect with each other after overall rotations.
The direction angle was defined to quantitatively describe the rotation of the layers (Figure \ref{fig:cellulose_nanocrystal_fractures}), as the projection angle on the cross section between the horizontal axis and the two carbon atoms on a residue ring.
Furthermore, the direction angle of a frame is the average of all residues.
The recovery at the end of the direction angle curve, corresponded to the rebound after the final fracture of the slant model.

Subsequently, to reduce the impact of individual cases and simulation randomness and to consider the effect of stretch speed, 100 replica simulations were performed for all three models at varying stretch speeds.
The stretch speeds considered include 0.1, 1.0, 2.0, 4.0, 10.0, 20.0, and 40.0~nm/ns, and the replica number of simulations at 0.1~nm/ns was 10 owing to the computational expenses.
Although the stretch speed of 0.1~nm/ns remains significantly higher than the experimental strain rates, this tested range spans three orders of magnitude.
Slower speeds are inaccessible owing to computational resource constraints.
Nevertheless, the consistent trends suggest that the structural mechanics are largely independent of the loading rate within this speed range.
As shown by the stress-strain curves and strain energy curves of 100 replica stretches at 10.0~nm/ns (Figure \ref{fig:cellulose_nanocrystal_stretch_collection_and_speed_dependency}), the mechanical performance of the CNCs in the three characteristic directions was stable.
The brittleness of the vertical and horizontal models and toughness of the slant model were confirmed through replica simulations.
After considering all the strength and toughness performances across varying stretch speeds, it can be confirmed that the mechanical properties of the Type I CNCs are stable within the considered range of stretch speeds.

\subsubsection{Cellulose nanocrystals Type II}\indent

Another widely available CNC configuration with better thermal stability is Type II\cite{malaspina2019molecular}.
In addition to the lattice dimensions, the Type II CNCs are significantly different from Type I: the orientation of the cellulose residues.
In Type I, all side chains are tg type pointing in one direction; whereas in Type II, tg and gt types with opposite orientations are alternately connected.
As shown in Figure \ref{fig:cellulose_nanocrystal_ii_structure_and_models}, the side chains of the tg type point to the left, and those of the gt type point to the right.
The core differences between the tg and gt types lie in the orientation of the side chain and the arrangement pattern of hydrogen bonds between the side chain hydroxyl groups; the tg type forms more intra chain hydrogen bonds, whereas the gt type emphasizes more on inter chain hydrogen bonds.
Type II does not have a significant laminar structure and has only two characteristic directions corresponding to the vertical and slant models.

\begin{figure}[htbp]
    \centering
    \scriptsize
    \begin{subfigure}[b]{0.96\textwidth}
        \centering
        \begin{minipage}[b]{0.15\textwidth}
            \centering
            \includegraphics[width=0.96\textwidth]{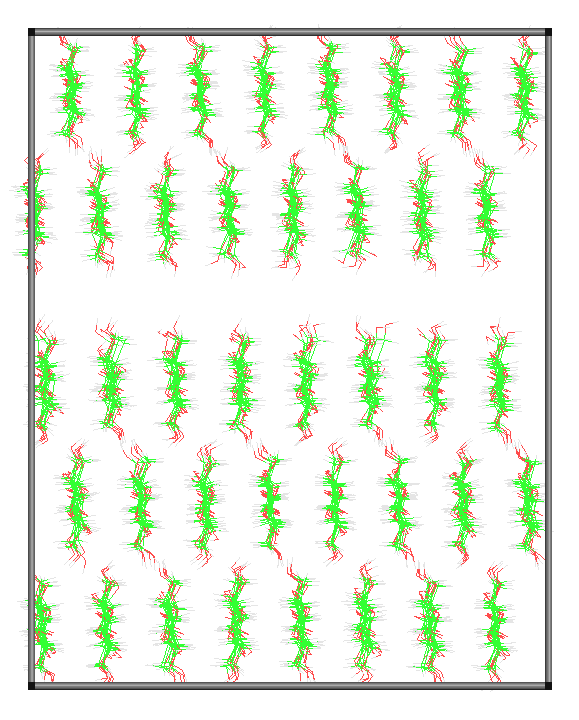}
            \\
            Vertical
        \end{minipage}
        \hspace{0.13\textwidth}
        \begin{minipage}[b]{0.16\textwidth}
            \centering
            \includegraphics[width=0.96\textwidth]{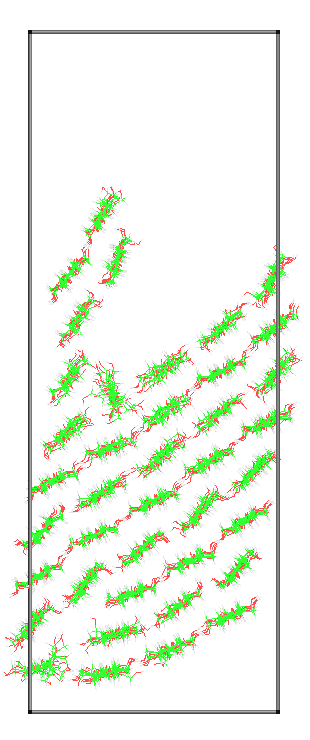}
            \\
            Slant
        \end{minipage}
        \hspace{0.17\textwidth}
        \begin{minipage}[b]{0.12\textwidth}
            \centering
            \includegraphics[width=0.96\textwidth]{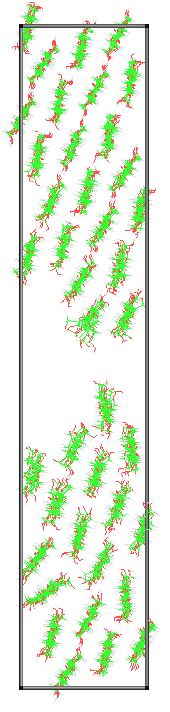}
            \\
            Slant Tough
        \end{minipage}
        \subcaption{}
    \end{subfigure}
    \\
    \begin{subfigure}[b]{0.44\textwidth}
        \centering
        \includegraphics[width=0.24\textwidth]{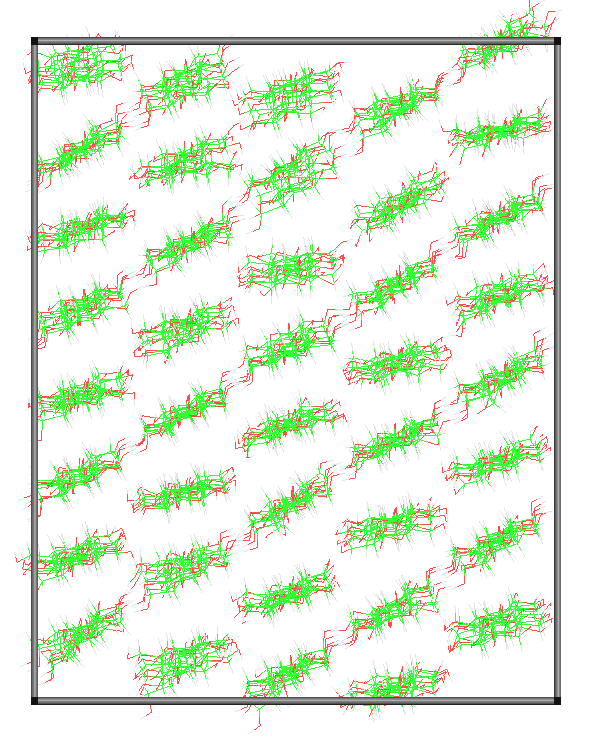}
        \includegraphics[width=0.24\textwidth]{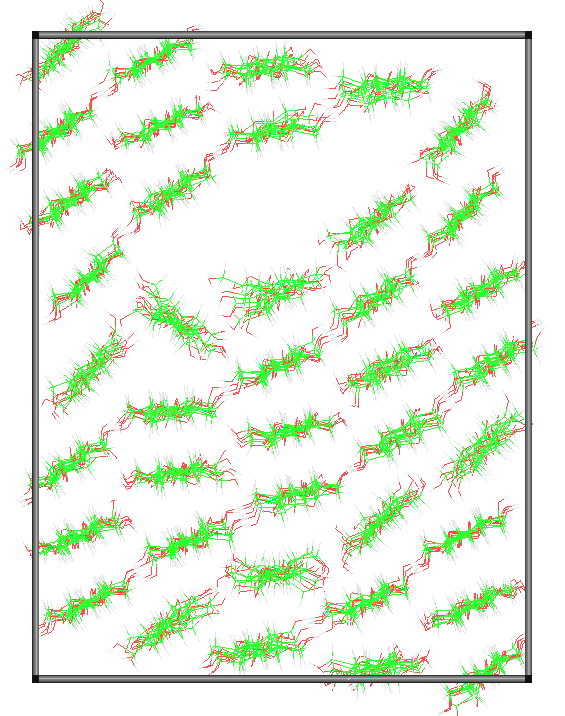}
        \includegraphics[width=0.24\textwidth]{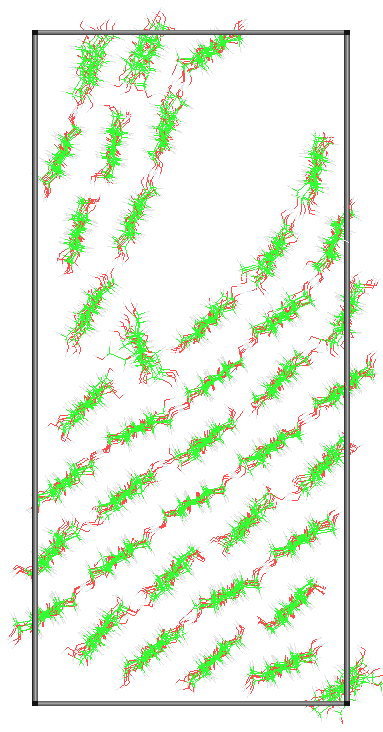}
        \includegraphics[width=0.24\textwidth]{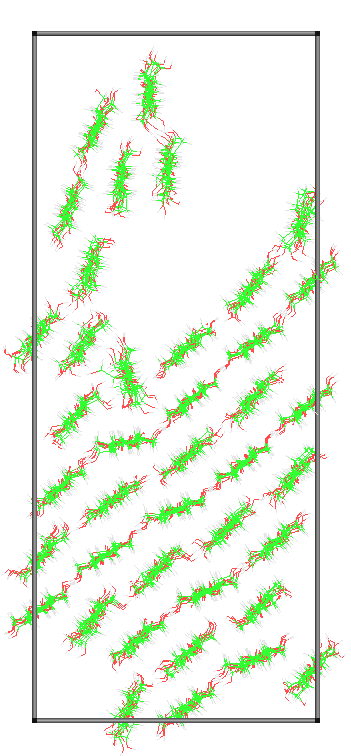}
        \subcaption{}
    \end{subfigure}
    \hspace{0.10\textwidth}
    \begin{subfigure}[b]{0.44\textwidth}
        \centering
        \includegraphics[width=0.24\textwidth]{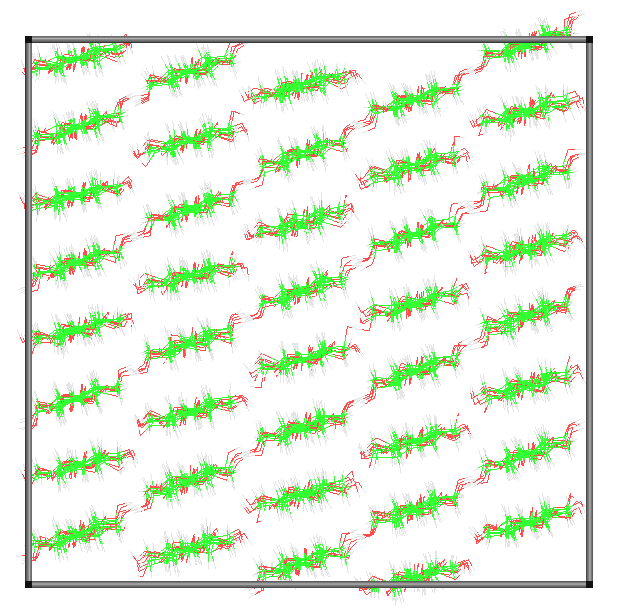}
        \includegraphics[width=0.24\textwidth]{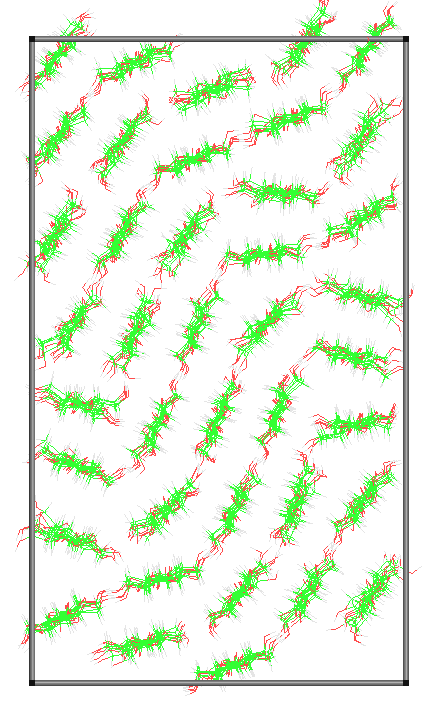}
        \includegraphics[width=0.24\textwidth]{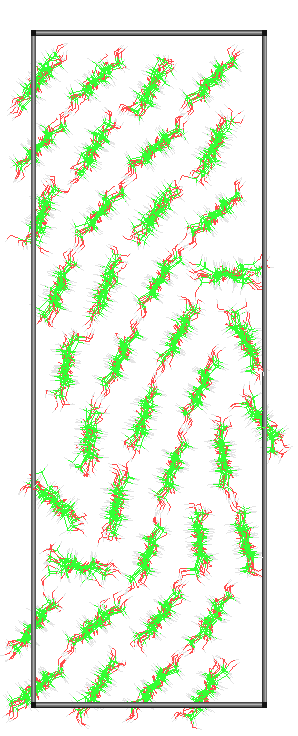}
        \includegraphics[width=0.24\textwidth]{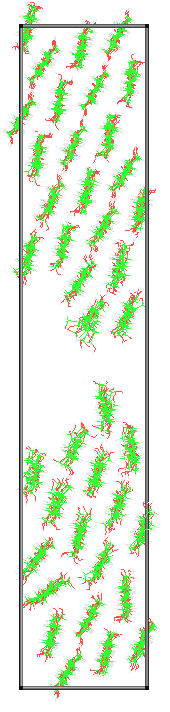}
        \subcaption{}
    \end{subfigure}
    \captionsetup{font=scriptsize}
    \caption{
    Fracture behaviors and structural transitions of the Type II CNCs.
    (a) Fracture behaviors.
    The vertical models exhibited brittle fractures, whereas the slant models may undergo structural reorganization during stretching.
    The slant model may also exhibit brittle fractures or form a laminar structure similar to Type I with a higher toughness (marked as ``Slant Tough'').
    This phenomenon is a configuration restructuring mechanism induced by external loads, which may be helpful for the extraction and material design of CNCs.
    (b) (c) Key frames of the slant models from two typical fracture behaviors.
    In contrast to the stable inter layer frictional sliding behavior of Type I, the slant model of Type II may undergo brittle fracture or restructuring into a laminar arrangement similar to Type I.
    The restructuring induced by mechanical loads with higher toughness corresponded to overall frictions and rotations, rather than localized fractures.
    }
    \label{fig:cellulose_nanocrystal_ii_fractures}
\end{figure}

\begin{figure}[htbp]
    \centering
    \scriptsize
    \begin{subfigure}[b]{0.96\textwidth}
        \centering
        \begin{minipage}[b]{0.32\textwidth}
            \centering
            \includegraphics[width=0.96\textwidth]{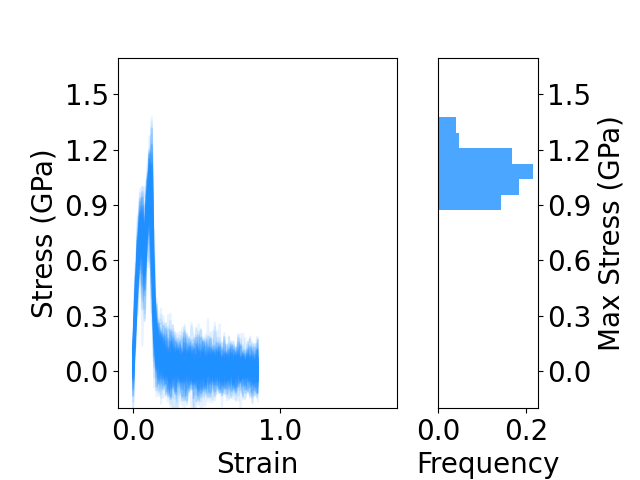}
        \end{minipage}
        \begin{minipage}[b]{0.32\textwidth}
            \centering
            \includegraphics[width=0.96\textwidth]{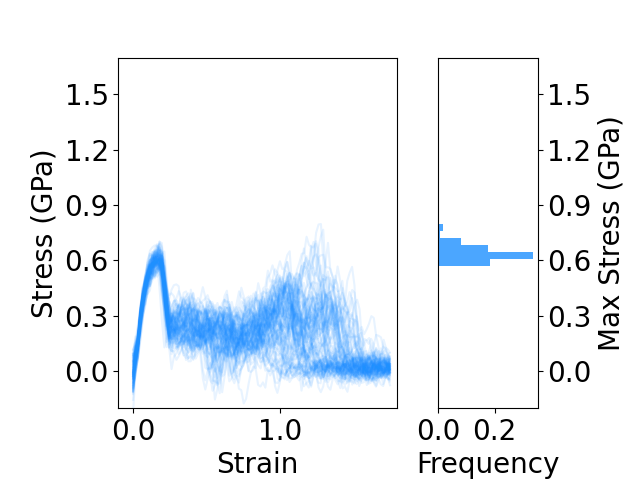}
        \end{minipage}
        \begin{minipage}[b]{0.32\textwidth}
            \centering
            \includegraphics[width=0.96\textwidth]{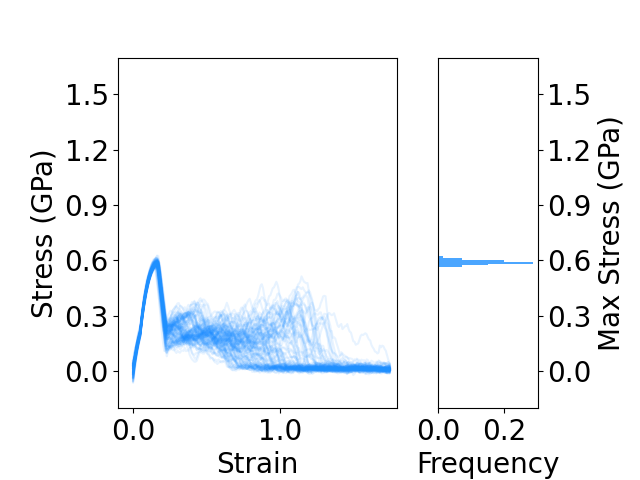}
        \end{minipage}
        \subcaption{}
    \end{subfigure}
    \\
    \begin{subfigure}[b]{0.96\textwidth}
        \centering
        \begin{minipage}[b]{0.32\textwidth}
            \centering
            \includegraphics[width=0.96\textwidth]{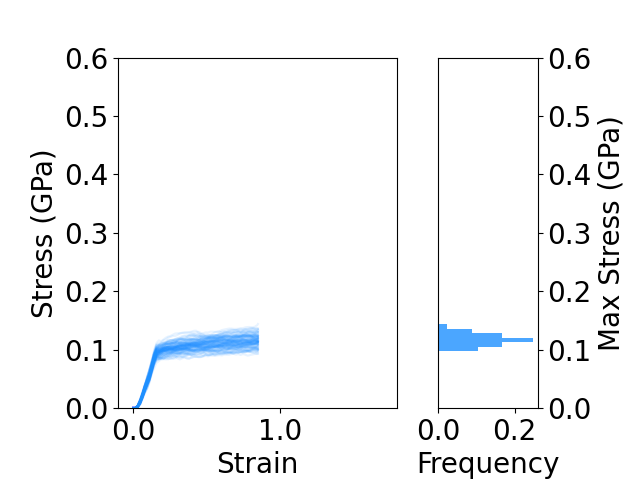}
            \\
            Vertical
        \end{minipage}
        \begin{minipage}[b]{0.32\textwidth}
            \centering
            \includegraphics[width=0.96\textwidth]{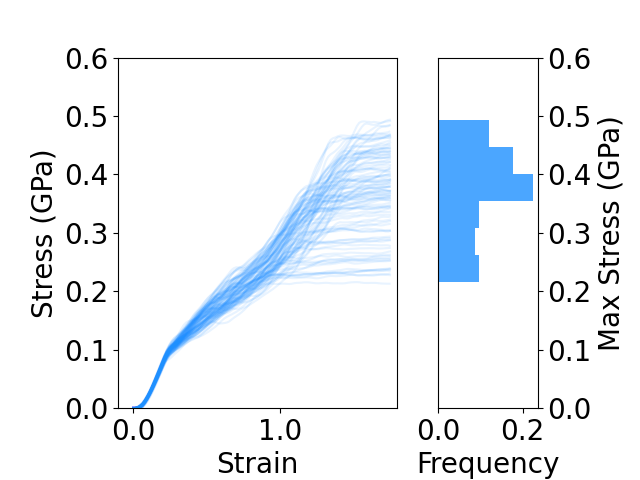}
            \\
            Slant
        \end{minipage}
        \begin{minipage}[b]{0.32\textwidth}
            \centering
            \includegraphics[width=0.96\textwidth]{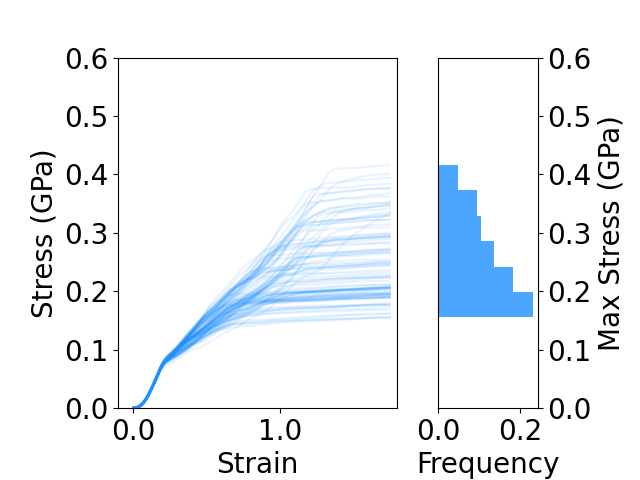}
            \\
            Slant Larger
        \end{minipage}
        \subcaption{}
    \end{subfigure}
    \\
    \begin{subfigure}[b]{0.96\textwidth}
        \includegraphics[width=0.48\textwidth]{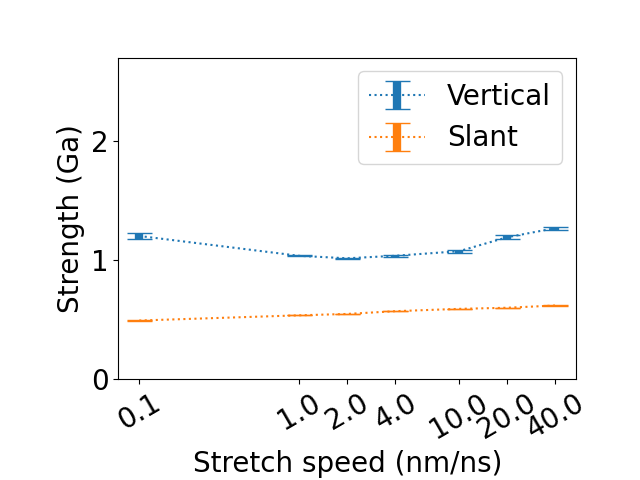}
        \includegraphics[width=0.48\textwidth]{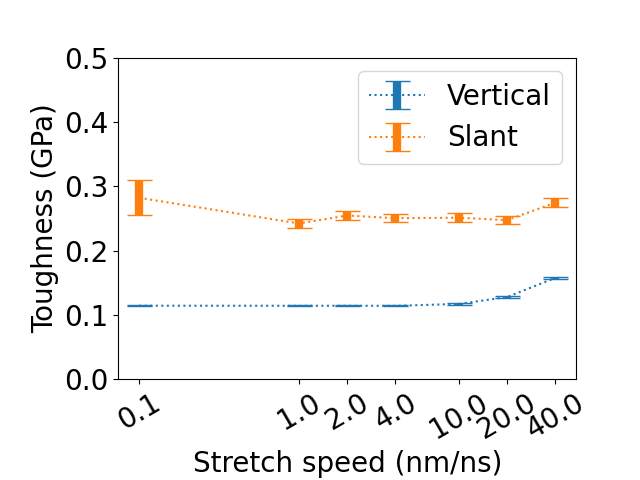}
        \subcaption{}
    \end{subfigure}
    \captionsetup{font=scriptsize}
    \caption{
    Stress curves and deviations of the Type II CNCs mechanical properties under varying stretch speeds.
    (a) Stress-strain curves and corresponding maximum stress distributions for the vertical, slant, and larger slant models under 100 replica simulations at 10.0~nm/ns.
    (b) Strain energy curves and corresponding maximum strain energy distributions for the vertical, slant, and larger slant models under 100 replica simulations at 10.0~nm/ns.
    (c) Mechanical properties of the vertical, slant, and larger size slant models with respect to stretch speed.
    Each data point represents the mean value plus or minus one standard error.
    Considering the deviation of the stress-strain curves of the slant model, simulation data from the larger (double the width and height in the cross section) ware added.
    Except for the 0.1~nm/ns which was repeated 10 times, all other stretch speed data were the results of 100 replica simulations with distributions similar to those at 10.0~nm/ns.
    These data indicate that the mechanical properties of the Type II CNCs do not change significantly with stretch speed within the considered range.
    }
    \label{fig:cellulose_nanocrystal_ii_stretch_collection_and_speed_dependency}
\end{figure}

Type II fracture behaviors (Figure \ref{fig:cellulose_nanocrystal_ii_fractures}) and mechanical property deviations (Figure \ref{fig:cellulose_nanocrystal_ii_stretch_collection_and_speed_dependency}) differed significantly from those of Type I.
The slant model of Type II nanocrystals can exhibit brittle behavior and lower toughness, or they can undergo restructuring, thereby exhibiting higher toughness.
During the stretching process, the structure of Type II cellulose crystals may transform into a laminar structure similar to that of Type I and achieve a higher toughness.
This is a phenomenon of configuration transformation that is mediated by mechanical loading.

Figure \ref{fig:cellulose_nanocrystal_ii_fractures} provides the key frames of the two typical stretch transitions for slant models, and the permutation with higher toughness is marked as ``Slant Tough''.
From the perspective of restructuring, the difference between the cases of lower and higher toughness in the slant models is whether the early fractures occurred only in a specific area.
For the overall restructuring, the cellulose chains would restructure to be laminar, similar to that of the Type I slant models.
This special deformation was also confirmed in larger models (the length and width of the cross section were both doubled, as shown in Figure \ref{fig:cellulose_nanocrystal_ii_larger_stretch_transitions}).

The specific data for stress, strain energy, hydrogen bond number, and potential energy during the stretching of the vertical and slant models are shown in Figure \ref{fig:cellulose_nanocrystal_ii_stretch_collection_and_speed_dependency} and Figure \ref{fig:cellulose_nanocrystal_ii_stretch_hydrogen_bond_number_and_energy}.
Based on the data in the figures, the vertical model of Type II is dominated by hydrogen bonding and possesses a higher strength than the slant model.
A higher strength corresponds to a greater increase in the Coulomb potential energy.
In contrast to Type I, the number of hydrogen bonds all decreased and the Coulomb potential energies of Type II all increased, illustrating the crucial role of hydrogen bonds.
Although the van der Waals potential all increased, the slant model with restructuring and higher toughness increased the most.
In Type II, both Coulomb and van der Waals interactions play a pivotal role in resisting stretches.
The unique restructuring phenomenon of the Type II slant model is dependent on the ideal crystal structure and periodicity, which is further illustrated in subsequent simulations with surrounding solvents.

Considering the uniqueness of the possible restructuring, replica simulations of speed dependency were also performed for Type II, as shown in Figure \ref{fig:cellulose_nanocrystal_ii_stretch_collection_and_speed_dependency}.
The brittleness of the vertical model and the possible restructuring of the slant model were confirmed through replica simulations.
After considering all the strength and toughness performances across varying stretch speeds, it can be confirmed that the mechanical properties of Type II are stable within the considered range of stretch speeds.

\subsection{Size dependency and toughening mechanism}\indent

\subsubsection{Size dependency}\indent

\begin{figure}[htbp]
    \centering
    \begin{subfigure}[b]{0.96\textwidth}
        \centering
        \includegraphics[width=0.12\textwidth]{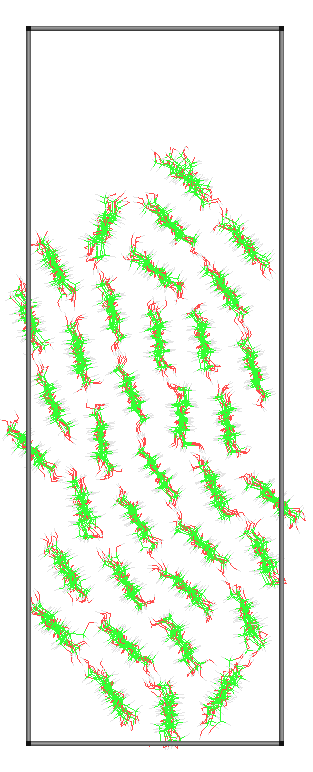}
        \includegraphics[width=0.12\textwidth]{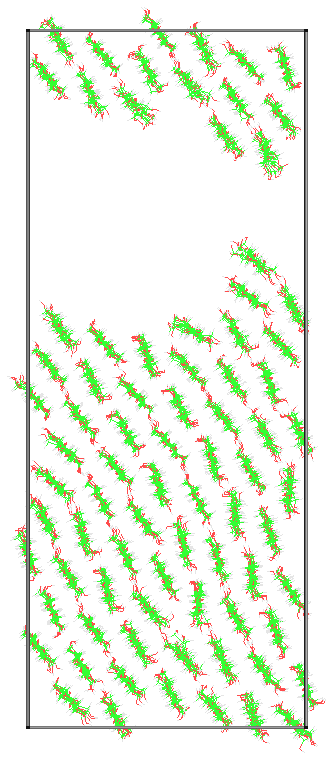}
        \includegraphics[width=0.12\textwidth]{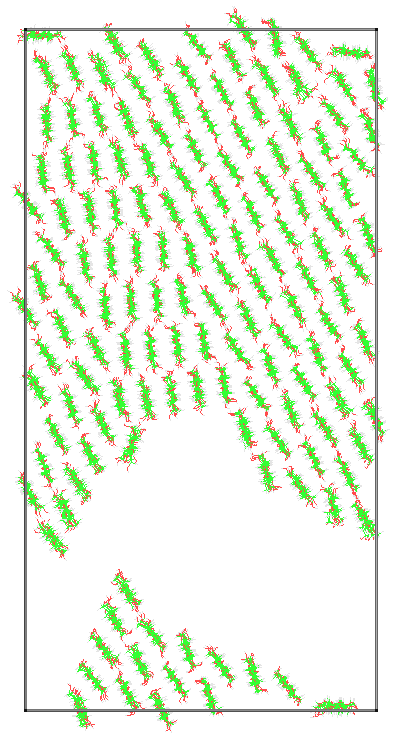}
        \includegraphics[width=0.12\textwidth]{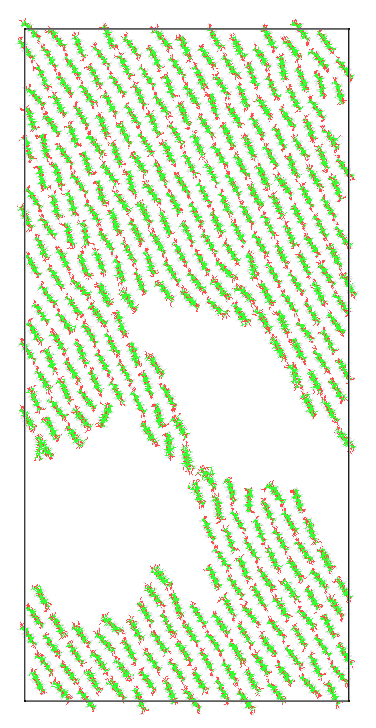}
        \includegraphics[width=0.12\textwidth]{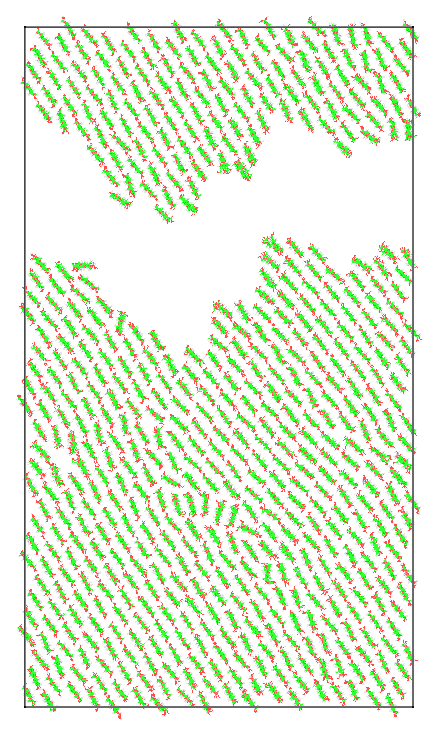}
        \includegraphics[width=0.12\textwidth]{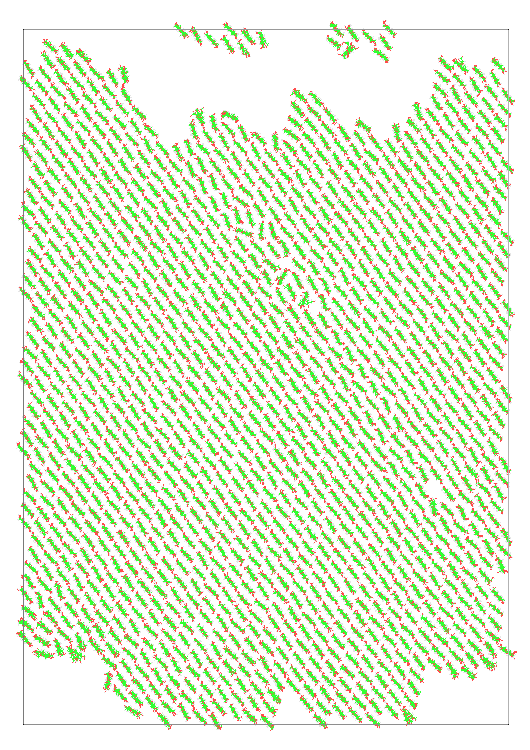}
        \includegraphics[width=0.12\textwidth]{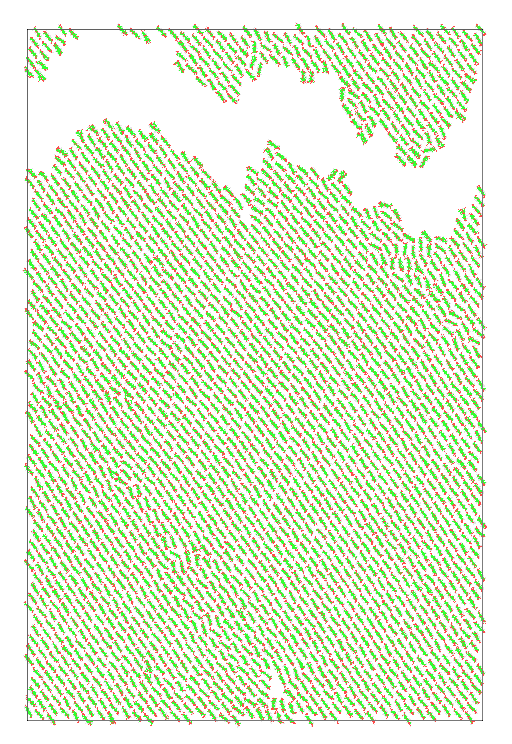}
        \includegraphics[width=0.12\textwidth]{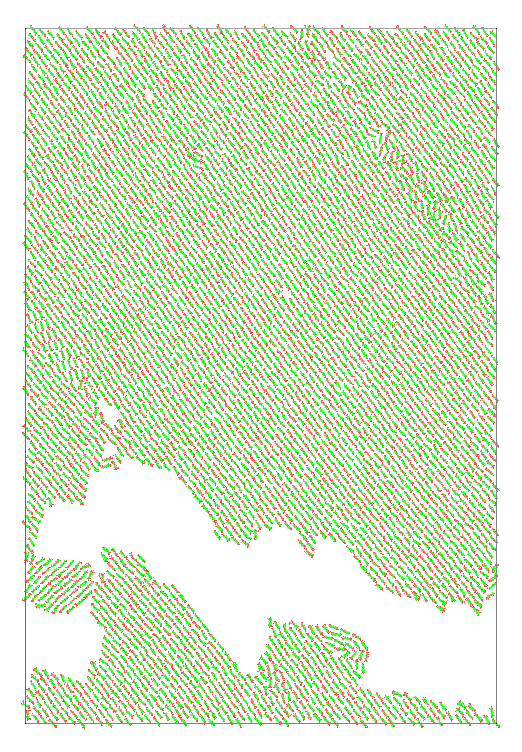}
        \subcaption{}
    \end{subfigure}
    \\
    \begin{subfigure}[b]{0.96\textwidth}
        \centering
        \begin{minipage}[b]{0.32\textwidth}
            \centering
            \includegraphics[width=0.96\textwidth]{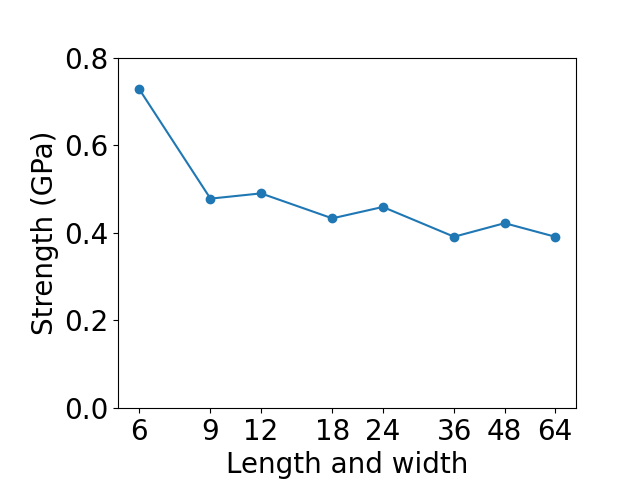}
        \end{minipage}
        \begin{minipage}[b]{0.32\textwidth}
            \centering
            \includegraphics[width=0.96\textwidth]{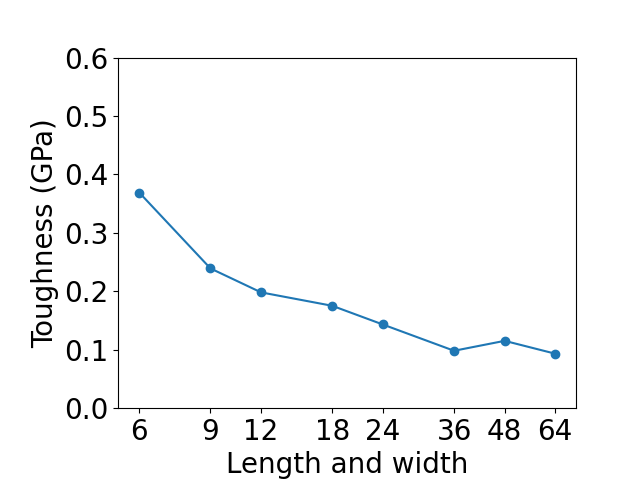}
        \end{minipage}
        \begin{minipage}[b]{0.32\textwidth}
            \centering
            \includegraphics[width=0.96\textwidth]{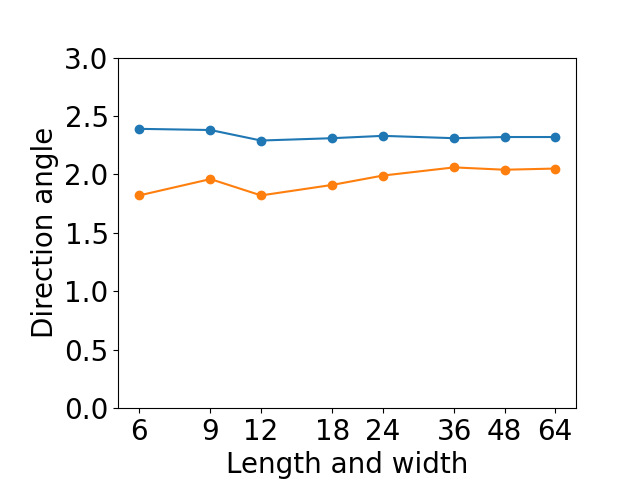}
        \end{minipage}
        \subcaption{}
    \end{subfigure}
    \captionsetup{font=scriptsize}
    \caption{
    Fractures and performances of the Type I CNCs with size.
    (a) Fractures of different sizes.
    The models considered here are Type I CNCs with identical number of rows and columns, such as 6$\times$6 and 12$\times$12.
    6, 9, 12, 18, 24, 36, and 64 were the considered lengths.
    From the perspective of fracture strain, it can be directly observed that the ductility tended to decrease as the nanocrystal size increased.
    (b) Strength, toughness and direction angle with respect to size.
    As the nanocrystal size increased, the strength, toughness, and direction angle decreased.
    }
    \label{fig:cellulose_nanocrystal_size_dependency}
\end{figure}

Although the anisotropy of CNCs was systematically validated, the model sizes in these simulations were manually determined and fixed.
According to the data provided by Robert Sinko et al.\cite{sinko2014dimensions}, the mechanical properties of Type I CNCs are closely related to their sizes.
When the size reached a certain point, the mechanical performance of Type I tended to stabilize.
Our simulation results confirmed this, as shown in Figure \ref{fig:cellulose_nanocrystal_size_dependency}.
These models featured a square cross section, meaning that the length and width of the cross section were identical in terms of row and column number, such as 6$\times$6 and 12$\times$12.
The lengths considered here were 6, 9, 12, 18, 24, 36, and 64.
With increases in size, the fracture strain (inferred from the fractures), strength, toughness, and direction angle exhibited monotonically decreasing trends.

\begin{figure}[htbp]
    \centering
    \begin{subfigure}[b]{0.48\textwidth}
        \includegraphics[width=0.96\textwidth]{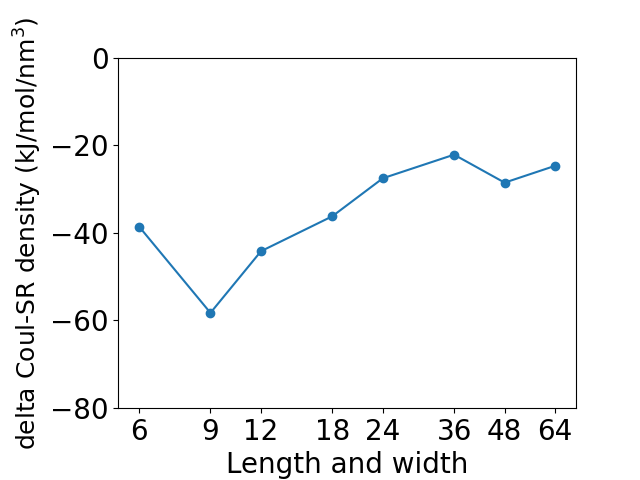}
        \subcaption{}
    \end{subfigure}
    \begin{subfigure}[b]{0.48\textwidth}
        \includegraphics[width=0.96\textwidth]{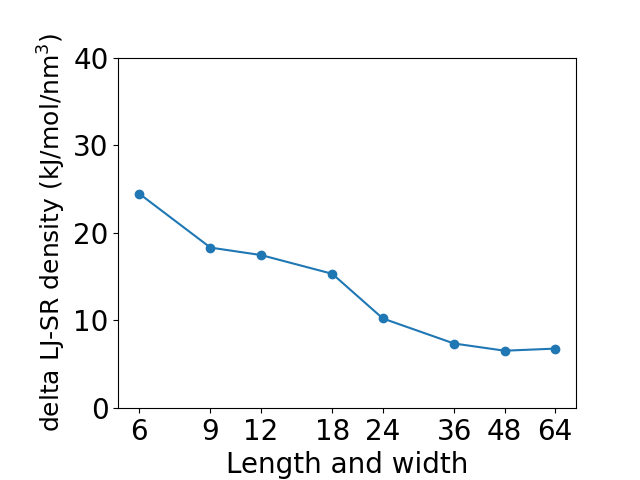}
        \subcaption{}
    \end{subfigure}
    \captionsetup{font=scriptsize}
    \caption{
    Delta potential energy density of the Type I CNCs with size.
    (a) Delta Coulomb potential energy density and (b) delta van der Waals potential energy density with size.
    The delta potential energy density was defined as the ratio of the delta potential energy during stretching relative to the nanocrystal volume before stretching, that is, the delta potential energy per unit volume.
    As the size of the nanocrystal increased, the absolute value of the delta potential energy density during stretching gradually decreased, particularly for the van der Waals interaction resisting stretches.
    From the perspective of structure, larger nanocrystals have a stronger ability to maintain themselves.
    Even though inter layer frictional sliding was observed in all the models, in larger nanocrystals, the parts away from the inter layer frictional sliding interface can maintain a structure similar to that of the crystal.
    The structural changes were not significant compared with those in the smallest 6$\times$6 model.
    Based on the delta potential energy density, the damage to the CNCs is a localized interaction at the sliding interface, therefore the delta potential energy density does not increase linearly with the crystal volume.
    The volumes of the nanocrystal increased quadratically with the length, whereas the delta potential energy of the sliding interface increased at a lower power.
    Therefore, the delta potential energy density and mechanical performance of the Type I CNCs weakened as the nanocrystal size increased.
    }
    \label{fig:cellulose_nanocrystal_delta_energy_density_with_size}
\end{figure}

Sinko et al. used the root mean square fluctuation (RMSF) to explain the variation in mechanical properties with size for the Type I CNCs.
Here, we use the delta potential energy density to explain the size dependency of the mechanical properties, as shown in Figure \ref{fig:cellulose_nanocrystal_delta_energy_density_with_size}.
The delta potential energy density was defined as the ratio of the delta value of the potential energy to the volume of the nanocrystal before stretching, or the amount of delta potential energy per unit volume.
As the size of the nanocrystal increased, the absolute value of the delta potential energy density during the stretching tended to decrease,  particularly for the van der Waals interaction resisting the stretch loads.
The special fracture behavior of the Type I slant model is a local deformation morphology of the inter layer frictional sliding.
Therefore, the overall volume and potential energy increased quadratically with the length, but the potential energy of the interfaces of the friction sliding increased at a lower order power of length, approximately linearly.
Consequently, the absolute value of the delta potential energy per unit volume during stretching gradually decreased as the model size increased, and the corresponding mechanical properties gradually decreased.

\subsubsection{Toughening mechanism inspired by anisotropy and size dependency}\indent

\begin{figure}[htbp]
    \centering
    \scriptsize
    \begin{subfigure}[b]{0.96\textwidth}
        \centering
        \begin{minipage}[b]{0.24\textwidth}
            \centering
            \includegraphics[width=0.96\textwidth]{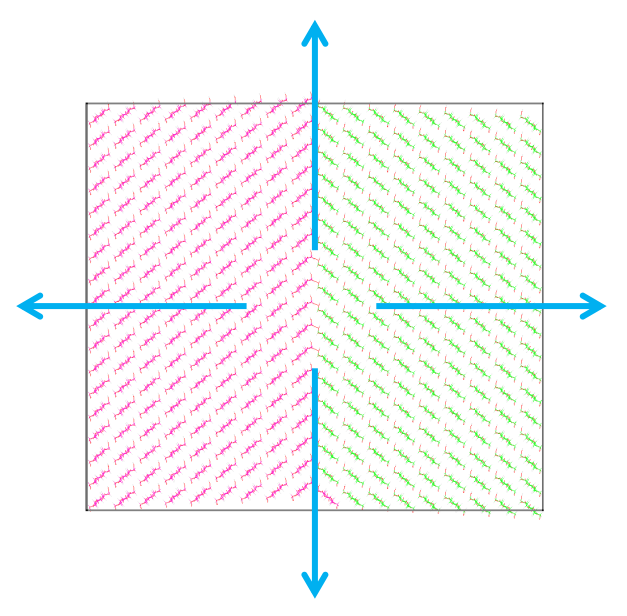}
            \\
            Symmetric
        \end{minipage}
        \begin{minipage}[b]{0.24\textwidth}
            \centering
            \includegraphics[width=0.96\textwidth]{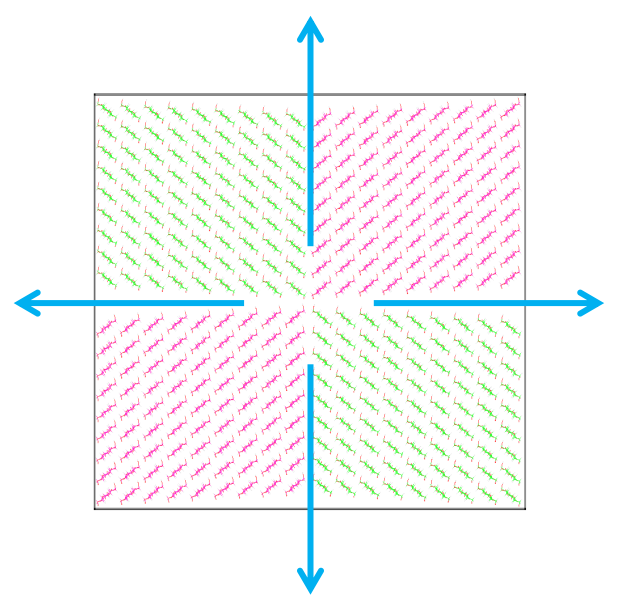}
            \\
            Cross
        \end{minipage}
        \subcaption{}
    \end{subfigure}
    \\
    \begin{subfigure}[b]{0.96\textwidth}
        \centering
        \begin{minipage}[b]{0.36\textwidth}
            \centering
            \includegraphics[width=0.24\textwidth]{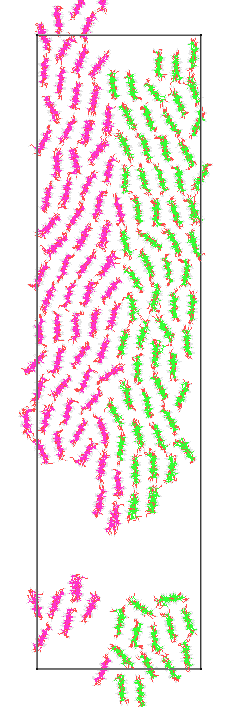}
            \includegraphics[width=0.24\textwidth]{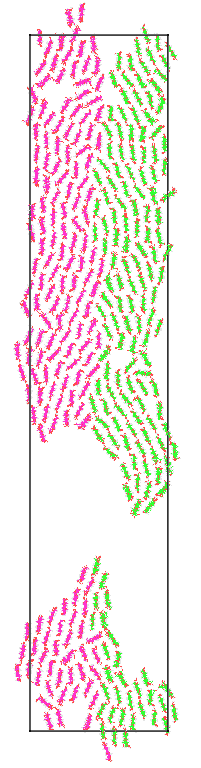}
            \includegraphics[width=0.24\textwidth]{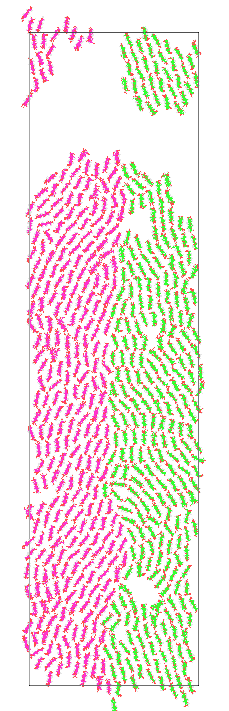}
            \\
            Symmetric
        \end{minipage}
        \hspace{0.12\textwidth}
        \begin{minipage}[b]{0.36\textwidth}
            \centering
            \includegraphics[width=0.24\textwidth]{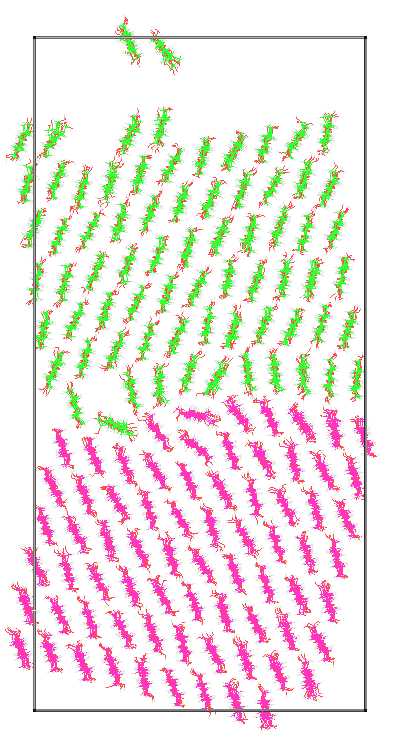}
            \includegraphics[width=0.24\textwidth]{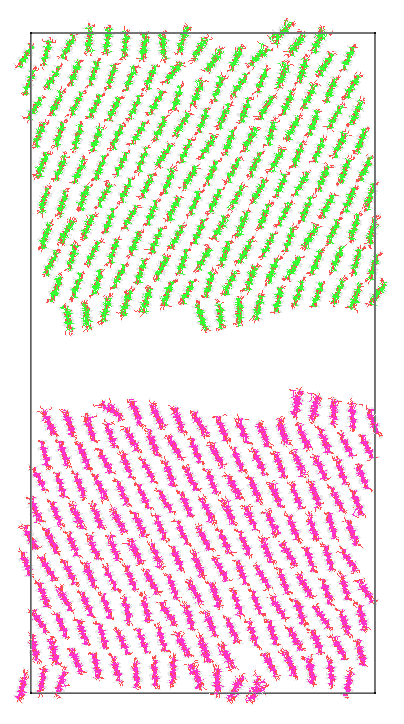}
            \includegraphics[width=0.24\textwidth]{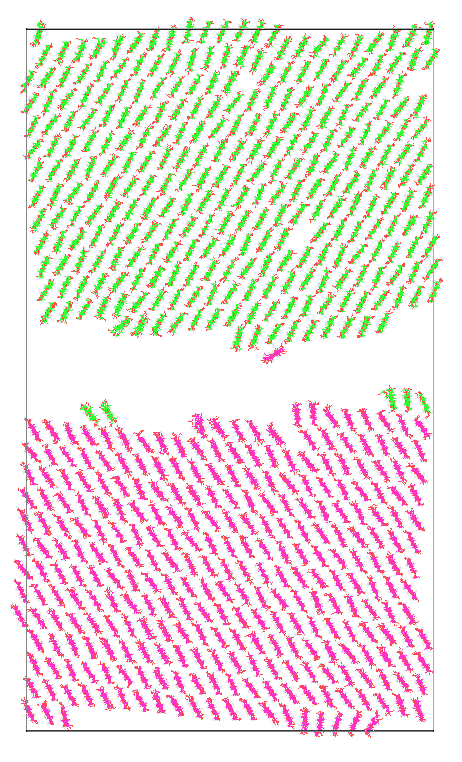}
            \\
            Symmetric Horizontal
        \end{minipage}
        \subcaption{}
    \end{subfigure}
    \\
    \begin{subfigure}[b]{0.96\textwidth}
        \centering
        \begin{minipage}[b]{0.36\textwidth}
            \centering
            \includegraphics[width=0.24\textwidth]{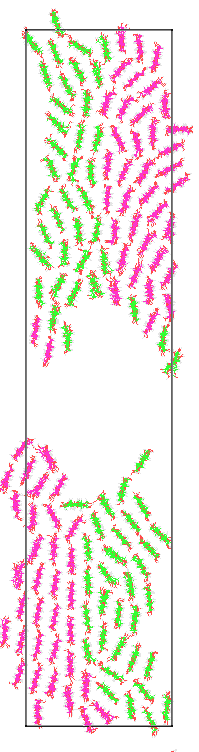}
            \includegraphics[width=0.24\textwidth]{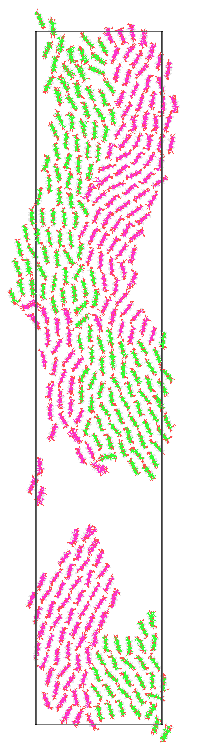}
            \includegraphics[width=0.24\textwidth]{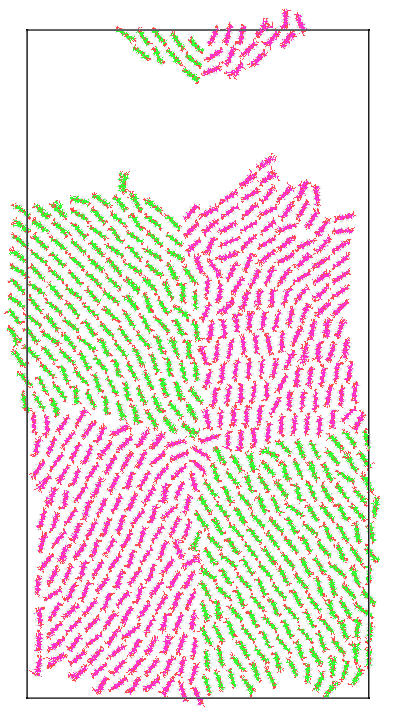}
            \\
            Cross
        \end{minipage}
        \hspace{0.12\textwidth}
        \begin{minipage}[b]{0.36\textwidth}
            \centering
            \includegraphics[width=0.24\textwidth]{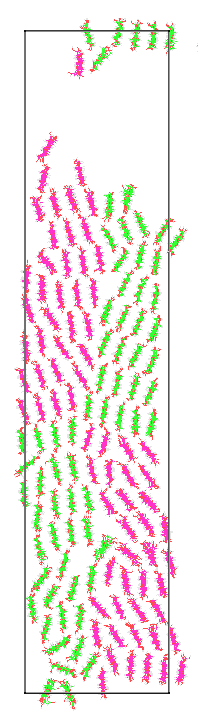}
            \includegraphics[width=0.24\textwidth]{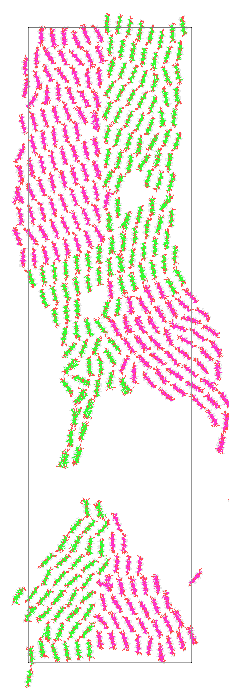}
            \includegraphics[width=0.24\textwidth]{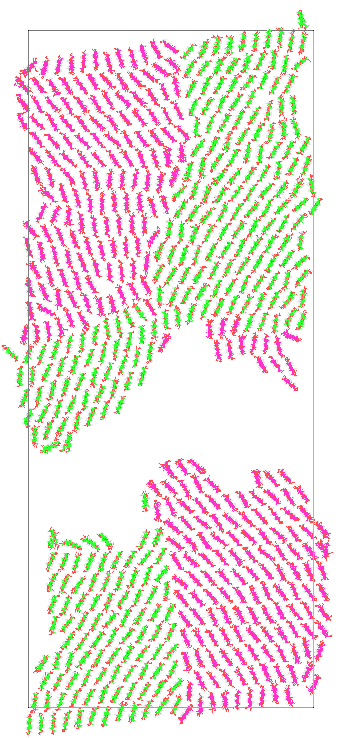}
            \\
            Cross Horizontal
        \end{minipage}
        \subcaption{}
    \end{subfigure}
    \captionsetup{font=scriptsize}
    \caption{
    Transverse arrangement patterns and corresponding fractures of the Type I CNCs.
    (a) Two transverse arrangement patterns: symmetric and cross.
    The cross pattern represents a biaxial permutation of the symmetric arrangement.
    Based on the structure, the performances of the two arrangement patterns differed in the vertical and horizontal directions, and the models were stretched in both directions.
    The structures stretched under horizontal loads were rotated counter clockwise, including models composed of different unit block sizes (6$\times$6, 9$\times$9, and 12$\times$12 rows and columns; one model was composed of 4 of unit blocks).
    (b) Fractures of the symmetric arrangements under vertical and horizontal stretching.
    The symmetric arrangements exhibited good toughness and ductility under vertical loading but were weak when subjected to horizontal loading, owing to the weak interfaces.
    (c) Fractures of the cross arrangements under vertical and horizontal stretching.
    The cross arrangements exhibited good toughness and ductility under both vertical and horizontal loading, serving as an effective toughening mechanism.
    In both the symmetric and cross arrangements, the inter layer frictional sliding of each unit was hindered and led to localized rotation becoming the dominant deformation mechanism, which resulted in improved toughness and ductility.
    Additionally, these arrangements were affected by the size dependency of the units.
    The structures composed of smaller units exhibited better toughness and ductility.
    }
    \label{fig:cellulose_nanocrystal_transverse_arrangement_fractures}
\end{figure}

Considering the size dependency of the mechanical performance of CNCs, we propose manipulating the transverse arrangement of the unit blocks to tune the overall mechanical performance of the structure.
A structure formed by smaller unit blocks in a certain arrangement pattern may exhibit a better mechanical performance than a CNC with the same overall dimensions.
We propose a symmetric pattern and a cross pattern (where the cross pattern represents a biaxial permutation of the symmetric arrangement), as shown in Figure \ref{fig:cellulose_nanocrystal_transverse_arrangement_fractures}.
These two patterns still have differences in structures and mechanical properties in the horizontal and vertical directions;
and models composed of unit nanocrystal blocks of different sizes were constructed to examine the size dependency of the patterns.
The considered unit block sizes were 6$\times$6, 9$\times$9, and 12$\times$12, and 4 units were arranged to form the final structure.
The models and fractures of the arranged structure are shown in Figure \ref{fig:cellulose_nanocrystal_transverse_arrangement_fractures}.

The mechanical properties of the slant model of the Type II CNCs depend significantly on the crystal structure and periodicity (this is further explained in the simulation containing solvents in the following sections).
Otherwise, it is difficult to restructuring and exhibit good toughness, and only the symmetric pattern under vertical stretching can exhibit a higher toughness than the crystal structure.
The size dependency and transverse toughening simulation data of Type II are shown in Figure \ref{fig:cellulose_nanocrystal_ii_size_dependency} and Figure \ref{fig:cellulose_nanocrystal_ii_delta_energy_density_with_size} for reference.

\begin{figure}[htbp]
    \centering
    \begin{subfigure}[b]{0.96\textwidth}
        \includegraphics[width=0.48\textwidth]{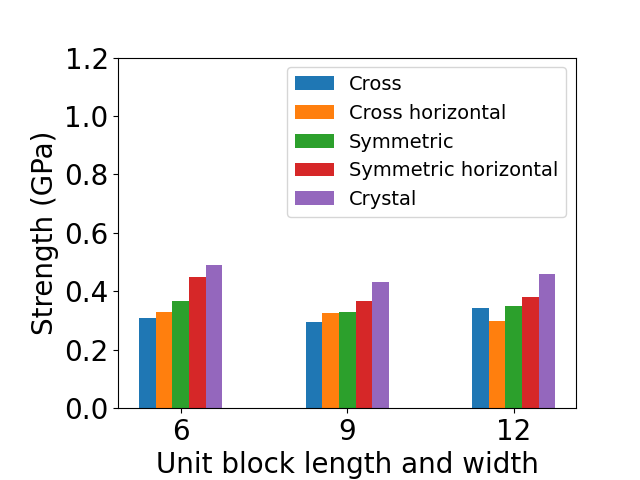}
        \includegraphics[width=0.48\textwidth]{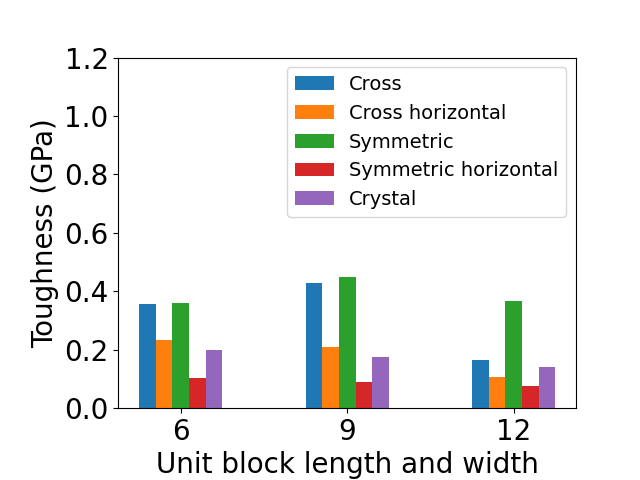}
        \subcaption{}
    \end{subfigure}
    \\
    \begin{subfigure}[b]{0.96\textwidth}
        \includegraphics[width=0.48\textwidth]{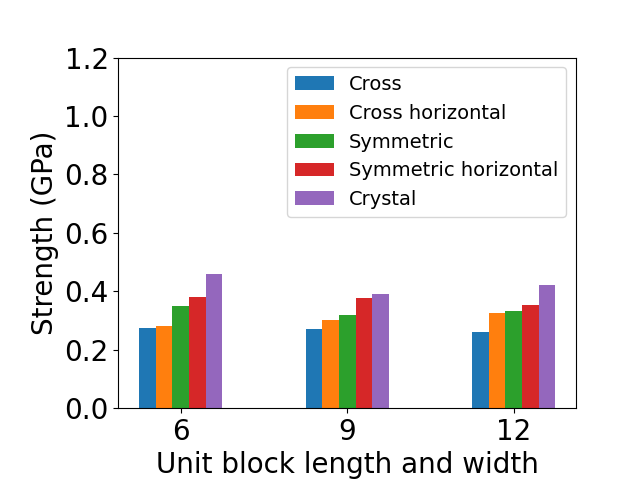}
        \includegraphics[width=0.48\textwidth]{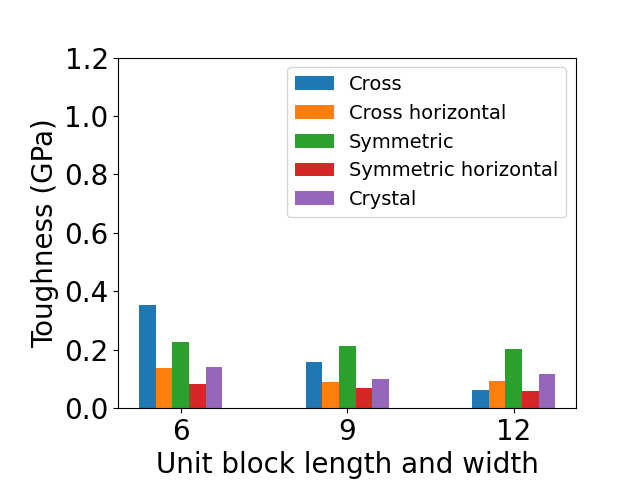}
        \subcaption{}
    \end{subfigure}
    \captionsetup{font=scriptsize}
    \caption{
    Mechanical performance of transverse arrangement patterns of the Type I CNCs.
    (a) Strength and toughness of the transverse patterns.
    (b) Strength and toughness of the transverse patterns with more unit blocks.
    By comparing with crystal models of the same sizes, it is clear that both symmetric and cross patterns can significantly improved the ductility and toughness with a slight sacrifice in strength.
    The toughening mechanisms of the transverse patterns and the dependency on the small unit block size were further represented by models composed of 16 unit blocks.
    }
    \label{fig:cellulose_nanocrystal_transverse_arrangement_performance_with_patterns}
\end{figure}

The simulation results showed that the symmetric and cross patterns can improve the ductility and toughness of the structure, which is influenced by the size of their unit blocks.
In the symmetric and cross patterns, the inter layer friction sliding was hindered, and local rotation replaced it as the dominant deformation phenomenon.
The structures subjected to horizontal stretching were rotated counter clockwise for ease of comparison.
The internal interfaces of the symmetric and cross patterns unit blocks were weaker in the horizontal direction;
correspondingly, the ductility and toughness when subjected to horizontal stretching were also worse, particularly for the symmetric pattern.
In summary, the cross pattern exhibited better ductility and toughness when subjected to both vertical and horizontal stretching, and the symmetric pattern preferred vertical loads.
The ductility of the arranged structure was also significantly related to the size of the unit blocks, and the structure formed by 6$\times$6 unit blocks exhibited better ductility.
Specific quantitative performance comparisons of the arrangement patterns and crystal structures of the same size are shown in Figure \ref{fig:cellulose_nanocrystal_transverse_arrangement_performance_with_patterns}.

To reduce the influence of artificial settings on the simulation results, arranged structures composed of 16 unit blocks were constructed and tested (Figure \ref{fig:cellulose_nanocrystal_more_transverse_arrangement_fractures} and Figure \ref{fig:cellulose_nanocrystal_ii_more_transverse_arrangement_fractures}).
With more unit blocks of arrangements, the local rotation still replaced the hindered inter layer friction sliding.
Because the interfacial interaction strength between unit blocks is weak in cross patterns, when the number of unit block increased, the arranged structure was more significantly affected by the size of the unit crystals and still outperform when using 6$\times$6 unit blocks.
Considering the quantitative performances and fracture behaviors of the two patterns, arranging small unit blocks in these patterns rather than forming a single larger nanocrystal, is an effective toughening mechanism.

Figure \ref{fig:cellulose_nanocrystal_transverse_arrangement_performance_with_patterns} provides quantitative data on the influences of arrangement pattern, loading direction, and unit block size on strength and toughness.
These data confirmed previous conclusions based on the morphological results.
Compared with nanocrystal models of the same size, the strengths of the symmetric and cross patterns decreased, but the toughness significantly improved, particularly in the case of small unit blocks.
With further requirements for material design and technical development, the transverse arrangement of CNCs should be emphasized in the development of cellulose nanomaterials.

\subsection{Mechanical properties when surrounded by solvents on both sides}\indent

Under natural (such as cellulose crystals in plants)\cite{kim2015review} and non-natural (such as artificially prepared composites) conditions\cite{peng2021cellulose,zhang2019non,liu2009supramolecular}, solvents such as water often coexist with CNCs.
Analyzing the influence of the solvent environment is crucial for understanding CNCs in real environments and can assist in the development of cellulose materials.
Therefore, comparisons of the solvent effects on the CNCs anisotropy and fracture behavior could further extend and enhance this study.
Eight solvent conditions were selected based on the relevant literature: N,N-dimethylformamide (DMF)\cite{bui2022synthesis}, dimethyl sulfoxide (DMSO)\cite{dong2024preparation}, ethylenediamine (EDA)\cite{li2024mild}, ethanol (ETA)\cite{yang2025evaluating}, isopropanol (IPA)\cite{yeo2022robust}, methanol (MTA)\cite{przybylek2023use}, and water (SOL)\cite{etale2023cellulose}, and a solvent-free (NONE) control group.
Some abbreviations for the solvents used here are not standard.
In a solvent environment, both the CNCs, and the interactions between cellulose and the solvent are crucial factors.

\begin{figure}[htbp]
    \centering
    \begin{subfigure}[b]{0.96\textwidth}
        \centering
        \begin{minipage}[b]{0.12\textwidth}
            \centering
            \includegraphics[height=0.57\textwidth]{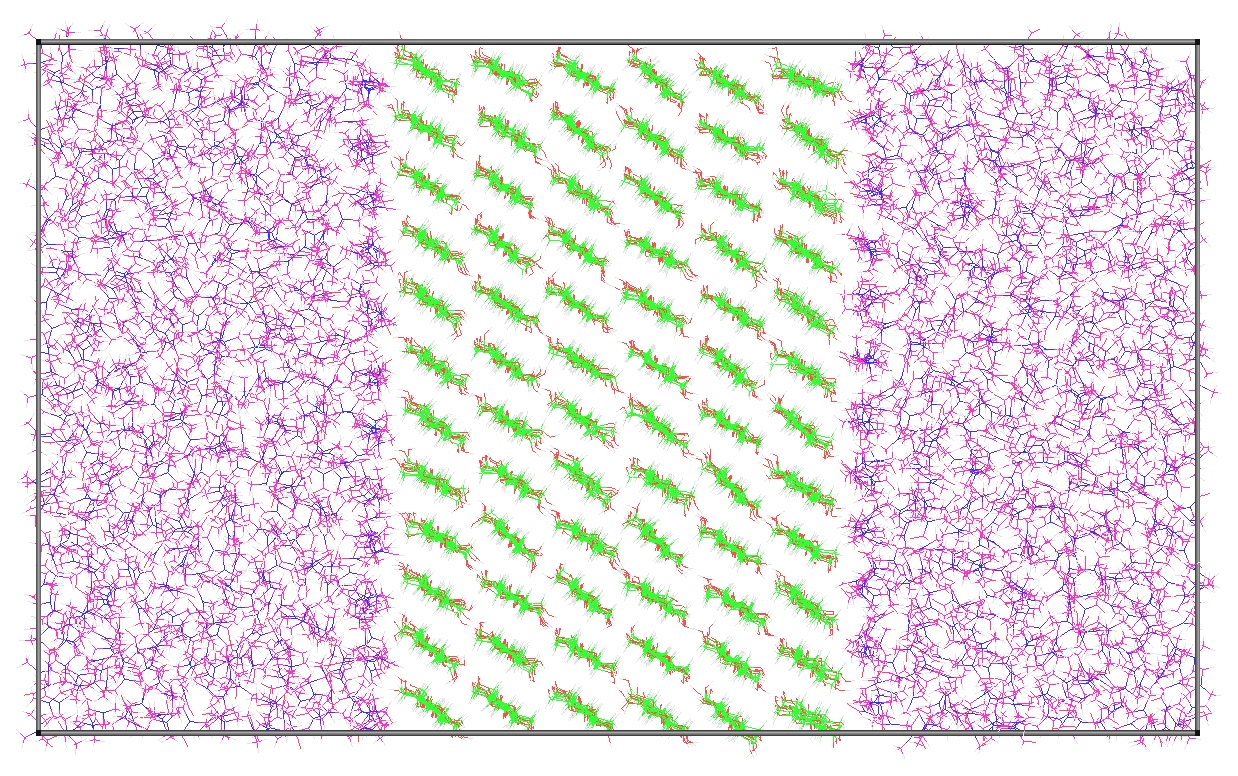}
        \end{minipage}
        \begin{minipage}[b]{0.12\textwidth}
            \centering
            \includegraphics[height=0.57\textwidth]{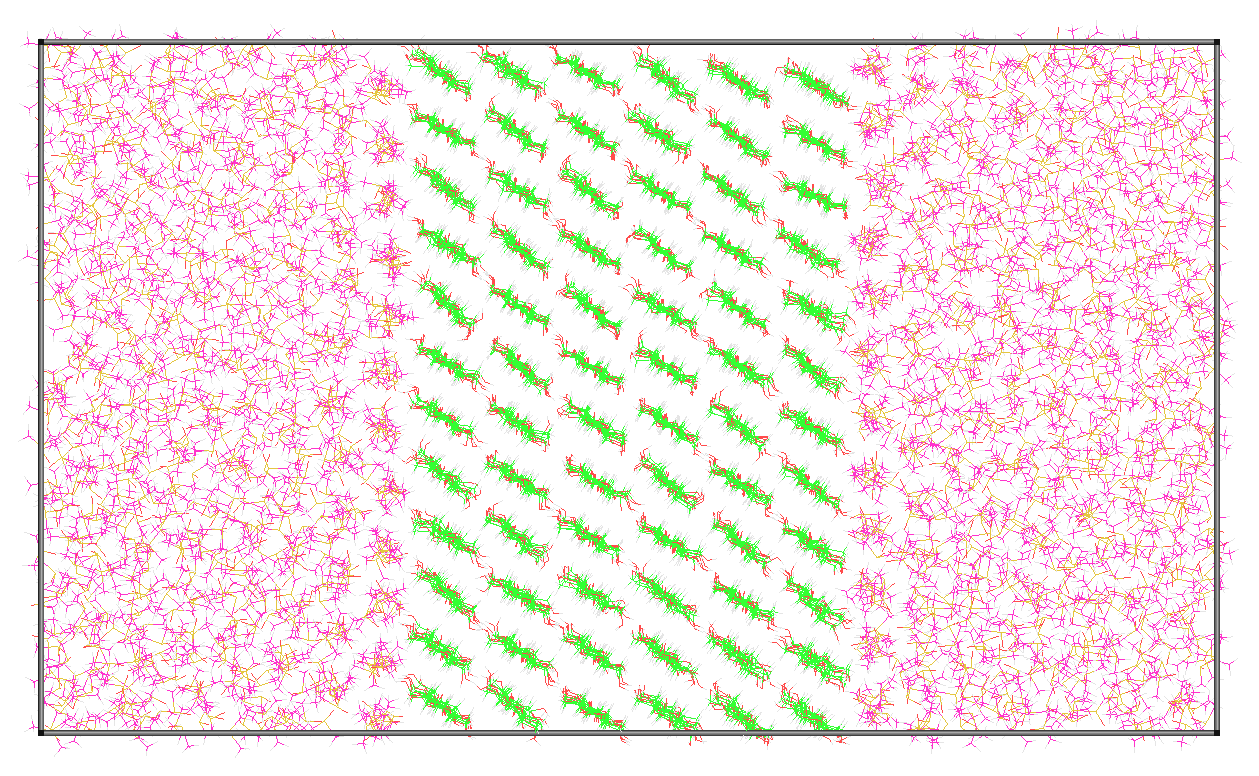}
        \end{minipage}
        \begin{minipage}[b]{0.12\textwidth}
            \centering
            \includegraphics[height=0.57\textwidth]{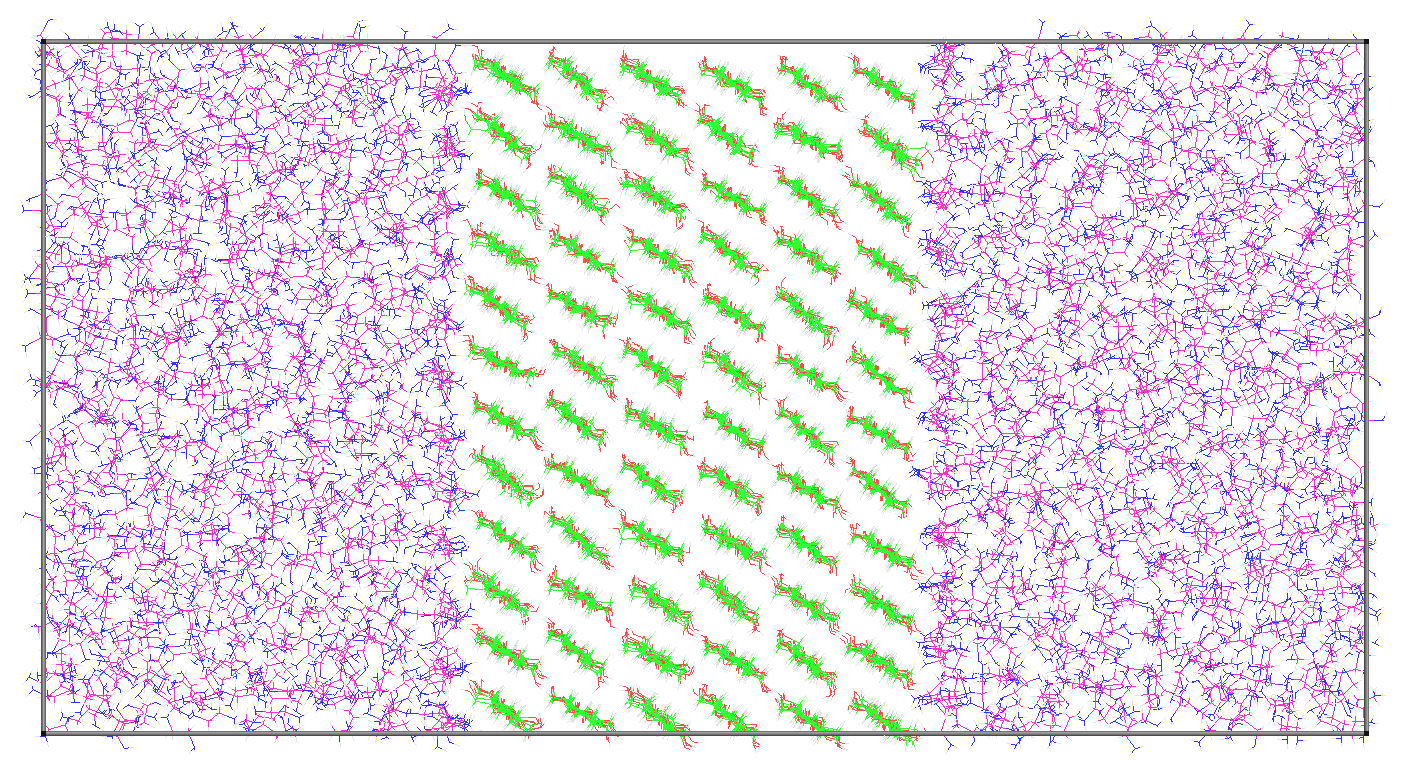}
        \end{minipage}
        \begin{minipage}[b]{0.12\textwidth}
            \centering
            \includegraphics[height=0.57\textwidth]{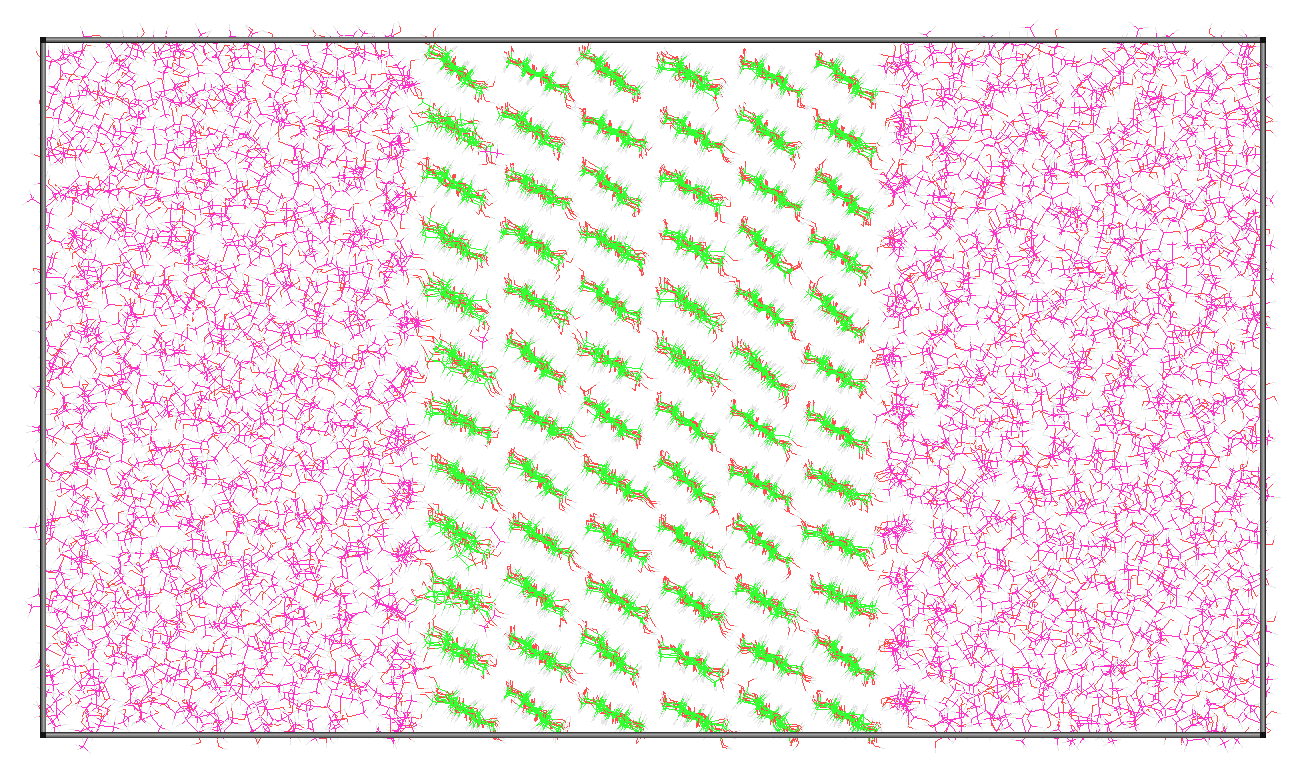}
        \end{minipage}
        \begin{minipage}[b]{0.12\textwidth}
            \centering
            \includegraphics[height=0.57\textwidth]{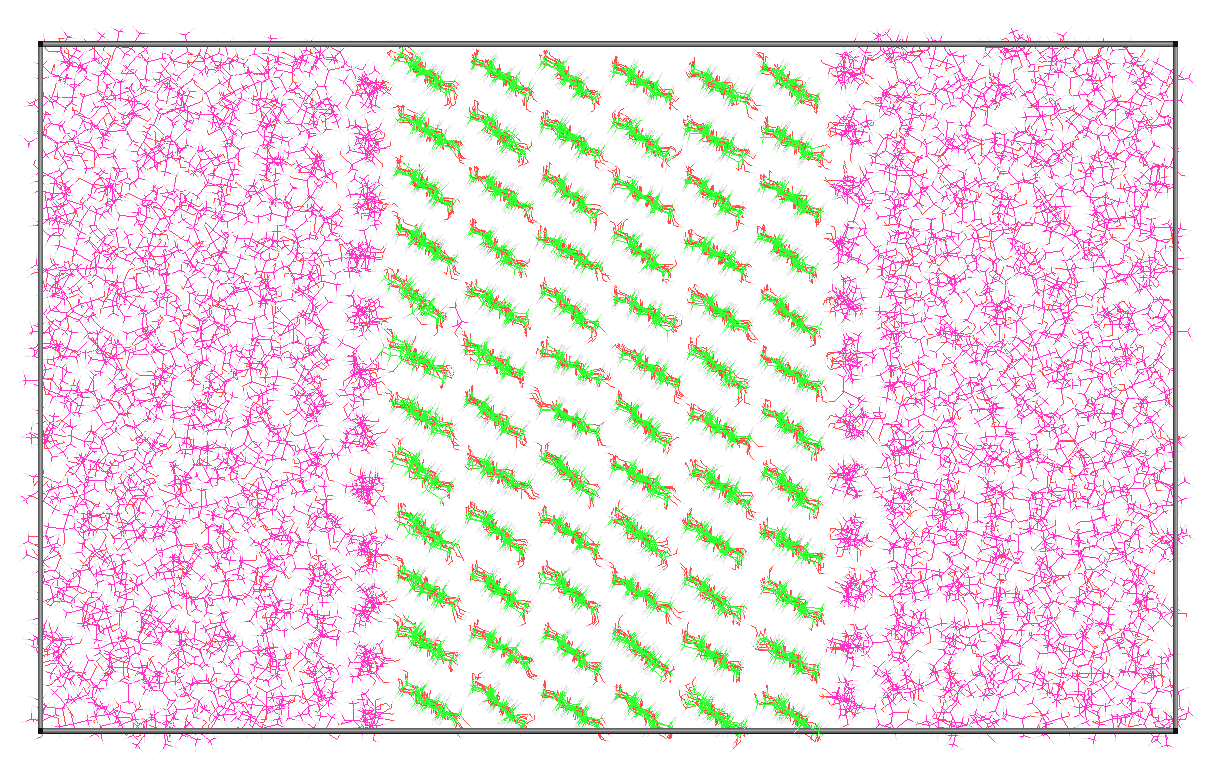}
        \end{minipage}
        \begin{minipage}[b]{0.12\textwidth}
            \centering
            \includegraphics[height=0.57\textwidth]{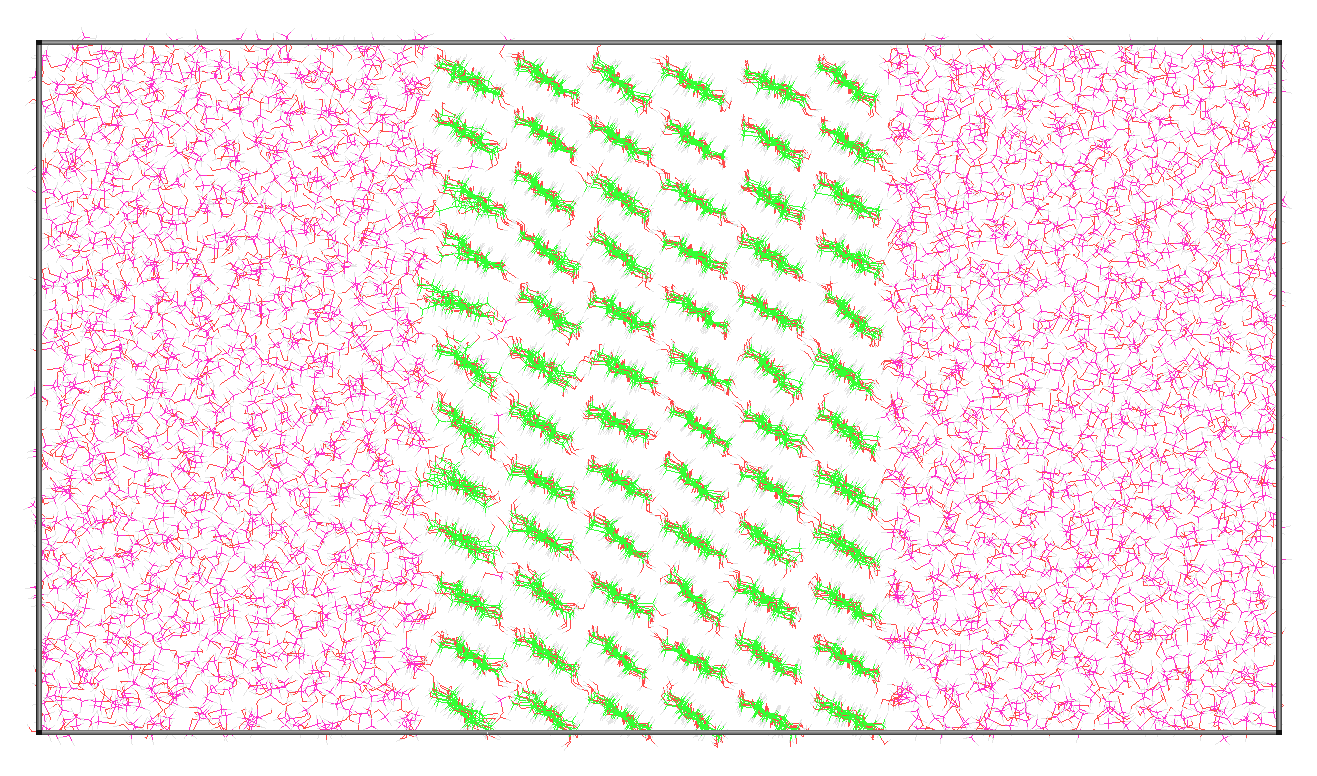}
        \end{minipage}
        \begin{minipage}[b]{0.12\textwidth}
            \centering
            \includegraphics[height=0.57\textwidth]{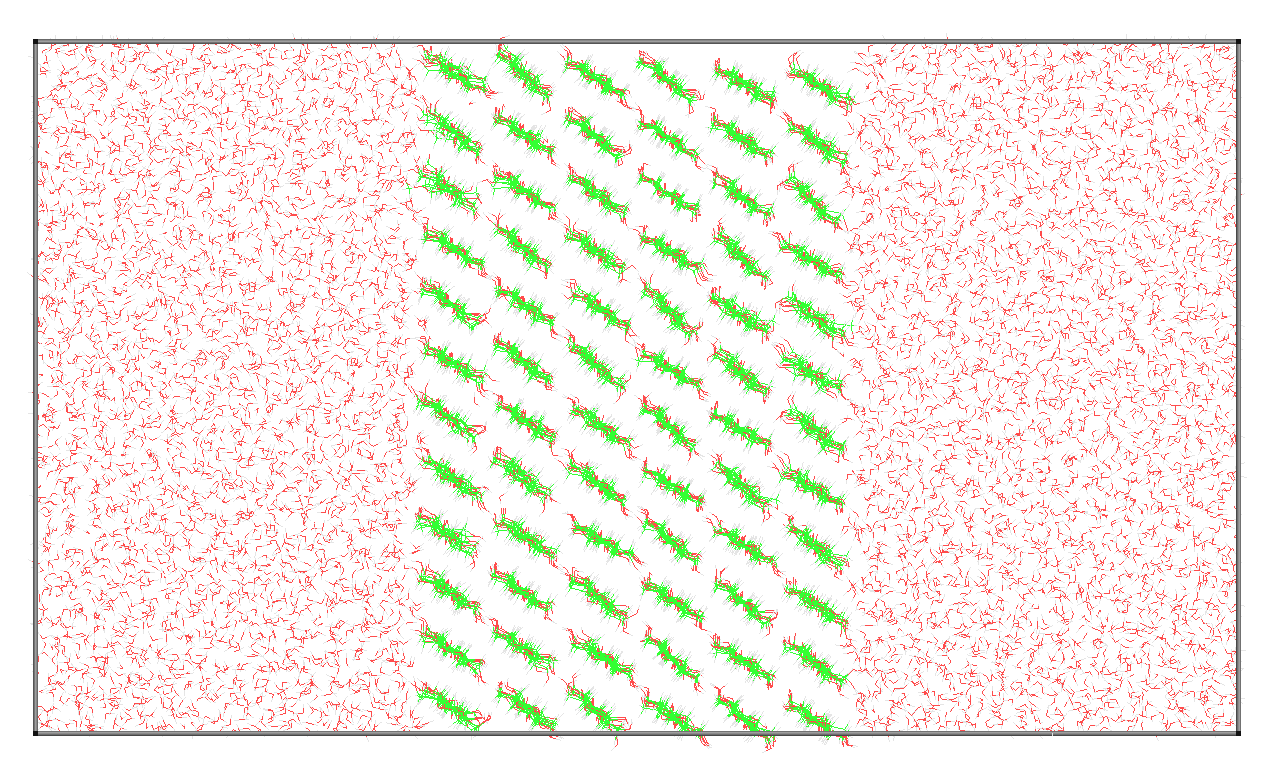}
        \end{minipage}
        \begin{minipage}[b]{0.12\textwidth}
            \centering
            \includegraphics[height=0.57\textwidth]{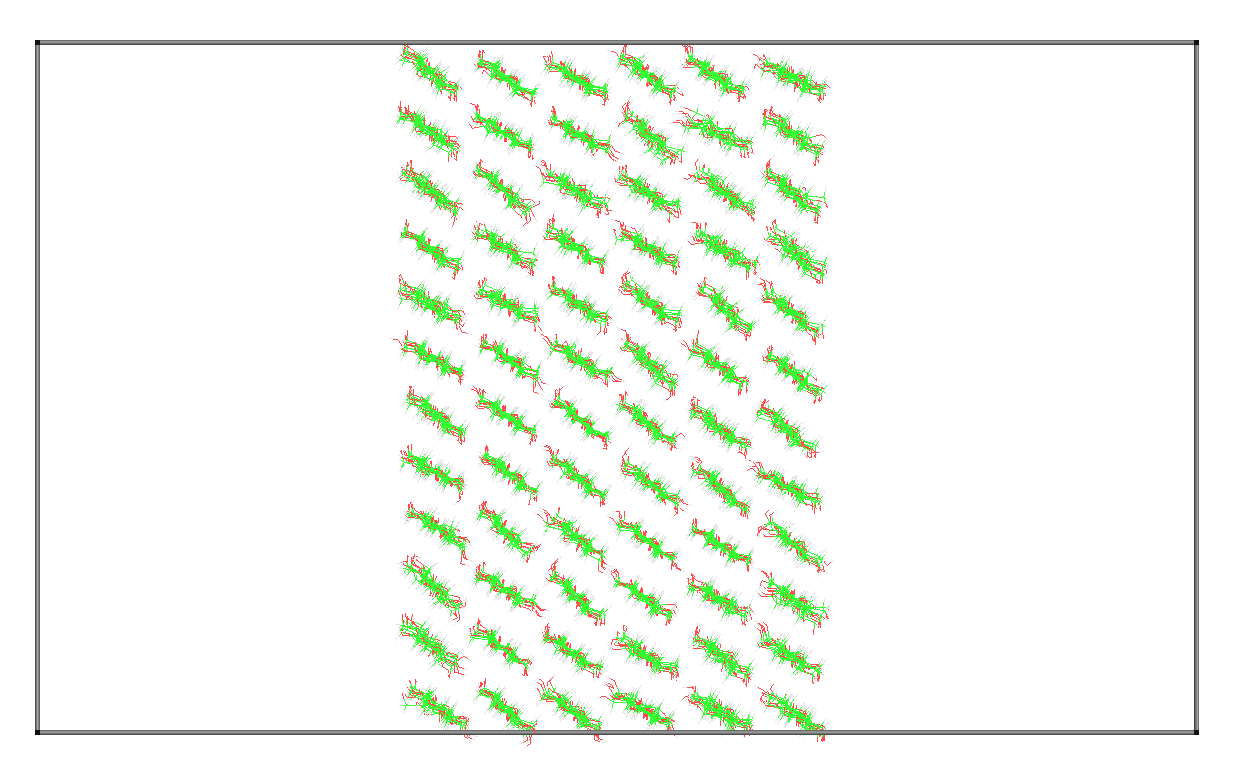}
        \end{minipage}
        \subcaption{}
    \end{subfigure}
    \\
    \begin{subfigure}[b]{0.96\textwidth}
        \centering
        \begin{minipage}[b]{0.12\textwidth}
            \centering
            \includegraphics[height=0.75\textwidth]{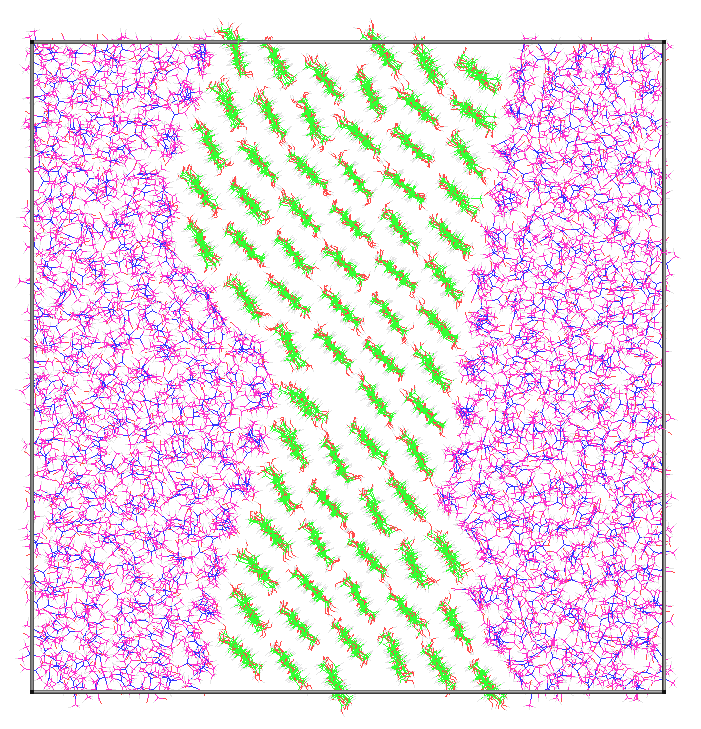}
        \end{minipage}
        \begin{minipage}[b]{0.12\textwidth}
            \centering
            \includegraphics[height=0.75\textwidth]{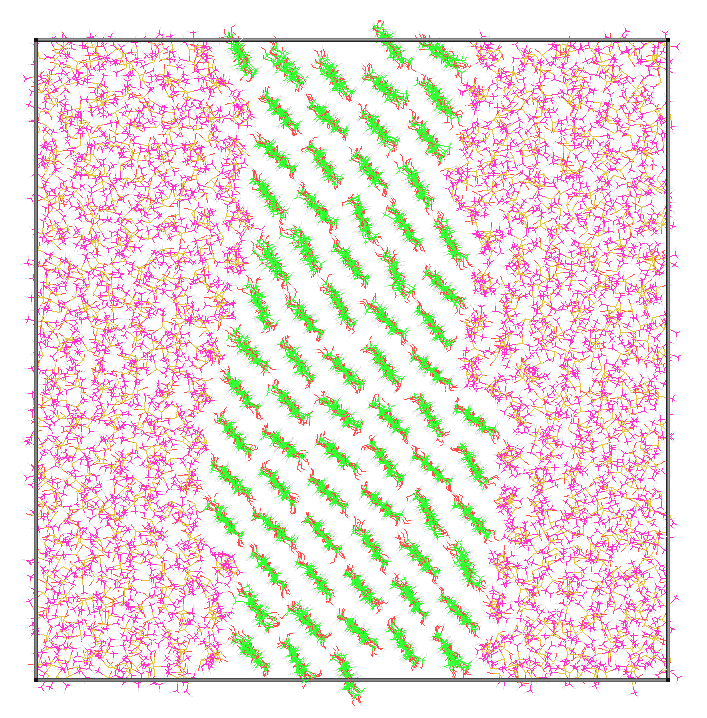}
        \end{minipage}
        \begin{minipage}[b]{0.12\textwidth}
            \centering
            \includegraphics[height=0.75\textwidth]{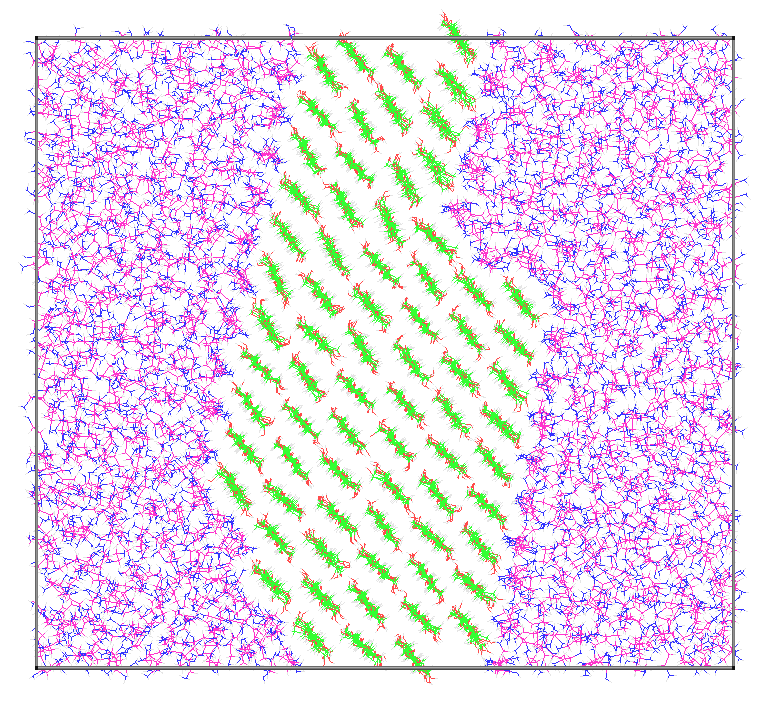}
        \end{minipage}
        \begin{minipage}[b]{0.12\textwidth}
            \centering
            \includegraphics[height=0.75\textwidth]{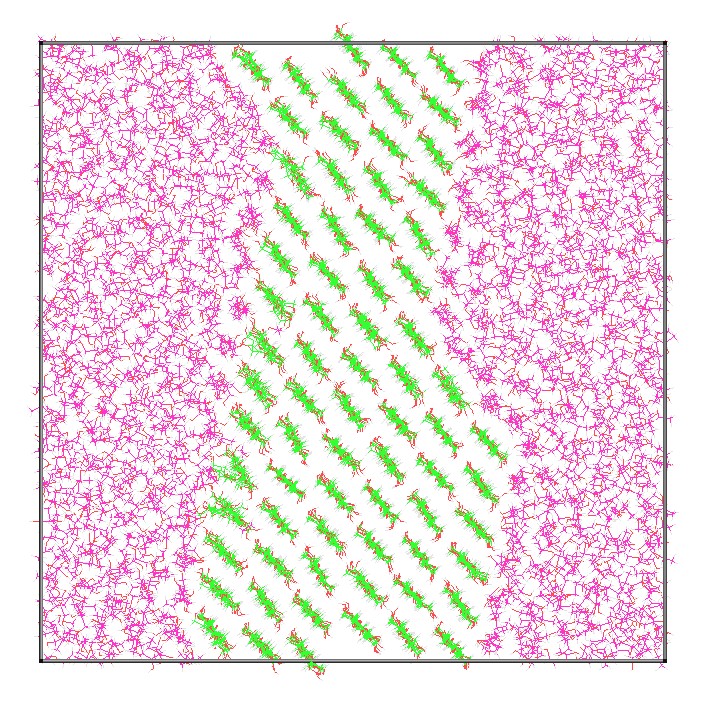}
        \end{minipage}
        \begin{minipage}[b]{0.12\textwidth}
            \centering
            \includegraphics[height=0.75\textwidth]{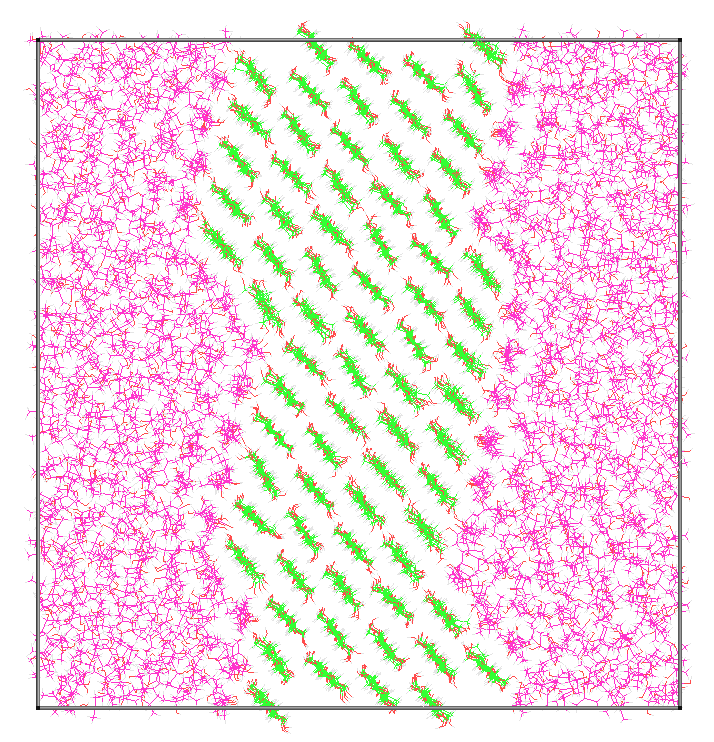}
        \end{minipage}
        \begin{minipage}[b]{0.12\textwidth}
            \centering
            \includegraphics[height=0.75\textwidth]{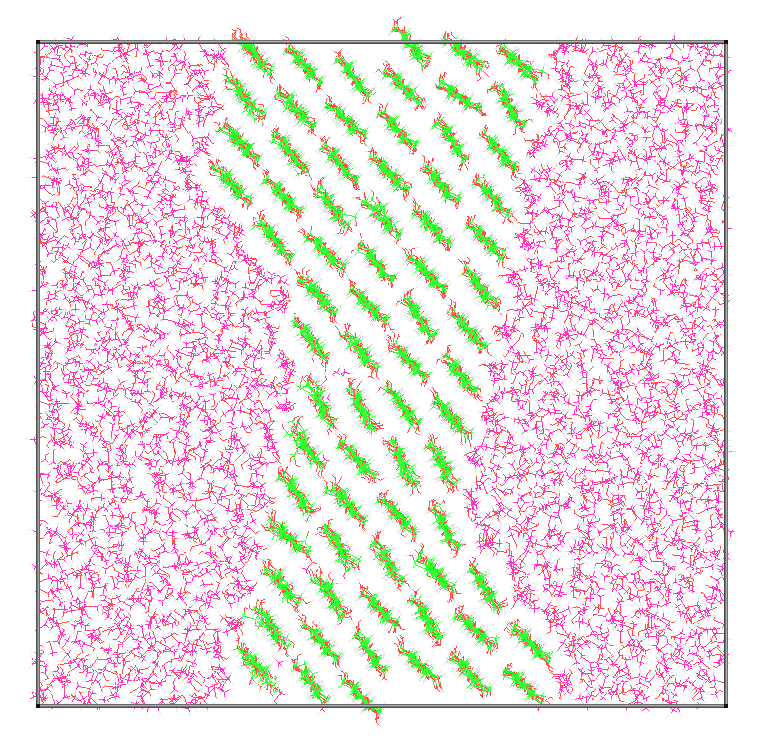}
        \end{minipage}
        \begin{minipage}[b]{0.12\textwidth}
            \centering
            \includegraphics[height=0.75\textwidth]{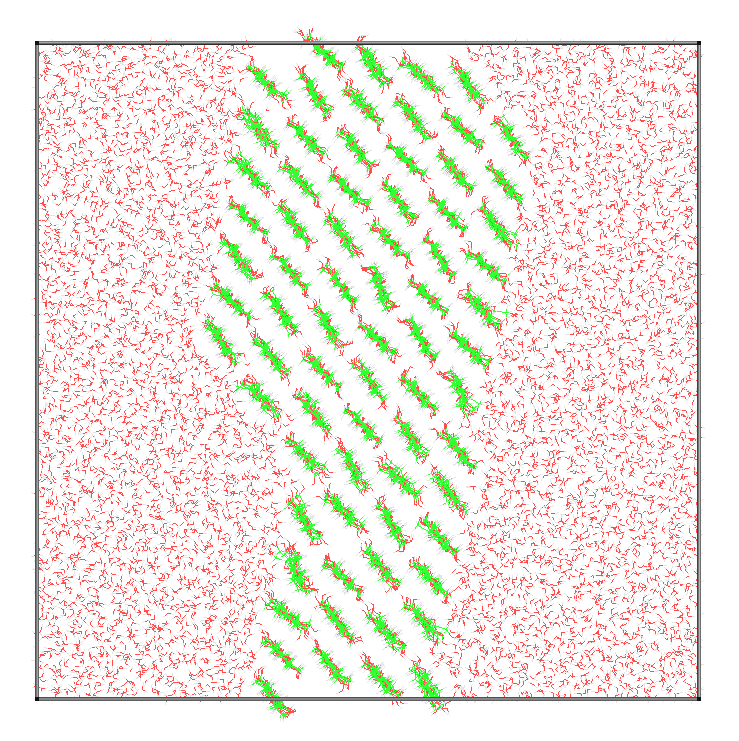}
        \end{minipage}
        \begin{minipage}[b]{0.12\textwidth}
            \centering
            \includegraphics[height=0.75\textwidth]{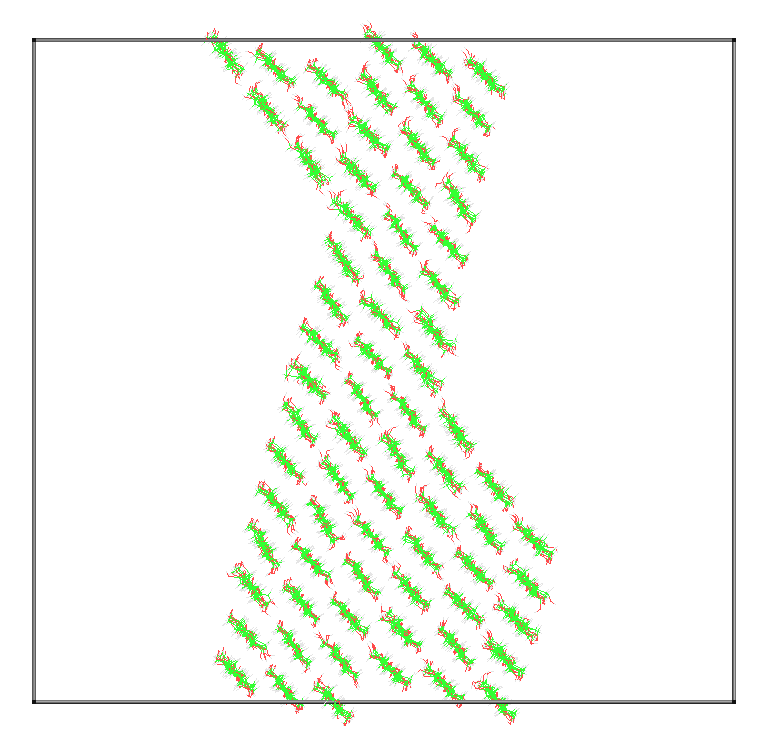}
        \end{minipage}
        \subcaption{}
    \end{subfigure}
    \\
    \begin{subfigure}[b]{0.96\textwidth}
        \centering
        \begin{minipage}[b]{0.12\textwidth}
            \centering
            \includegraphics[height=0.93\textwidth]{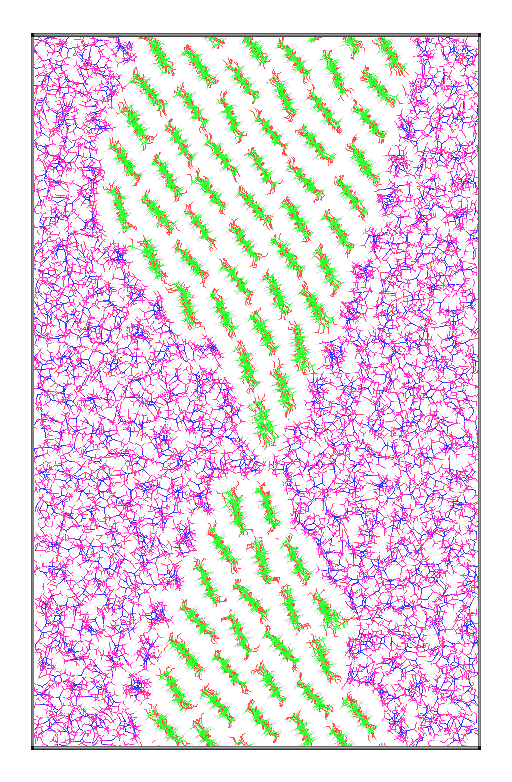}
            \\
            DMF
        \end{minipage}
        \begin{minipage}[b]{0.12\textwidth}
            \centering
            \includegraphics[height=0.93\textwidth]{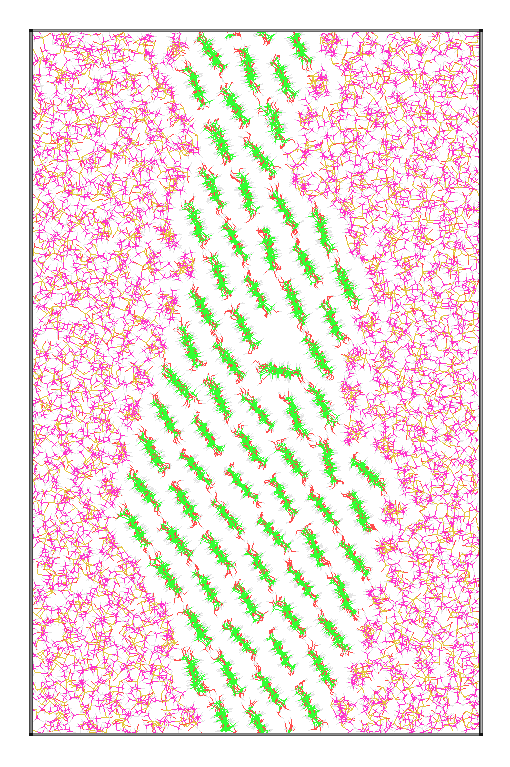}
            \\
            DMSO
        \end{minipage}
        \begin{minipage}[b]{0.12\textwidth}
            \centering
            \includegraphics[height=0.93\textwidth]{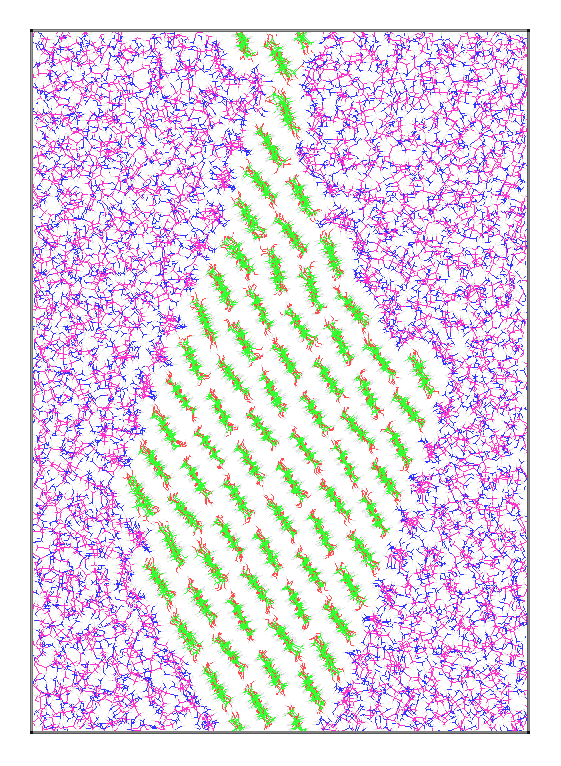}
            \\
            EDA
        \end{minipage}
        \begin{minipage}[b]{0.12\textwidth}
            \centering
            \includegraphics[height=0.93\textwidth]{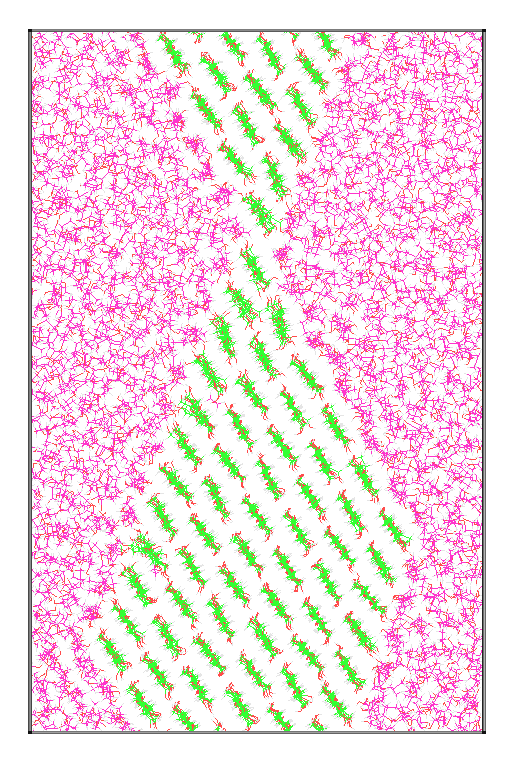}
            \\
            ETA
        \end{minipage}
        \begin{minipage}[b]{0.12\textwidth}
            \centering
            \includegraphics[height=0.93\textwidth]{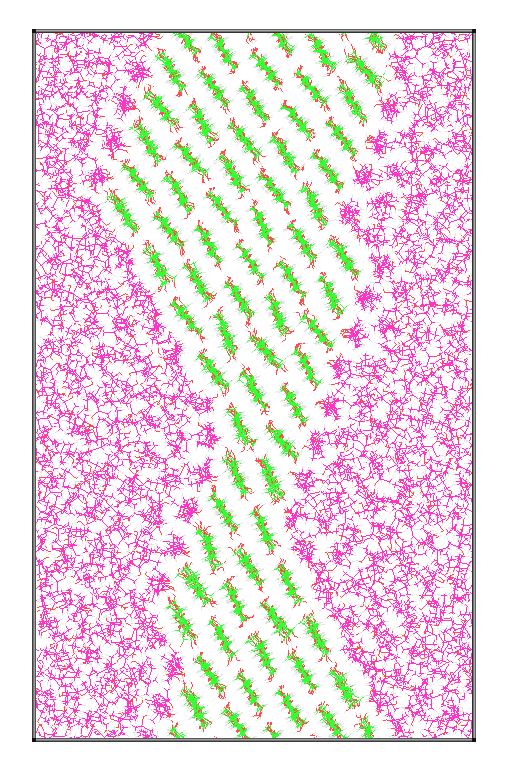}
            \\
            IPA
        \end{minipage}
        \begin{minipage}[b]{0.12\textwidth}
            \centering
            \includegraphics[height=0.93\textwidth]{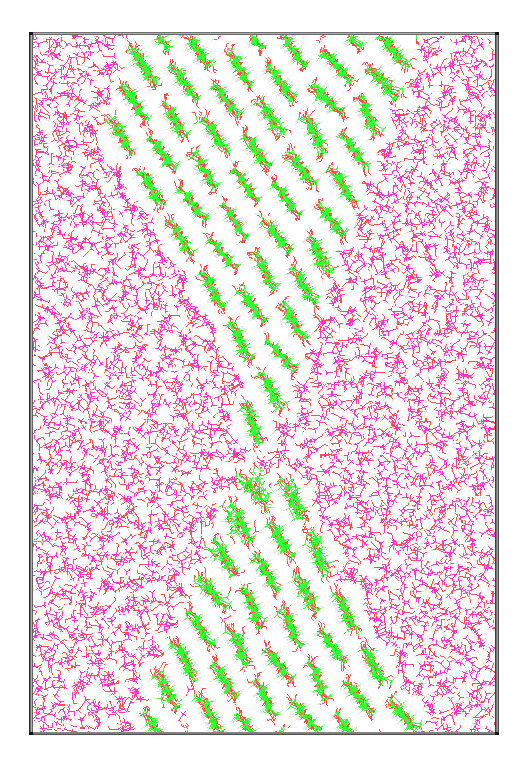}
            \\
            MTA
        \end{minipage}
        \begin{minipage}[b]{0.12\textwidth}
            \centering
            \includegraphics[height=0.93\textwidth]{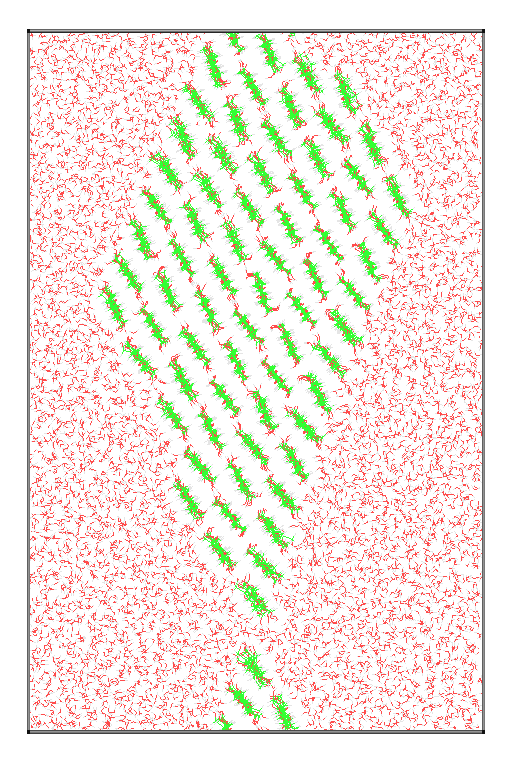}
            \\
            SOL
        \end{minipage}
        \begin{minipage}[b]{0.12\textwidth}
            \centering
            \includegraphics[height=0.93\textwidth]{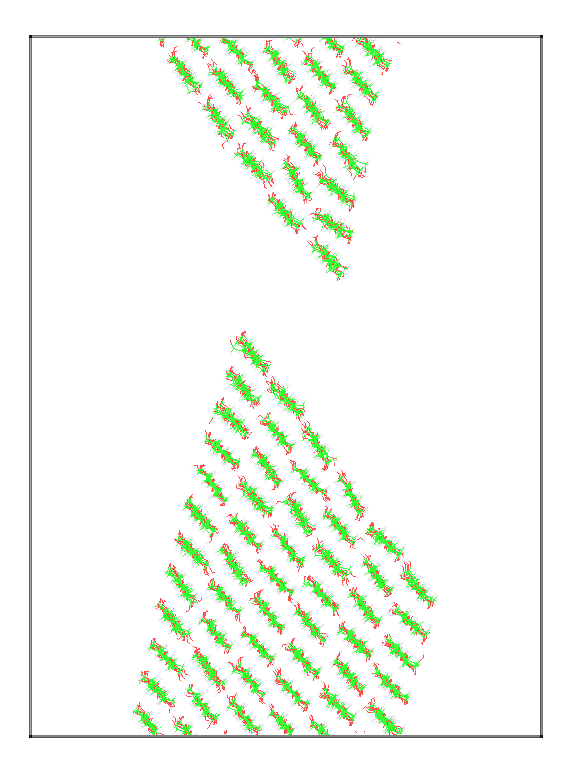}
            \\
            NONE
        \end{minipage}
        \subcaption{}
    \end{subfigure}
    \captionsetup{font=scriptsize}
    \caption{
    Models and stretches of the Type I CNCs surrounded by solvents on both sides.
    (a) Models surrounded by solvents on both sides.
    (b) Intermediate structures.
    (c) Structures after stretching under the same strain.
    Models with DMF, DMSO, EDA, ETA, IPA, MTA, SOL, and a solvent-free (NONE) control were constructed.
    All were simulated under the NPT ensemble in the xyz directions, except for the solvent-free group with only pressure control in the yz directions.
    The results indicated that friction sliding remained important in solvent environments.
    Small polar molecules such as EDA, MTA, and SOL exhibited strong erosive abilities.
    }
    \label{fig:cellulose_nanocrystal_with_solvent_around_fractures}
\end{figure}

Considering the mechanical properties of the CNCs in the previous sections, the CNCs were simulated under fully xyz periodic boundary conditions.
The emphasis of this section is as follows: first, to consider the mechanical properties of CNCs under non-fully periodic boundary conditions; and second, to consider the impact of solvents on the mechanical properties of CNCs.
The solvent surrounded model considered was defined such that the overall model still maintained xyz periodic boundary conditions, but the left and right sides of the nanocrystal were non-continuous and surrounded by solvents.
In this situation, the CNCs were only periodic in the y and z directions, emulating edge exposure to the solvents.
The eight solvent conditions were DMF, DMSO, EDA, ETA, IPA, MTA, SOL, and NONE, and the same vertical strain was applied to these models, as shown in Figure \ref{fig:cellulose_nanocrystal_with_solvent_around_fractures}.
For the solvent-free control group, pressure control was performed in the yz directions only, while all other solvent groups were controlled in the xyz directions.

\begin{figure}[htbp]
    \centering
    \begin{subfigure}[b]{0.48\textwidth}
        \includegraphics[width=0.96\textwidth]{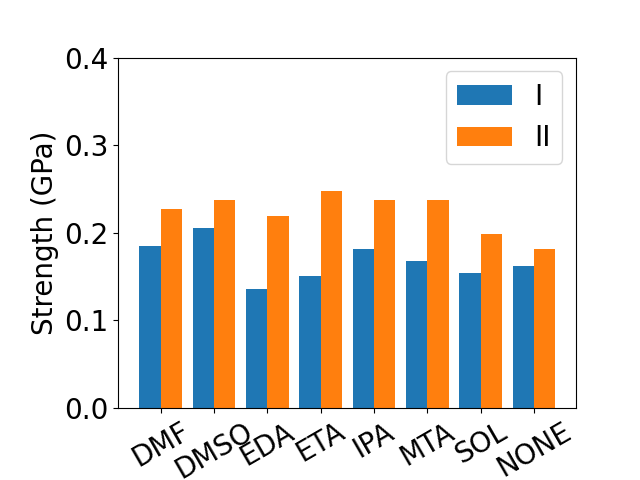}
        \subcaption{}
    \end{subfigure}
    \begin{subfigure}[b]{0.48\textwidth}
        \includegraphics[width=0.96\textwidth]{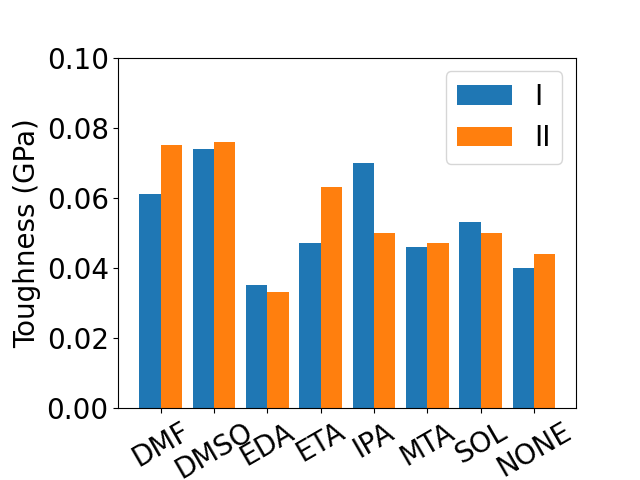}
        \subcaption{}
    \end{subfigure}
    \captionsetup{font=scriptsize}
    \caption{
    Mechanical properties of CNCs surrounded by solvents on both sides.
    (a) Strength and (b) toughness with solvent type and CNC configuration.
    Overall, Type I nanocrystals still exhibited friction sliding and had lower strength in this specific solvent scenario than Type II.
    The restructuring toughening mechanism of Type II is heavily dependent on the ideal crystal structure, which is not presented here.
    Small polar molecules such as ethylenediamine (EDA), methanol (MTA), and water (SOL) exhibit a strong erosive ability, corresponding to the lowest structural toughness.
    Other molecules represented by DMSO, which are slightly larger and less polar, can enhance toughness.
    }
    \label{fig:cellulose_nanocrystal_with_solvent_around_performance}
\end{figure}

The solvent-free control group showed that even without full periodicity, friction sliding remained the primary fracture behavior for Type I and this behavior persisted in the solvent environments.
Although the final morphologies of all the models were similar, by comparing the intermediate structures during loading,
it was observed that small polar molecules (such as EDA, MTA, and SOL) had stronger erosive abilities on the CNCs, promoting the friction sliding at the interfaces.
As a molecule with a relatively large size and weak polarity, DMSO has a weaker erosive ability on the Type I CNCs, and improves the ductility and toughness compared with the other solvents.
The quantitative strength and toughness data are presented in Figure \ref{fig:cellulose_nanocrystal_with_solvent_around_performance}.
These simulation data indicate that the size and polarity of the solvent molecules are key factors that affect the performances and phenomena of Type I CNCs in solvent environments.
This is related to the crucial influences of hydrogen bonding and Coulomb interactions during the stretching of Type I.

\begin{figure}[htbp]
    \centering
    \begin{subfigure}[b]{0.96\textwidth}
        \centering
        \begin{minipage}[b]{0.12\textwidth}
            \centering
            \includegraphics[height=0.60\textwidth]{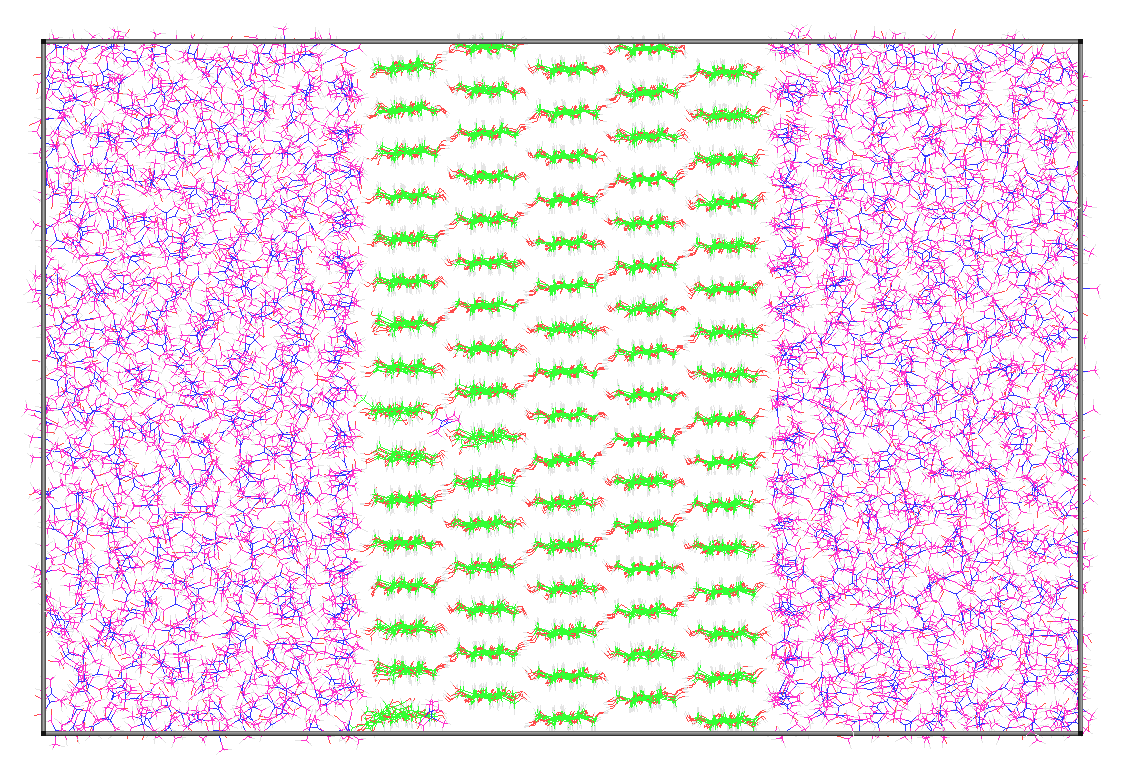}
        \end{minipage}
        \begin{minipage}[b]{0.12\textwidth}
            \centering
            \includegraphics[height=0.60\textwidth]{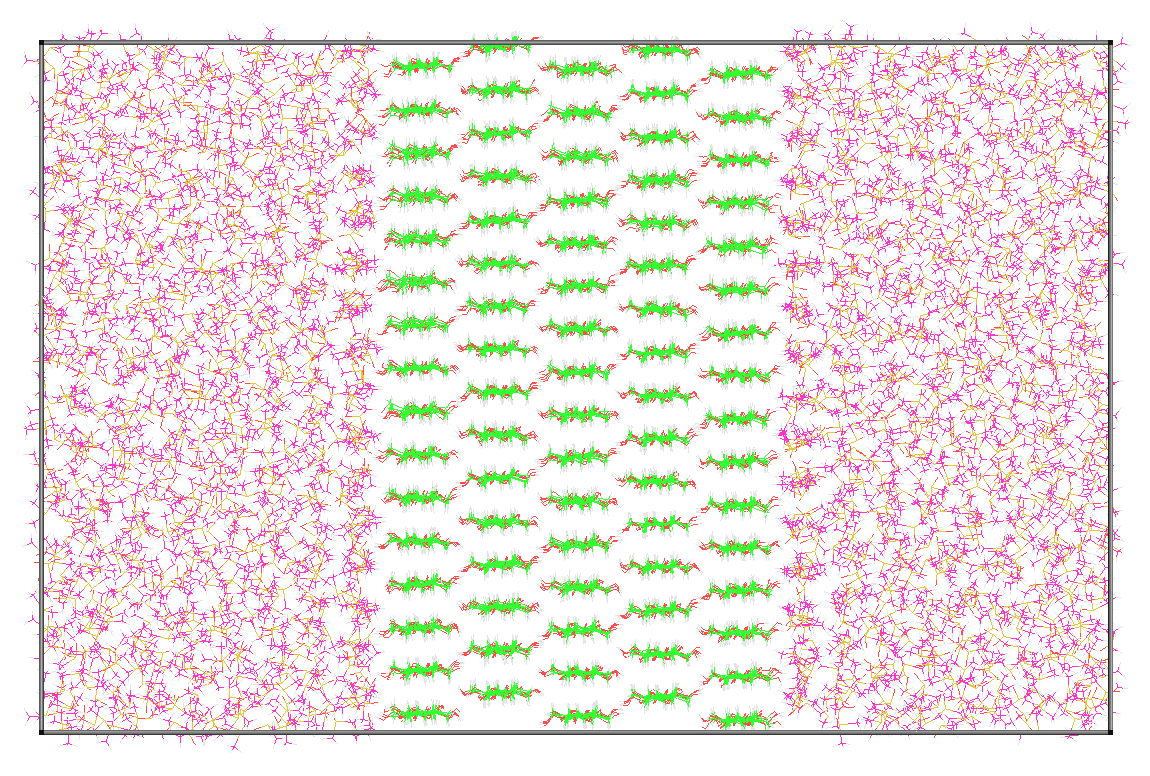}
        \end{minipage}
        \begin{minipage}[b]{0.12\textwidth}
            \centering
            \includegraphics[height=0.60\textwidth]{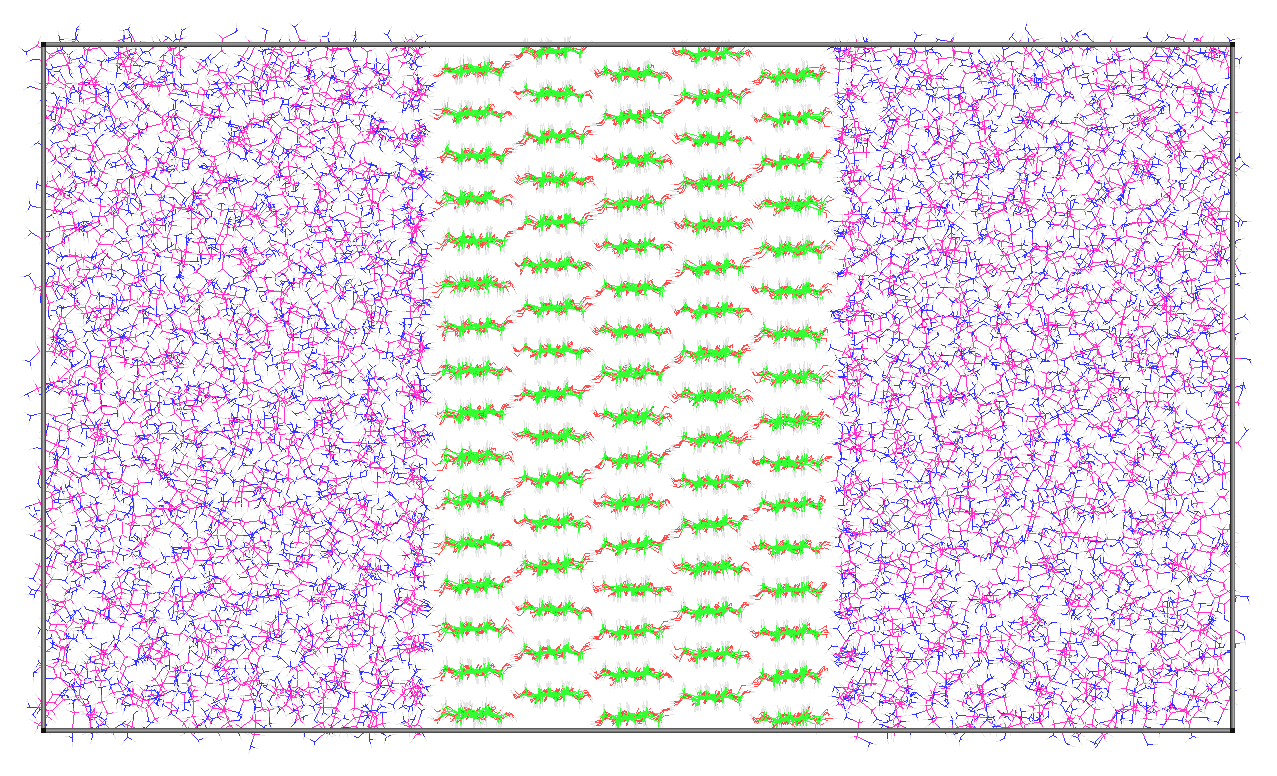}
        \end{minipage}
        \begin{minipage}[b]{0.12\textwidth}
            \centering
            \includegraphics[height=0.60\textwidth]{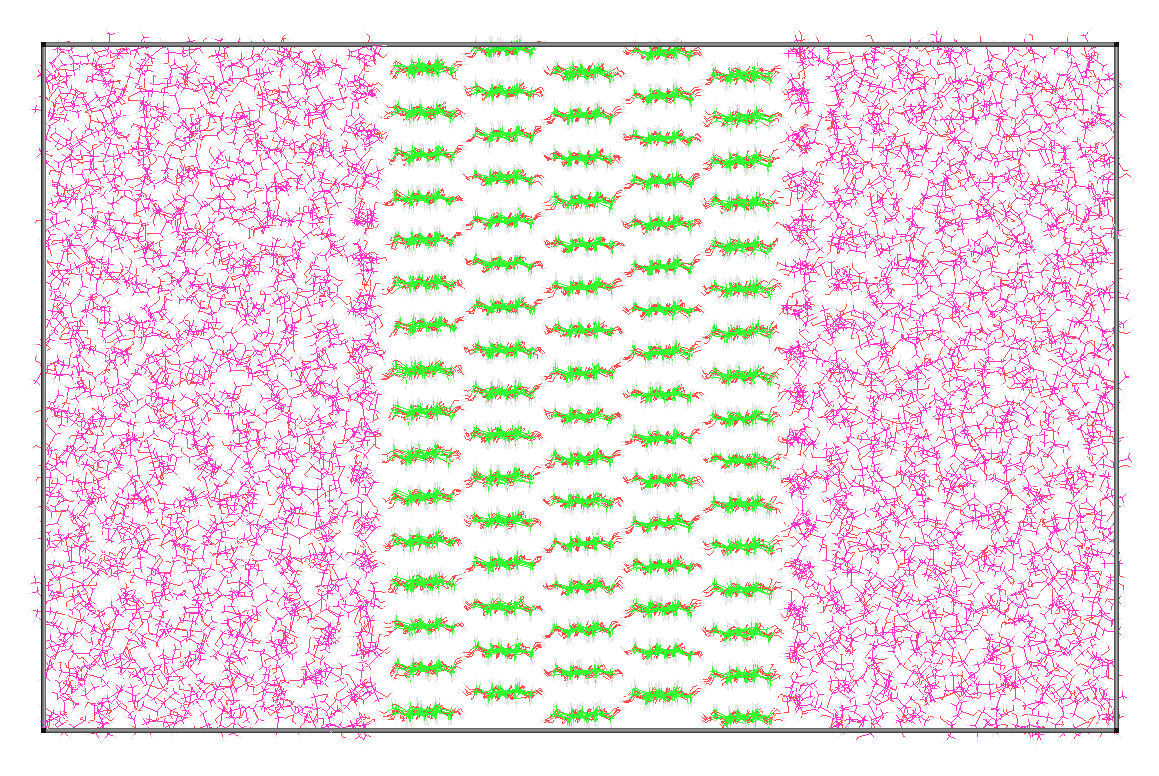}
        \end{minipage}
        \begin{minipage}[b]{0.12\textwidth}
            \centering
            \includegraphics[height=0.60\textwidth]{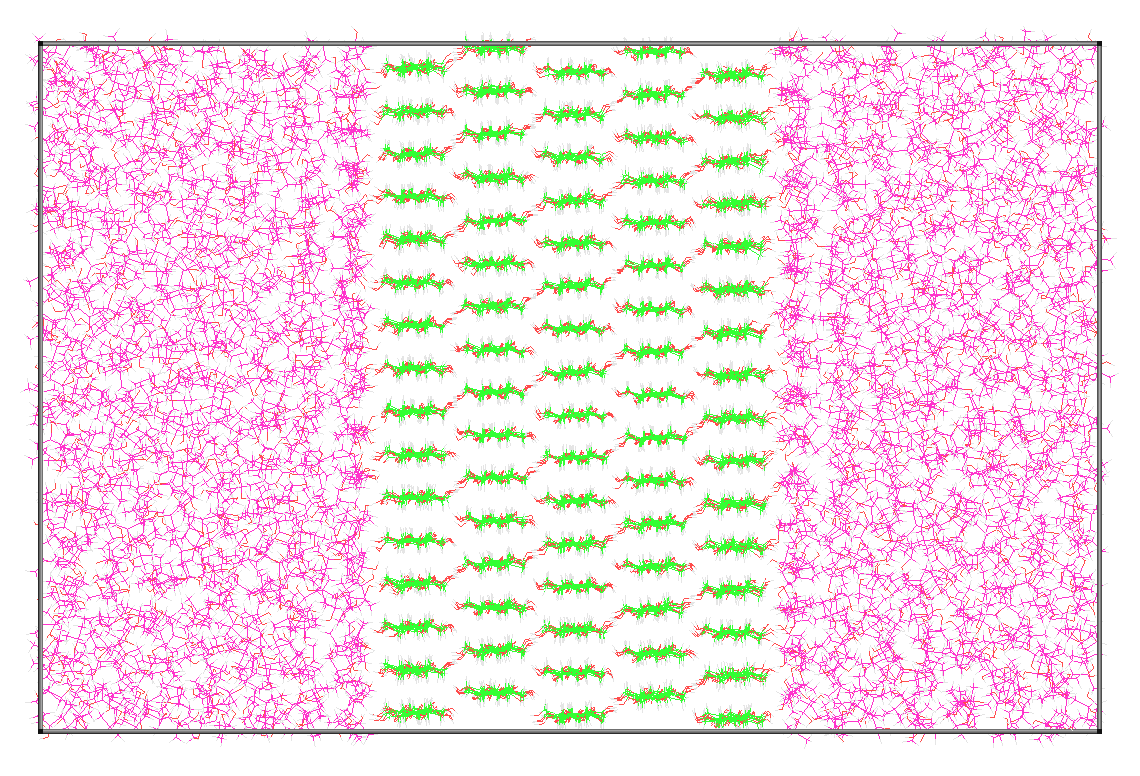}
        \end{minipage}
        \begin{minipage}[b]{0.12\textwidth}
            \centering
            \includegraphics[height=0.60\textwidth]{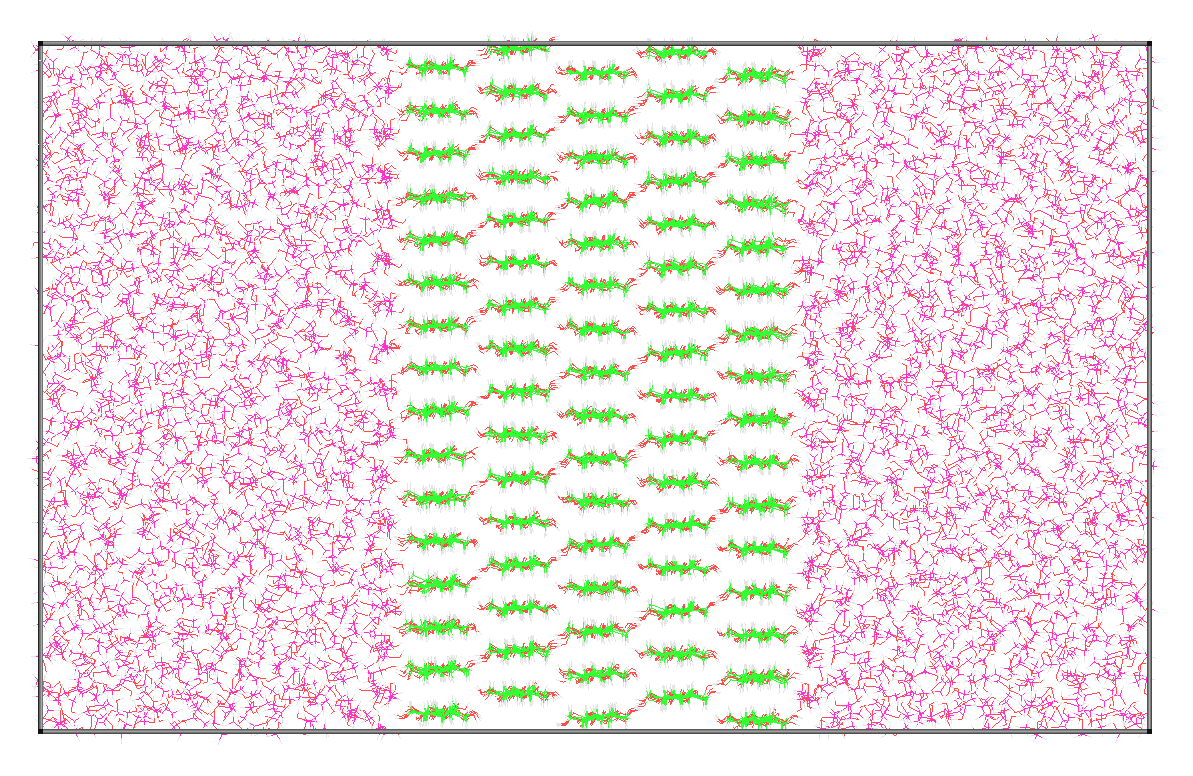}
        \end{minipage}
        \begin{minipage}[b]{0.12\textwidth}
            \centering
            \includegraphics[height=0.60\textwidth]{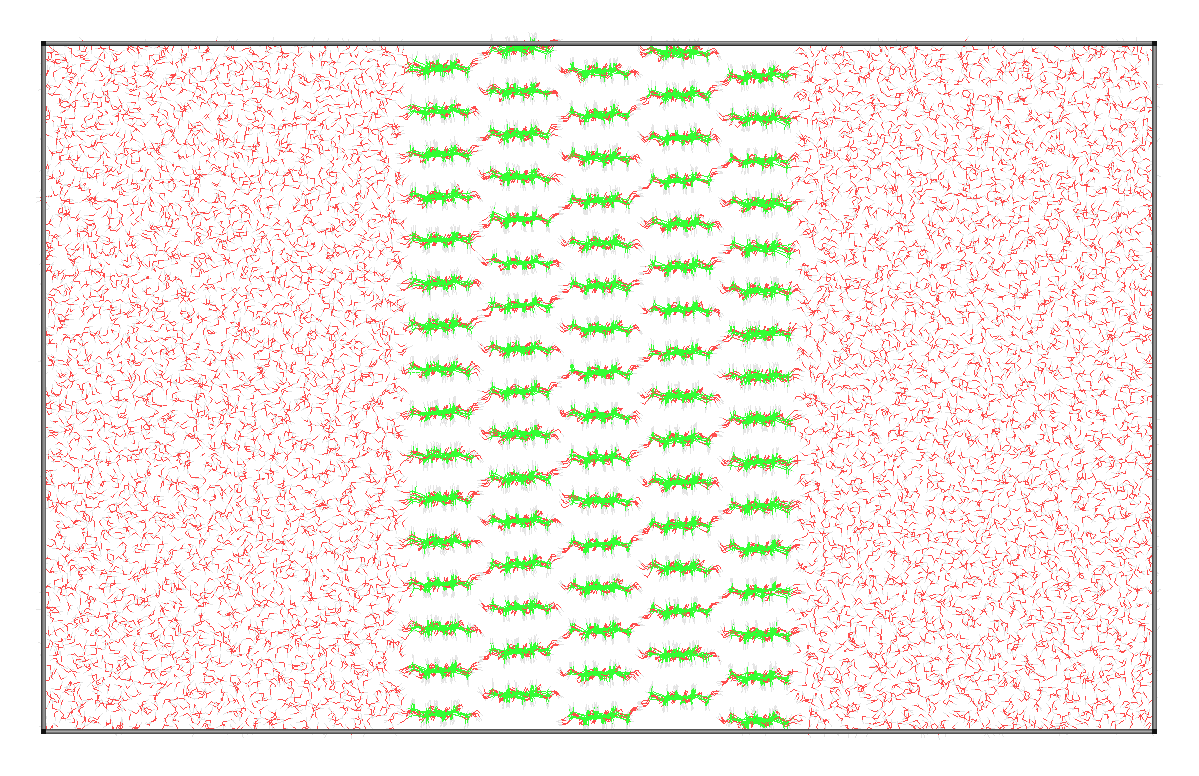}
        \end{minipage}
        \begin{minipage}[b]{0.12\textwidth}
            \centering
            \includegraphics[height=0.60\textwidth]{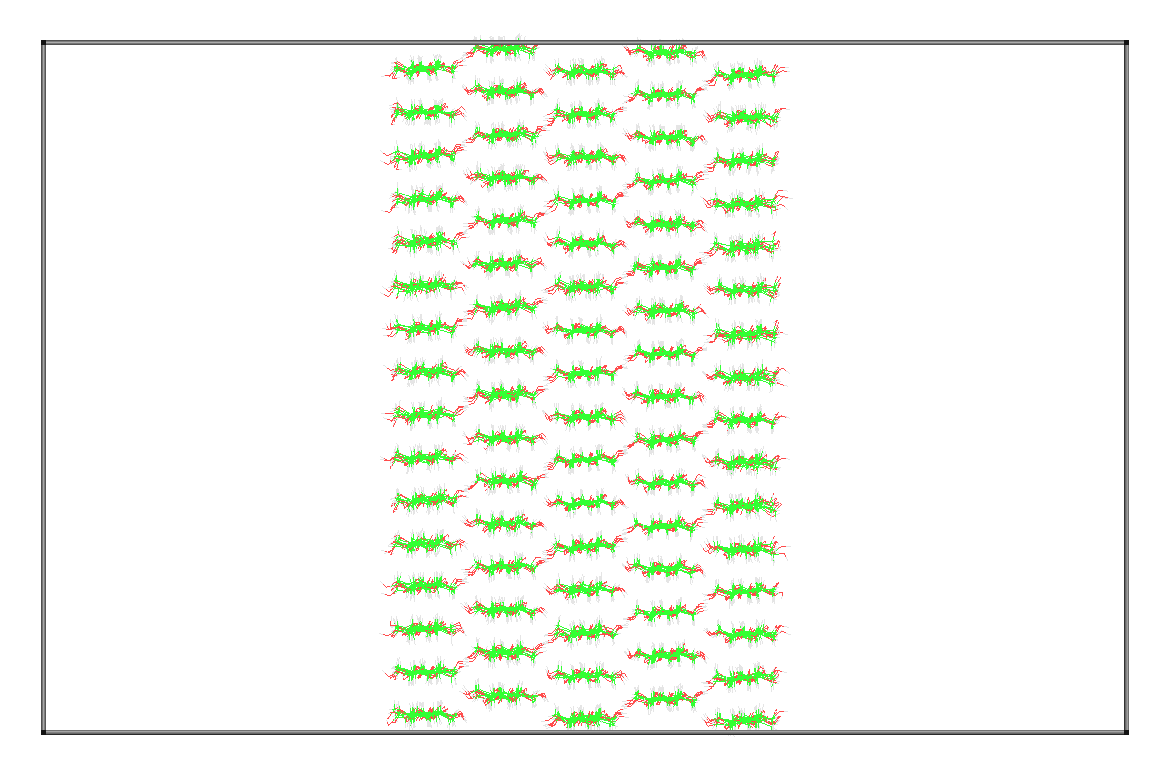}
        \end{minipage}
        \subcaption{}
    \end{subfigure}
    \\
    \begin{subfigure}[b]{0.96\textwidth}
        \centering
        \begin{minipage}[b]{0.12\textwidth}
            \centering
            \includegraphics[height=0.77\textwidth]{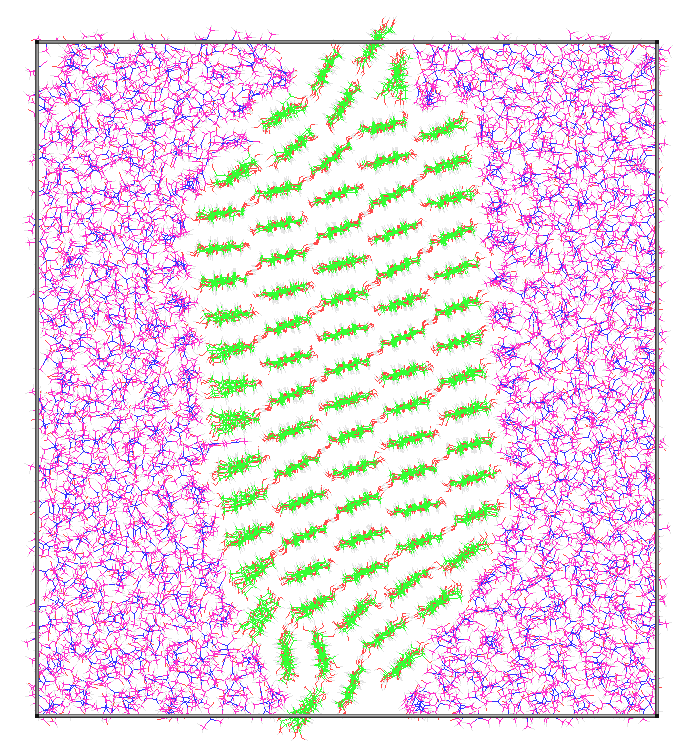}
        \end{minipage}
        \begin{minipage}[b]{0.12\textwidth}
            \centering
            \includegraphics[height=0.77\textwidth]{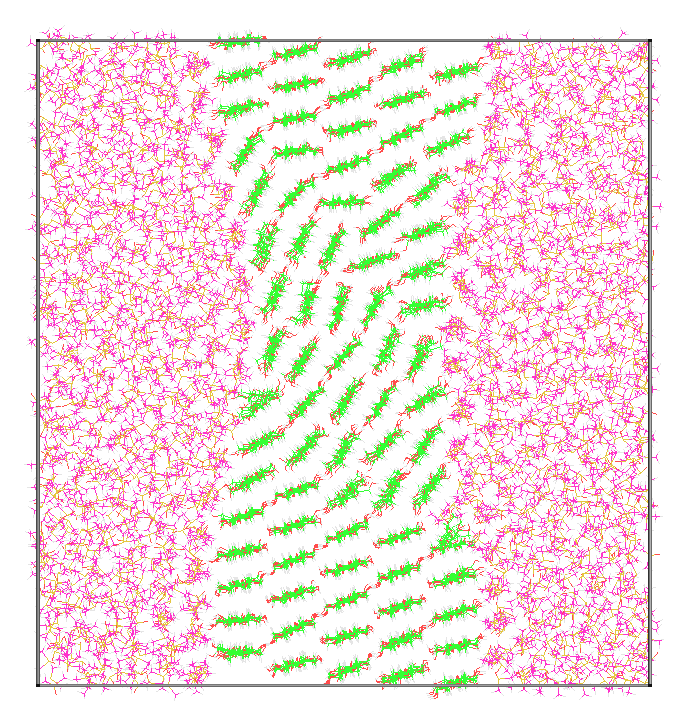}
        \end{minipage}
        \begin{minipage}[b]{0.12\textwidth}
            \centering
            \includegraphics[height=0.77\textwidth]{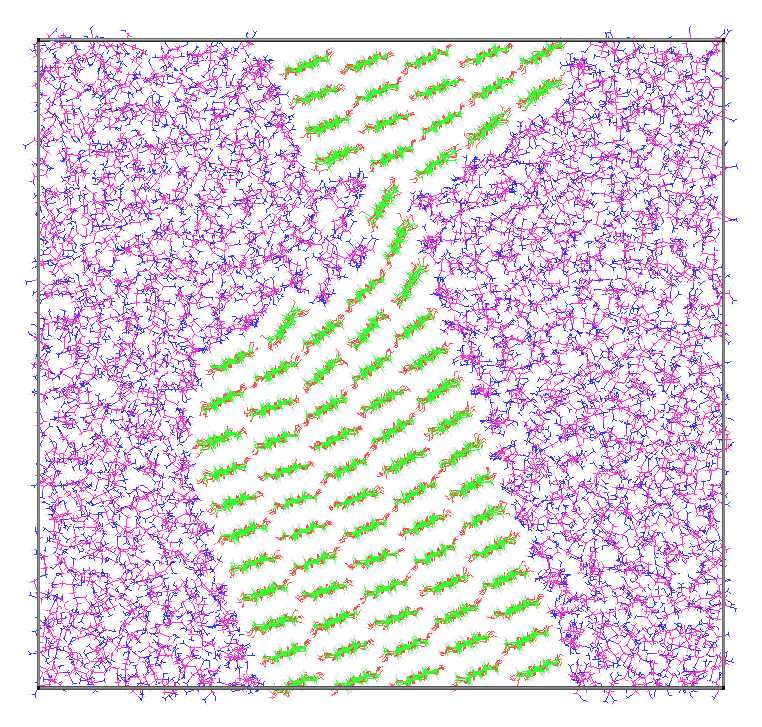}
        \end{minipage}
        \begin{minipage}[b]{0.12\textwidth}
            \centering
            \includegraphics[height=0.77\textwidth]{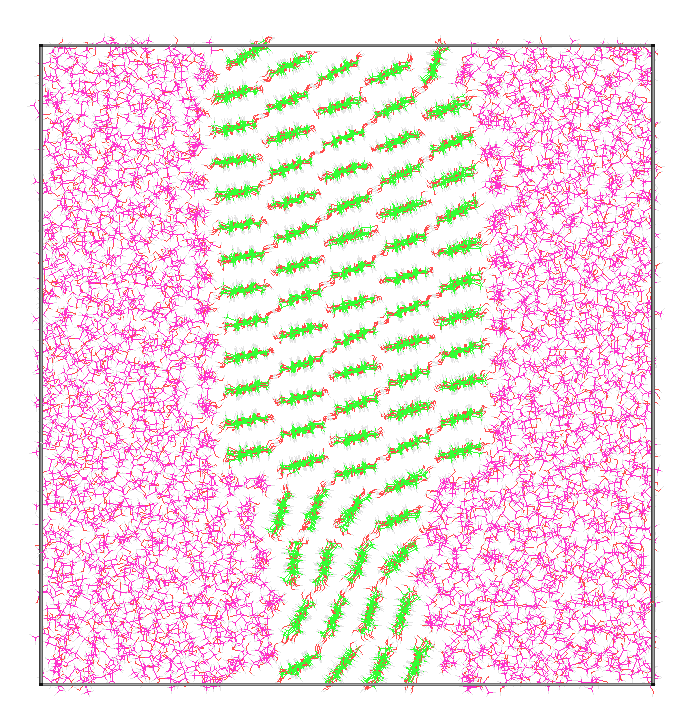}
        \end{minipage}
        \begin{minipage}[b]{0.12\textwidth}
            \centering
            \includegraphics[height=0.77\textwidth]{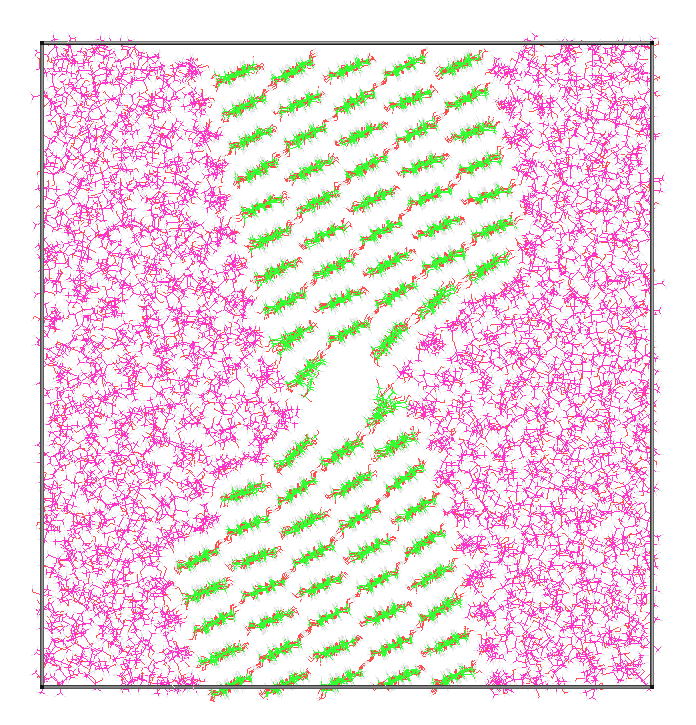}
        \end{minipage}
        \begin{minipage}[b]{0.12\textwidth}
            \centering
            \includegraphics[height=0.77\textwidth]{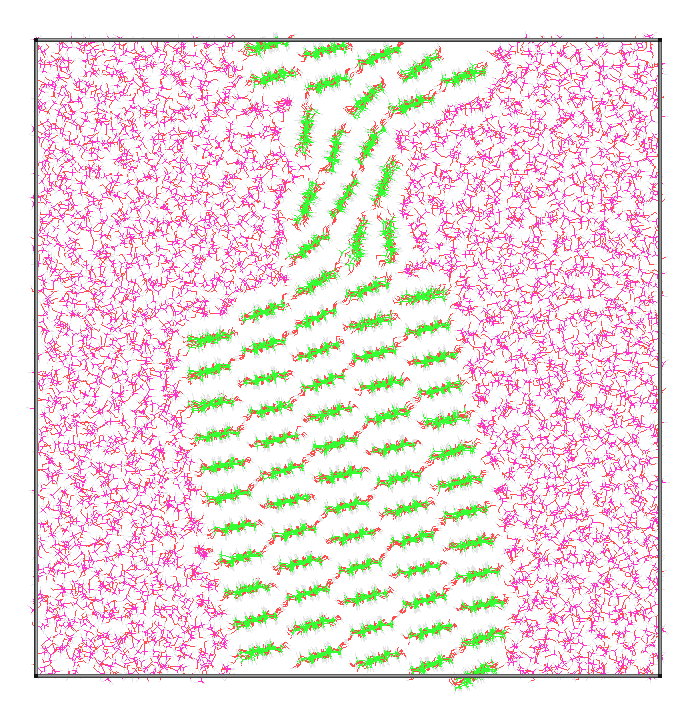}
        \end{minipage}
        \begin{minipage}[b]{0.12\textwidth}
            \centering
            \includegraphics[height=0.77\textwidth]{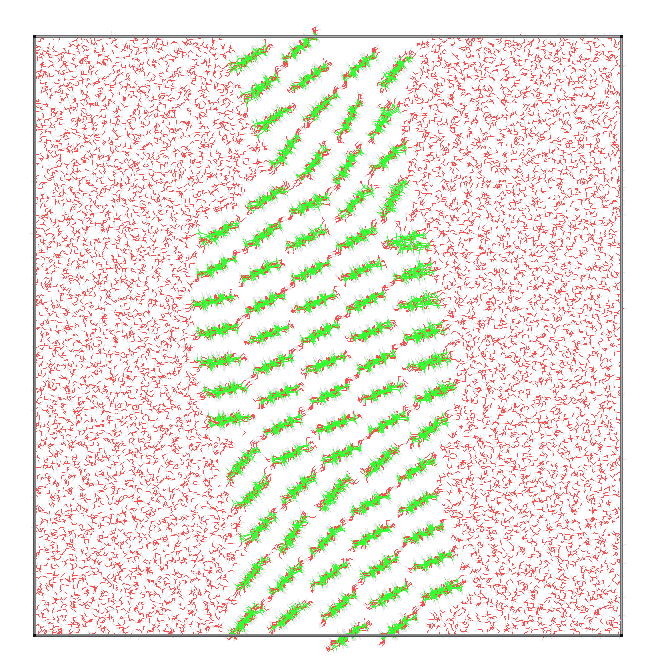}
        \end{minipage}
        \begin{minipage}[b]{0.12\textwidth}
            \centering
            \includegraphics[height=0.77\textwidth]{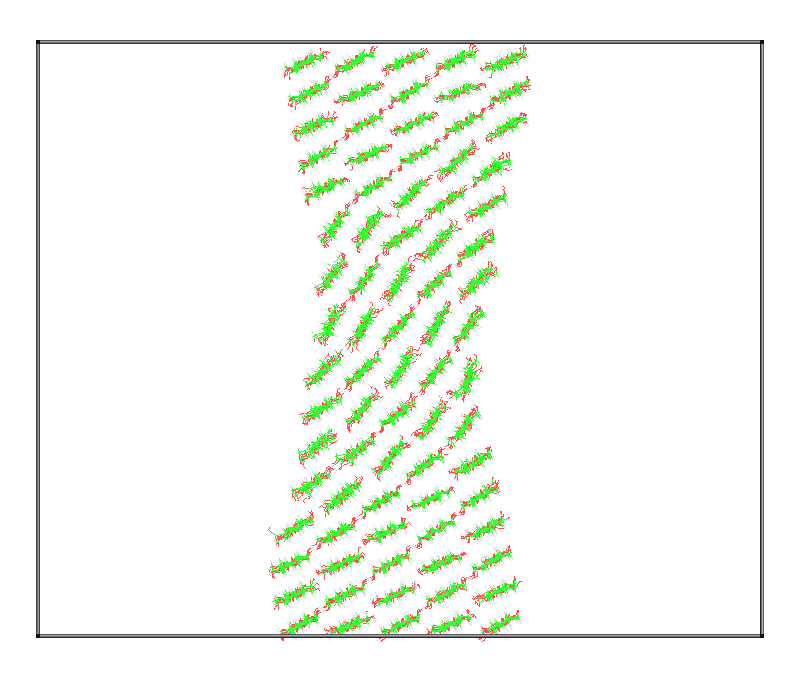}
        \end{minipage}
        \subcaption{}
    \end{subfigure}
    \\
    \begin{subfigure}[b]{0.96\textwidth}
        \centering
        \begin{minipage}[b]{0.12\textwidth}
            \centering
            \includegraphics[height=0.93\textwidth]{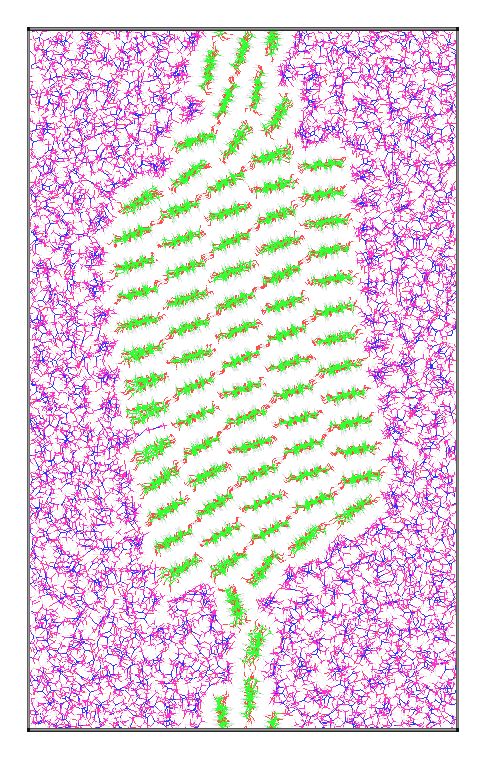}
            \\
            DMF
        \end{minipage}
        \begin{minipage}[b]{0.12\textwidth}
            \centering
            \includegraphics[height=0.93\textwidth]{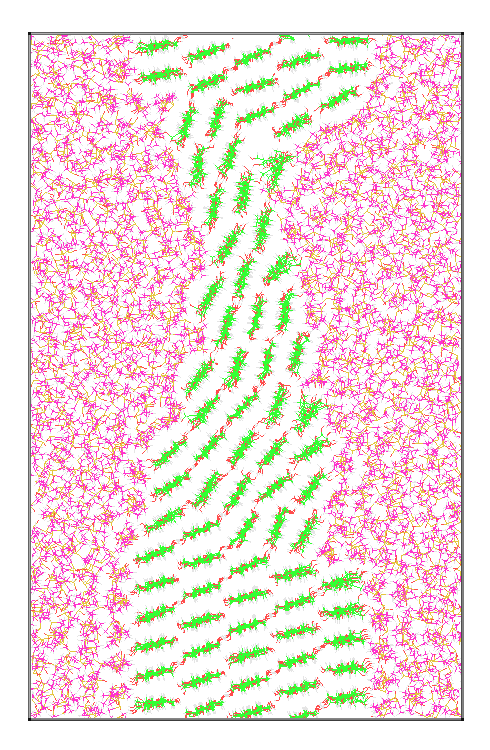}
            \\
            DMSO
        \end{minipage}
        \begin{minipage}[b]{0.12\textwidth}
            \centering
            \includegraphics[height=0.93\textwidth]{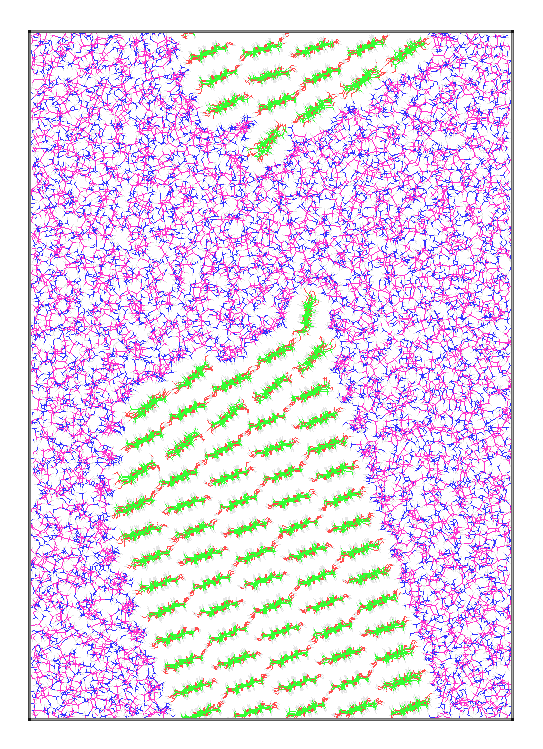}
            \\
            EDA
        \end{minipage}
        \begin{minipage}[b]{0.12\textwidth}
            \centering
            \includegraphics[height=0.93\textwidth]{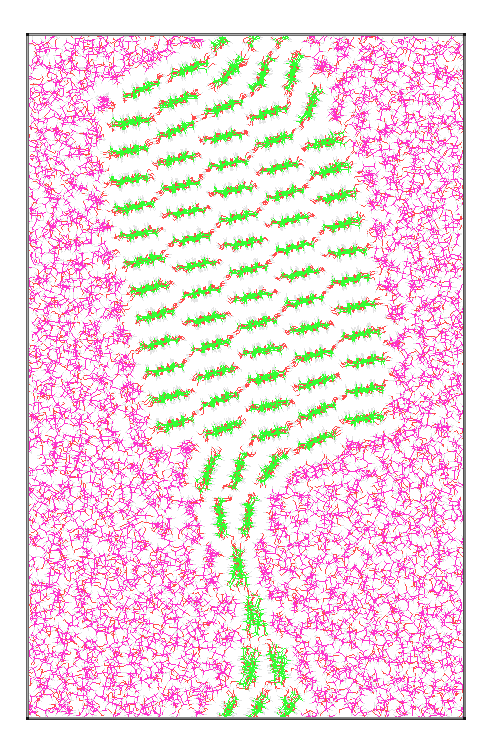}
            \\
            ETA
        \end{minipage}
        \begin{minipage}[b]{0.12\textwidth}
            \centering
            \includegraphics[height=0.93\textwidth]{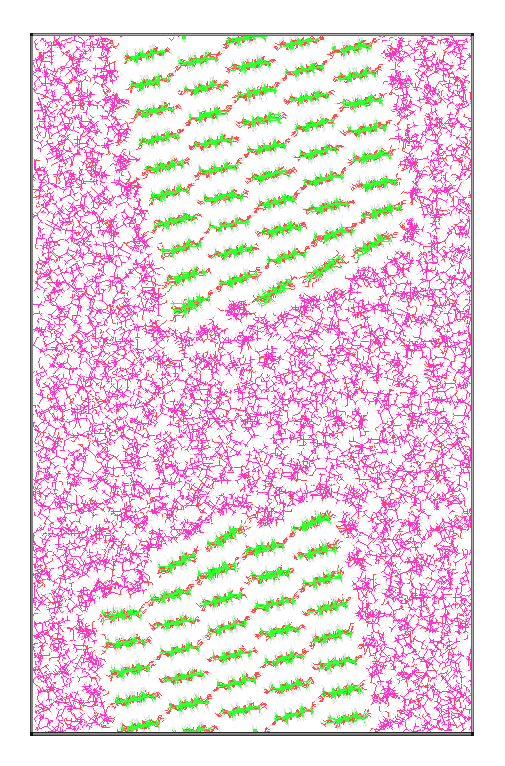}
            \\
            IPA
        \end{minipage}
        \begin{minipage}[b]{0.12\textwidth}
            \centering
            \includegraphics[height=0.93\textwidth]{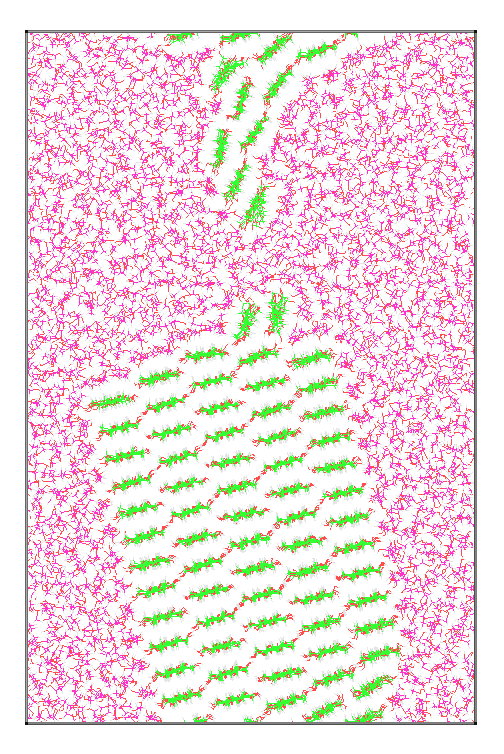}
            \\
            MTA
        \end{minipage}
        \begin{minipage}[b]{0.12\textwidth}
            \centering
            \includegraphics[height=0.93\textwidth]{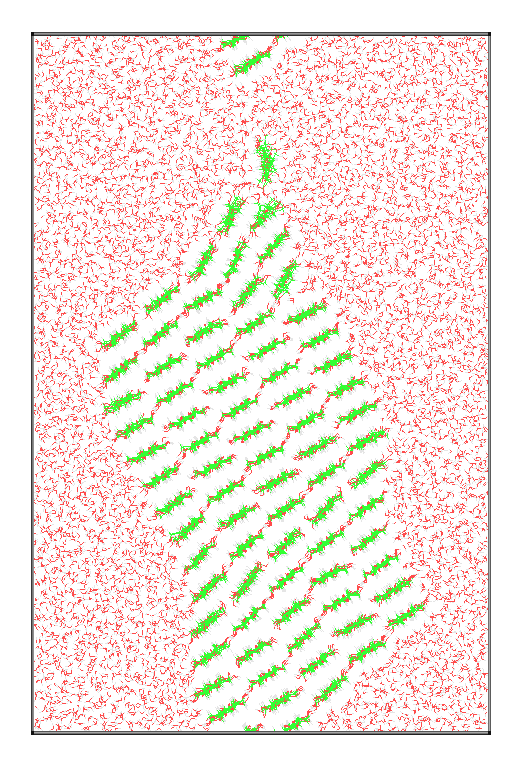}
            \\
            SOL
        \end{minipage}
        \begin{minipage}[b]{0.12\textwidth}
            \centering
            \includegraphics[height=0.93\textwidth]{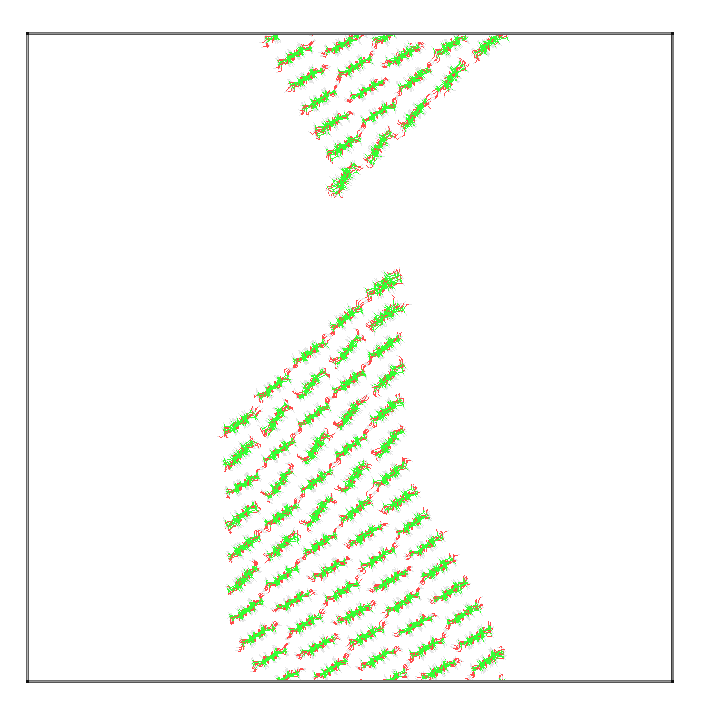}
            \\
            NONE
        \end{minipage}
        \subcaption{}
    \end{subfigure}
    \captionsetup{font=scriptsize}
    \caption{
    Models and stretches of the Type II CNCs surrounded by solvents on both sides.
    (a) Models surrounded by solvents on both sides.
    (b) Intermediate structures.
    (c) Structures after stretching under the same strain.
    Models with DMF, DMSO, EDA, ETA, IPA, MTA, SOL, and a solvent-free (NONE) control were constructed.
    All were simulated under the NPT ensemble in the xyz directions, except for the solvent-free group with only pressure control in the yz directions.
    In solvent environments, the restructuring toughening of non-fully periodic Type II nanocrystals decreased significantly.
    The dependency of the periodicity and crystal structure on the restructuring and toughness of Type II was confirmed.
    Small polar molecules such as EDA, MTA, and SOL exhibit strong erosive ability.
    }
    \label{fig:cellulose_nanocrystal_ii_with_solvent_around_fractures}
\end{figure}

Using the same modelling procedure, pressure control, and stretch speed, the Type II CNCs were simulated under the same treatments (Figure \ref{fig:cellulose_nanocrystal_ii_with_solvent_around_fractures}).
The results of the solvent-free control group shows that in a non-fully periodic situation, Type II cannot undergo complex restructuring or exhibit high toughness.

The intermediate structures and fractures shown in Figure \ref{fig:cellulose_nanocrystal_ii_with_solvent_around_fractures} further illustrate the impact of the solvent polarity and molecule size on the fracture behaviors.
EDA, MTA, SOL, and the slightly larger IPA still exhibited strong erosive ability, capable of promoting horizontal dislocations and fractures of Type II nanocrystals, as shown in the intermediate structures.
Molecules represented by DMSO, which are weaker in polarity and slightly larger in volume, still contributed to an increase in the ductility of Type II CNCs, protecting rather than promoting fractures.
However, the behavior of water differed from that observed for Type I.
As a typical small polar molecule, water did not promote dislocations in Type II; instead, it tended to promote local rotation and overall tension.
This could be related to the greater exposure of hydroxyl groups to these molecules.
Therefore, the local horizontal dislocations were not evident in the intermediate state caused by water, and the final fracture structure exhibited more dislocations.

By quantitatively comparing the mechanical properties of Type I and Type II surrounded by solvents on both sides,
it was noted that small polar molecules such as ethylenediamine (EDA), methanol (MTA), and water (SOL) can reduce the toughness during stretching, whereas molecules with weaker polarity and larger volumes, such as dimethyl sulfoxide (DMSO), can increase the toughness.
When the CNCs were surrounded by the solvent on both sides, the strength of the CNCs was not significantly affected.
The Type I CNCs, influenced by their unique friction sliding, exhibited lower strength compared with Type II.

\section*{Discussion}
\addtocounter{section}{1}
\setcounter{subsection}{0}
\addcontentsline{toc}{section}{\protect\numberline{}Discussion}\indent

As an emerging class of cellulose materials, CNCs are green, widely available, and exhibit excellent performances.
To design advanced cellulose materials, a full understanding of their properties of cellulose materials on a microscopic scale is necessary.
The research topics in this study, anisotropy\cite{wu2014tensile}, fracture behaviors\cite{wu2013atomistic,zhang2021hydrogen}, configuration transformation\cite{chundawat2011restructuring,el2011crystal,kobayashi2011crystal,miyamoto2015molecular,kugo2024elucidating}, size dependency\cite{sinko2014dimensions}, solvent effects\cite{wang2020highly,he2021effects,fink2001structure}, and structural design\cite{tu2021superior,chen2020electric,li2021ionic} of CNCs have been partially addressed in previous studies.
Owing to the stability of the very strong covalent bonds along the chain, this study focused on transverse anisotropy, stressing the three-dimensional structures of CNCs.
This study emphasized the mechanical performance of nanocrystals and further extended these topics.
Systematic comparisons of the characteristic directions and varying solvents were performed.
Based on the anisotropy and size dependency, we propose a toughening mechanism using transverse arrangements.

The extensions included the following points:
1. For configuration, except for those involving configuration transformations, previous studies usually only focused on the Type I CNCs.
However, the Type II CNCs were equally emphasized in this study.
2. For anisotropy and fracture behaviors, this study bypassed the technical difficulty of applying shearing loads in molecular dynamics by constructing specific models for the characteristic directions, and the cell deformation method with full pressure control made it possible to better address the mechanical performance of CNCs.
As a result, the mechanical performance and inter layer frictional sliding phenomenon of Type I slant models were presented, and the possible restructuring of the Type II slant models was also illustrated.
These special fracture behaviors were the basis for the following toughening designs.
For the vertical and horizontal models of the Type I, their reference strengths from previous all atom simulations are 0.9 GPa and 0.5 GPa\cite{wu2014tensile} respectively, and the measured strengths (1.2~GPa and 0.6~GPa) well agreed with them even using completely different force fields.
These simulation strength data are also within the range of the experimental results\cite{diddens2008anisotropic,moon2011cellulose}.
3. For configuration transformations, we found that the structure of Type II CNCs could be induced to a laminar structure similar to that of the Type I only by external mechanical loads.
Considering the configuration transformation requirement for extraction from plants and material performance tuning, this transformation mechanism may be helpful for future studies.
4. For size dependency, we provided a quantitative explanation of the delta potential energy per unit volume.
Inspired by the morphological phenomenon, we realized that the fracture behavior was localized.
As an assumption, the contribution of localized fracture behaviors in larger models should be lower, as proven by the delta potential energy density.
The size dependency curves by Sinko et al.\cite{sinko2014dimensions} also support this conclusion, as they found a saturating phenomenon for the mechanical performance of CNCs, and the 6$\times$6 CNCs exhibit great strength and toughness.
5. For solvent effects, our perspective is unique in terms of periodicity and solvent exposure.
In previous studies, CNCs were usually accompanied and fully surrounded by solvent molecules, and only a few studies have emphasized the wetting (the solvents entered the gaps of chains and became part of the crystals)\cite{malaspina2019molecular,uto2019molecular}.
In this study, the solvents were placed only in the x direction (the left and right sides of the CNCs), while the y and z directions of the CNCs were still periodic.
Therefore, the simulations of solvents concentrated on the solvent-nanocrystal interactions for the mechanical performance of nanocrystals, whether erosive or protective.
The erosive effects of small polar molecules, such as water and EDA, were expected because they could interact with hydroxyl groups using hydrogen bonds and Coulomb interactions and promote dismantling.
However, larger and less polar molecules such as dimethyl sulfoxide (DMSO) exhibit protective phenomena as they can delay the sliding behaviors of both Type I and Type II.
These solvent effects are closely connected to the active hydroxyls and hydrogen bonds of cellulose, which have been widely researched and supported by previous studies\cite{wohlert2022cellulose,zhang2021hydrogen,nishiyama2003crystal,chundawat2011restructuring}.

These points are the only further demonstration of existing studies.
The transverse arrangement toughening patterns represent the primary contribution of this study.
After understanding the inter layer frictional sliding and size dependency of the Type I CNCs, we realized that arranging unit blocks in a non-crystal pattern may help improve the ductility and toughness of the entire structure.
Therefore, symmetric and cross patterns were designed and examined.
They hindered inter layer friction sliding and promoted localized rotations, thereby improved overall ductility.
This toughening mechanism may aid in the future design and utilization of cellulose nanomaterials.

However, this study has significant inherent shortcomings.
First, constrained by computational resources and a lack of experimental verification, the simulations were performed on a relatively small scale, and the applicability of solvent effects and toughening mechanisms may be more limited in practical cellulose materials.
Additionally, the solvent types considered may be insufficient.
On the other hand, although some delicate methods to operate cellulose chains such as cross-linking\cite{tu2021superior} and alignment\cite{chen2020electric,li2021ionic} have already been developed,
achieving precise manipulations in this study experimentally remains a significant fabrication challenge.
Nevertheless, we hope that, this fundamental and pioneering study will be helpful for the further understanding and development of CNCs.

\section*{Conclusions}
\addtocounter{section}{1}
\setcounter{subsection}{0}
\addcontentsline{toc}{section}{\protect\numberline{}Conclusions}\indent

In this study, a series of CNCs models was constructed and simulated to address the anisotropy, size dependency, toughening mechanisms, and solvent effects for both the Type I and Type II CNCs.
To overcome the technical difficulty of applying shearing loads and the well known strong covalent bonds along the chain, models were constructed based on the characteristic directions in the cross section.
The important inter layer frictional sliding of the Type I CNCs was presented, which is attributed to the laminar structure and hydrogen bonds by side chains.
Type II CNCs could exhibit both brittle behavior and restructuring with higher toughness.
The restructuring of Type II led to a laminar fracture similar to that of Type I.
Therefore, it may be exploited as a configuration transformation mechanism purely induced by mechanical loads.
These fracture behaviors were further confirmed by replica simulations at varying loading speeds.
Inspired by previous studies on the size effects on the Type I CNCs, we examined the mechanical performance size dependency of CNCs for both Type I and Type II, and found that small CNCs exhibited better ductility and toughness.
Subsequently, a quantitative explanation for the size dependency was provided by the delta potential energy density, as suggested by the localized fracture behaviors.
After understanding the fracture behaviors and size dependency, symmetric and cross arrangement patterns were proposed and proven for their toughening mechanisms.
We also examined the solvent effects from our own perspective, as the solvent is strictly outside and surrounds the CNCs on both sides, which is a specific load case concentrating on the solvent exposure effects on the mechanical performance of the nanocrystals.
Our results in the solvent environments confirmed that the small and polar molecule can interfere with the side chain hydroxyls and sliding fractures for both Type I and Type II, while the larger and less polar molecules could delay the fractures and protect the nanocrystals.
For limited comparisons, further simulations and experiments can enhance the scope and depth of this study.
With the development of material design and preparation, these results which require delicate manipulations may be practically implemented and will help future cellulose material development.

\section*{Conflict of Interest declaration}
\addtocounter{section}{1}
\setcounter{subsection}{0}
\addcontentsline{toc}{section}{\protect\numberline{}Conflict of Interest declaration}\indent

The author declares no competing financial interest.

\section*{Data availability}
\addtocounter{section}{1}
\setcounter{subsection}{0}
\addcontentsline{toc}{section}{\protect\numberline{}Data availability}\indent

The models, simulation codes, processed data, plot scripts, and latest manuscript are available at

$\textrm{https://github.com/EiPiFun/cll-transverse-data}$

\noindent
. The raw data, pre-processing codes, and post-processing codes are available upon requests.

\section*{Acknowledgments}
\addtocounter{section}{1}
\setcounter{subsection}{0}
\addcontentsline{toc}{section}{\protect\numberline{}Acknowledgments}\indent

Xu Dong thanks the Department of Engineering Mechanics of Zhejiang University for the funding and computational resources.

\section*{Author Contributions}
\addtocounter{section}{1}
\setcounter{subsection}{0}
\addcontentsline{toc}{section}{\protect\numberline{}Author Contributions}\indent

Xu Dong: Conceptualization, Data curation, Formal analysis, Investigation, Methodology, Project administration, Resources, Software, Validation, Visualization, Writing - original draft, Writing - review \& editing.

A professor from the Department of Engineering Mechanics of Zhejiang University provided some additional Methodology, Resources, and Writing - review \& editing helps and chose to be anonymous.
%\\\\
%\noindent\Large{\textbf{References}}
%\newsavebox{\tempbib}
%\savebox{\tempbib}{\parbox{\textwidth}{
%\bibliography{cll_transverse_references}
%}}

\linespread{1.0}\footnotesize\bibliography{cll_transverse_references}
\addtocounter{section}{1}
\setcounter{subsection}{0}
\addcontentsline{toc}{section}{\protect\numberline{}References}

\setcounter{section}{0}
\setcounter{subsection}{0}

\setcounter{table}{0}
\setcounter{figure}{0}
\setcounter{equation}{0}

\renewcommand{\thesection}{S\arabic{section}}
\renewcommand{\thesubsection}{S\arabic{subsection}}

\renewcommand{\thetable}{S\arabic{table}}
\renewcommand{\thefigure}{S\arabic{figure}}
\renewcommand{\theequation}{S\arabic{equation}}

\section*{Supplementary Information}
\addtocounter{section}{1}
\setcounter{subsection}{0}
\addcontentsline{toc}{section}{\protect\numberline{}Supplementary Information}\indent

\normalsize The following subsections and data are included in the Supplementary Information.
\\\indent

Models and relaxations of cellulose nanocrystals Type I

Hydrogen bond number and potential energy during stretching of the Type I

Models and relaxations of cellulose nanocrystals Type II

Hydrogen bond number and potential energy during stretching of the Type II

Stretch transitions of cellulose nanocrystals Type II larger model

Transverse arrangement patterns fractures of the Type I with more unit blocks

Simulations of different size models and size dependency of cellulose nanocrystals Type II

Transverse arrangement patterns of the Type II
\\\indent

\refstepcounter{figure}\label{fig:cellulose_nanocrystal_structure_and_models}\figurename\quad\thefigure,\quad
\refstepcounter{figure}\label{fig:cellulose_nanocrystal_relaxations}\figurename\quad\thefigure,\quad
\refstepcounter{figure}\label{fig:cellulose_nanocrystal_stretch_hydrogen_bond_number_and_energy}\figurename\quad\thefigure,\quad
\refstepcounter{figure}\label{fig:cellulose_nanocrystal_ii_structure_and_models}\figurename\quad\thefigure,\quad

\refstepcounter{figure}\label{fig:cellulose_nanocrystal_ii_relaxations}\figurename\quad\thefigure,\quad
\refstepcounter{figure}\label{fig:cellulose_nanocrystal_ii_larger_stretch_transitions}\figurename\quad\thefigure,\quad
\refstepcounter{figure}\label{fig:cellulose_nanocrystal_ii_stretch_hydrogen_bond_number_and_energy}\figurename\quad\thefigure,\quad
\refstepcounter{figure}\label{fig:cellulose_nanocrystal_more_transverse_arrangement_fractures}\figurename\quad\thefigure,\quad

\refstepcounter{figure}\label{fig:cellulose_nanocrystal_ii_size_dependency}\figurename\quad\thefigure,\quad
\refstepcounter{figure}\label{fig:cellulose_nanocrystal_ii_delta_energy_density_with_size}\figurename\quad\thefigure,\quad
\refstepcounter{figure}\label{fig:cellulose_nanocrystal_ii_transverse_arrangement_fractures}\figurename\quad\thefigure,\quad
\refstepcounter{figure}\label{fig:cellulose_nanocrystal_ii_more_transverse_arrangement_fractures}\figurename\quad\thefigure,\quad
\refstepcounter{figure}\label{fig:cellulose_nanocrystal_ii_transverse_arrangement_performance_with_patterns}\figurename\quad\thefigure\quad

\end{document}